\documentclass[9pt,shortpaper,twoside,web]{ieeecolor}
\usepackage{generic}
\usepackage{amsmath,mathtools,amssymb}
\let\labelindent\relax

\usepackage{algorithm,algcompatible}
\usepackage{cite,cleveref,enumitem}

\usepackage{graphicx}
\usepackage{epstopdf}
\usepackage{caption,subcaption}
\usepackage{comment}
\usepackage{array,setspace,multirow}
\usepackage{cleveref}
\usepackage{xr}

\crefname{app}{Appendix}{Appendices}

\Crefname{figure}{Fig.}{Figs.}
\Crefname{table}{Tab.}{Tabs.}
\Crefname{section}{Sec.}{Secs.}
\crefname{lemma}{Lemma}{Lemmas}

\newtheorem{lemma}{Lemma}

\newcommand{\bs}{\boldsymbol}
\newcommand{\bb}{\mathbb}
\newcommand{\mcal}{\mathcal}
\newcommand{\eye}{\bs{I}}
\newcommand{\zero}{\bs{0}}

\newcommand{\T}{\mathsf{T}}

\newcommand{\lb}{\left( }
\newcommand{\rb}{\right) }
\newcommand{\ls}{\left[}
\newcommand{\rs}{\right]}
\newcommand{\lc}{\left\{}
\newcommand{\rc}{\right\}}
\newcommand{\ld}{\left.}
\newcommand{\rd}{\right.}
\newcommand{\lv}{\left\vert}
\newcommand{\rv}{\right\vert}
\newcommand{\lV}{\left\Vert}
\newcommand{\rV}{\right\Vert}

\allowdisplaybreaks

\def\BibTeX{{\rm B\kern-.05em{\sc i\kern-.025em b}\kern-.08em
    T\kern-.1667em\lower.7ex\hbox{E}\kern-.125emX}}
\title{Joint State and Sparse Input Estimation in Linear Dynamical Systems}

\author{Rupam Kalyan Chakraborty$^{1}$, Geethu Joseph$^{1}$, and Chandra R. Murthy$^{2}$
	\thanks{$^{1}$Faculty of Electrical Engg., Mathematics, and Computer Science, Delft Univ. of Technology, Delft 2628 CD, The Netherlands. Emails:{ \{r.k.chakraborty, g.joseph\}@tudelft.nl}. Most of the work of R.K.C. was conducted while at IISc, Bangalore.}%
	\thanks{$^{2}$Department of Electrical Communication Engg., Indian Institute of Science (IISc), Bangalore 560012, India. 
		Email: {cmurthy@iisc.ac.in}.}%
}

\begin{document}
\maketitle
\begin{abstract}
Sparsity constraints on the control inputs of a linear dynamical system naturally arise in several practical applications 
such as networked control,  computer vision, seismic signal processing, and cyber-physical systems. 
In this work, we consider the problem of jointly estimating the states and sparse inputs of such systems from low-dimensional (compressive) measurements. Due to the low-dimensional measurements, conventional Kalman filtering and smoothing algorithms fail to accurately estimate the states and inputs. We present a Bayesian approach that exploits the input sparsity to significantly improve estimation accuracy. Sparsity in the input estimates is promoted by using different prior distributions on the input. We investigate two main approaches: regularizer-based MAP, and {Bayesian learning-based estimation}. We also extend the approaches to handle control inputs with common support and analyze the time and memory complexities of the presented algorithms. Finally, using numerical simulations, we show that our algorithms outperform the state-of-the-art methods in terms of accuracy and time/memory complexities, especially in the low-dimensional measurement regime. 
\end{abstract}
\vspace{-0.1cm}
\section{Introduction}
\label{eq:sec:intro}
Sparse control of linear dynamical systems (LDSs) has recently gained considerable interest from the control community~\cite{olshevsky2014minimal,siami2020deterministic,joseph2020controllability,
joseph2021controllability,joseph2022output}. This new research area deals with the optimum control of an LDS with sparsity constraints on the control inputs, i.e., the number of nonzero entries in each input vector (also called the number of active input elements) is small compared to the length of each input vector. An LDS with sparse control inputs\footnote{The inputs, when considered as having a sparse representation in a known dictionary, could account for redundancy to handle exigent situations.} models several practical applications such as networked control systems~\cite{olshevsky2014minimal,siami2020deterministic}, opinion dynamics manipulation~\cite{joseph2021ERcontrollability}, computer vision~\cite{bissacco2001recognition,raptis2010spike}, seismic signal processing~\cite{taylor1979deconvolution,o1994recovery}, and cyber-physical systems~\cite{cardenas2008research,slay2007lessons}.  In such applications, an important goal is to jointly estimate the states and sparse inputs of the LDS from its measurements or output. For example, human gait can be modeled as the state of an LDS whose input matrix is an overcomplete dictionary with its columns signifying the different features of the motion. A wide range of motions can be represented by the linear combination of a few columns of the input matrix. Thus, the problem of recognizing different types of human gait reduces to the recovery of states and sparse inputs of the underlying LDS~\cite{bissacco2001recognition}.
Similarly, sparse control inputs are desirable in networked control systems where the controller and plant communicate over a bandwidth-limited channel. Here, sparse control inputs take advantage of their compact representations in suitable bases~\cite{foucart2013math}. Malicious attacks on cyber-physical systems can also be modeled as sparse inputs~\cite{cardenas2008research,slay2007lessons}. Recovery of these sparse attacks is crucial to detecting and mitigating the attacks. 

Motivated by the above applications, this paper focuses on developing joint state and input recovery algorithms for observable LDSs with sparse control inputs. Specifically, we consider a discrete-time LDS, with state transition matrix $\bs{A}_k\in \mathbb{R}^{n\times n}$, input matrix $\bs{B}_k\in \mathbb{R}^{n\times m}$, measurement matrices $\bs{C}_k\in \mathbb{R}^{p\times n}$ and $\bs{D}_k\in \mathbb{R}^{p\times m}$ at  time instant $k$, whose dynamics are governed by
\begin{align}
\bs{x}_{k+1}&=\bs{A}_k\bs{x}_{k}+\bs{B}_k\bs{u}_{k}+\bs{w}_{k}\label{eq:state_eqn} \\
\bs{y}_{k}&=\bs{C}_k\bs{x}_{k}+\bs{D}_k\bs{u}_{k}+\bs{v}_{k}\label{eq:meas_eqn}.
\end{align}
Here, $\bs{u}_k \in \mathbb{R}^{m}$ is the input, $\bs{x}_k \in \mathbb{R}^{n}$ is the state, and $\bs{y}_k \in \mathbb{R}^{p}$ is the measurement at time $k$. Also, $\bs{w}_k$ and $\bs{v}_k $ are the process noise and measurement noise, respectively. We aim to simultaneously estimate the states and sparse inputs $\{\bs{x}_k, \bs{u}_k: \Vert\bs{u}_k\Vert_0\ll n\}_{k=1}^K$ from the low dimensional measurements $\{\bs{y}_k\}_{k=1}^K$ with $p<m$, for a given $K>0$. Here, $\Vert\cdot\Vert_0$ counts the number of nonzero entries in its argument. We note that the states need not be sparse since there are no constraints on the system matrices. We emphasize that the focus of this work is not on the design of sparse control inputs, but rather in  recovering the states and sparse inputs of the system.

Next, we review the existing literature on state and input estimation problems in LDSs and present the specific contributions of this paper.
\subsection{Related Work}
Joint recovery of states and input without assuming any specific structure on the inputs or states has been studied extensively, and several  algorithms exist in the control literature~\cite{friedland1969treatment,kitanidis1987unbiased,gillijns2007unbiased,gillijns2007unbiased2}. However, these algorithms do not account for any underlying sparsity structure that may exist in the system. 
Exploiting sparsity can potentially facilitate the recovery of states and inputs with far fewer measurements than conventional approaches, allowing one to reduce the dimension $p$ of the measurements $\bs{y}_k$. The limited existing studies on sparsity-constrained LDS consider one of three problems: 
1) the recovery of a sparse initial state $\bs{x}_0$; 2) the recovery of a sequence of sparse states $\{\bs{x}_k\}_{k=1}^N$; and 3) the joint recovery of (possibly non-sparse) states and sparse inputs $\{\bs{x}_k, \bs{u}_k\}_{k=1}^N$.

\subsubsection{Estimation of sparse initial state}
Estimating the sparse initial state of an LDS is mathematically equivalent to the standard sparse recovery problem and can be solved using algorithms like LASSO, orthogonal matching pursuit, or sparse Bayesian learning (SBL)~\cite{foucart2013math,joseph2018observability,joseph2019measurement}. It is known that systems that are unobservable using
classical control theory can become observable when the underlying
sparsity is exploited~\cite{dai2013observability,wakin2010observability,sanandaji2014observability, joseph2018observability,joseph2019measurement}. These works focus on estimating the sparse initial state, assuming knowledge of the inputs.

\subsubsection{Estimation of sparse states} 
Sparsity-aware Kalman filtering was proposed in \cite{vaswani2008kalman} to track  abrupt changes in the sequence of \emph{sparse states} of an LDS representing MRI images. This work presented a two-stage recovery algorithm that first estimated the support (indices of the nonzero entries) of the sparse state and then applied Kalman filtering to reconstruct the nonzero values. Further, joint sparse wireless channel estimation and symbol detection was discussed in \cite{prasad2014joint,ali2023}, assuming jointly-sparse states, i.e., they have common support. Here, the sparse states' support and nonzero entries were estimated using an SBL-based  algorithm. Joint-sparse state recovery algorithms were also developed by imposing an $\ell_1$ regularizer in the Kalman smoothing (KS) cost, followed by using the alternating direction method of multipliers (ADMM) method~\cite{angelosante2009compressed}. The recovery of sparse states without the joint sparsity assumption has also been studied via $\ell_1$ regularization-based dynamic filtering~\cite{charles2016dynamic}. Reference \cite{Kurisummoottil2019} derives a variational form of SBL to recover sparse states of an AR(1) process, while \cite{Shaughnessy2020} considers a general nonlinear state space model with linear measurements of sparse states, and develops an algorithm to propagate signal state prediction information from the dynamics model into the gamma hyperpriors of the states, which in turn results in accurate state estimates. However, in all these works, since sparse states are assumed, the system matrices and inputs are restricted to ensure that $\bs{x}_{k+1}=\bs{A}_k\bs{x}_{k}+\bs{B}_k\bs{u}_{k}+\bs{w}_{k}$ remains sparse for all~$k\geq 0$. 

\subsubsection{Joint estimation of states and sparse inputs}
Some works have solved the problem of jointly recovering the state and sparse input sequences as one of $\ell_1$ minimization using convex optimization methods~\cite{sefati2015linear}. The necessary and sufficient conditions for observability of sparse control inputs and the initial state for a noiseless LDS have also been investigated~\cite{poe2023necessary}. Reference \cite{FENG2020} developed a relevance vector machine based sparse Kalman filter and a fixed lag smoother. Similarly, blind deconvolution of a sparse spike sequence was studied using $\ell_1$ regularization~\cite{taylor1979deconvolution,o1994recovery}. However, these $\ell_1$ regularization-based methods involve solving for a large dimensional unknown sparse vector obtained by stacking the state/input vectors and 
do not exploit the temporal correlation, for example, in the state evolution model. Therefore, the resulting algorithms have high computational complexity and memory requirements, and there remains room for performance improvement by exploiting the temporal correlation. 

To address the above-mentioned gaps in the literature, this paper presents new sparsity-driven estimators with significantly better performance and low time and memory complexities. \vspace{-0.2cm}
\subsection{Contributions}
This paper focuses on the problem~3 mentioned above and develops two approaches that impose sparsity on the estimated control inputs using the exponential family of prior distributions. Our specific contributions are as follows:
\begin{itemize}
\item \emph{Regularizer-based approach:} In the first approach, presented in \Cref{sec:reg_RKS}, we integrate the sparsity-promoting priors in the maximum a posteriori (MAP) estimator of the system's state and inputs. We use two priors, leading to $\ell_1$-regularized and reweighted $\ell_2$-regularized optimization problems. Our method relies on Kalman filtering and smoothing, which results in a low-complexity algorithm for estimating the states and sparse inputs.
\item \emph{Bayesian learning-based approach:} Under the second approach, described in \Cref{sec:Bayesian} we use hierarchial Gaussian priors to induce sparsity. We explore two techniques here. In the first technique, we use Gaussian priors with unknown variances (as hyperparameters) to promote sparsity, inspired by the SBL framework. We then rely on the type-II maximum likelihood estimation combined with Kalman smoothing to determine the states and sparse inputs. In the second technique, we develop a variational Bayesian (VB)-based approach to group the prior parameters and unknown states and inputs as unobserved variables, which are inferred via their posterior distributions. 
\item \emph{Comparison and extension:} We analyze and derive the time and memory complexities of both approaches in \Cref{sec:comparison}. Our empirical studies in \Cref{sec:simulation}  indicate that the Bayesian learning-based algorithms outperform the regularizer-based approach. Further, unlike the classical sparse recovery setting, the SBL-based algorithm has lower complexity compared to the regularizer-based approach. Further, we extend the two approaches to the case of jointly sparse control inputs and present a similar analysis.
\end{itemize}
Overall, we present novel approaches for joint state and input estimation, exploiting the sparsity in the control inputs. Unlike the conventional approach, these sparsity-aware algorithms can accurately estimate the states and inputs even in the low-dimensional measurement regime, where the observation matrix $\bs{D}_k$ is column rank deficient. We also note that, in our problem, the input dimension is $m$ and the state dimension is $n$, and the system evolution is observed over $K$ time steps. At each time step, we collect an observation of dimension $p$, resulting in a total of $Kp$ observations. Therefore, we must recover $K(n+m)$ unknowns from $Kp$ equations, where $p \ll n+m$. Without structural assumptions like input sparsity, the problem is highly underdetermined and lacks a unique solution. By exploiting input sparsity using sparse recovery techniques and temporal correlation in the state sequence via KS, we achieve highly accurate state and input recovery, despite the system being underdetermined. The integration of sparse signal recovery techniques with the Kalman smoothing framework is the key innovation of our work. In summary, our paper bridges sparse signal processing and control theory, by employing different variants and techniques of sparse signal recovery within the KS framework and advancing both fields through new algorithmic developments. 
\subsubsection*{Notation} Boldface capital letters denote matrices, and boldface small letters denote vectors. For any vector $\bs{a}$, its $i^{\text{th}}$ entry is denoted by $\bs{a}(i)$ and for any matrix $\bs{A}$, its $(i,j)^{\text{th}}$ entry is denoted by $\bs{A}(i,j)$. $(\bs{A})_{\mathcal{I}}$ indicates the submatrix of $\bs{A}$ formed by choosing the columns of $\bs{A}$ indexed by the set of indices $\mathcal{I}$. The operator $\operatorname{diag}(\bs{a})$ denotes a diagonal matrix with the entries of $\bs{a}$ along the diagonal.  We denote a sequence of vectors $\{\bs{a}_1,\bs{a}_2,\ldots,\bs{a}_N\}$ by $\bs{A}^N_1$.
The $\ell_1$-norm is denoted by $\Vert \cdot\Vert_1$, and $\Vert \cdot \Vert$ denotes the induced $\ell_2$-norm for matrices and Euclidean norm for vectors. If $\bs{P}$ is a nonsingular matrix, we define the $\bs{P}^{-1}$-norm of vector $\bs{x}$ as $\Vert\bs{x}\Vert_{\bs{P}} = \bs{x}^{\T}\bs{P}^{-1}\bs{x}$. The identity matrix, all-zeros matrix (or vector), and all-ones matrix (or vector) are denoted by $\bs{I}$, $\bs{0}$, and $\bs{1}$, respectively. Finally, $\mathbb{R}$ denotes the set of real numbers.

\vspace{-0.2cm}
\section{Preliminaries: MAP Estimation of States and (Non-sparse) Inputs}
\label{sec:RKS}
In this section, we briefly discuss an extension of Kalman filtering and smoothing, which serves as the point of departure for presenting the new algorithms developed in this paper. Specifically, we consider the estimation of the states and inputs $\{\bs{x}_k, \bs{u}_k\}_{k=1}^K$ using measurements $\{\bs{y}_k\}_{k=1}^K$ using a Bayesian framework. In this section, we do not impose sparsity constraints on the inputs. The discussion below is based on the Bayesian approach presented in \cite{fang2013simultaneous, gillijns2007unbiased2}.

The Kalman filtering and smoothing framework is typically based on the following assumptions:
\begin{itemize}[leftmargin=0.8cm]
    \item [\textbf{A1:}] The noise $\bs{w}_k \sim \mcal{N}\lb\zero, \bs{Q}_{k}\rb $ and $\bs{v}_k \sim \mcal{N}\lb\zero, \bs{R}_{k}\rb$ are
independent, where $\bs{Q}_{k}\in\bb{R}^{n\times n}$ and $\bs{R}_{k}\in\bb{R}^{p\times p}$ are positive definite matrices. Here,  $\mathcal{N}(\bs{\mu}, \bs{\Sigma})$ denotes the Gaussian distribution with mean vector $\bs{\mu}$ and covariance matrix $\bs{\Sigma}$. 
    \item [\textbf{A2:}] The input $\bs{u}_k$ is independent of $\bs{x}_1$ and the other inputs\footnote{In general, the control inputs applied to the LDS need not be independent of the initial state. Here, we treat the inputs as unknown parameters independent of the state only for algorithm development.}. 
    \item [\textbf{A3:}] The posterior distribution of the state and input is \begin{equation}
        \hspace{-0.1cm}p\lb\bs{x}_t,\bs{u}_t|\bs{Y}^{k}_1\rb = \mcal{N}\lb\begin{bmatrix}\hat{\bs{x}}_{t|k}\\ \hat{\bs{u}}_{t|k}\end{bmatrix},\begin{bmatrix}
    \bs{P}_{t|k}^{\bs{x}} & \bs{P}_{t|k}^{\bs{xu}}\\
    \lb\bs{P}_{t|k}^{\bs{xu}}\rb^{\T} & \bs{P}_{t|k}^{\bs{u}}\end{bmatrix}\rb.\label{gaussian_approx1}
    \end{equation} 
    Here, $\hat{\bs{x}}_{t|k}\in\bb{R}^n$ and $\hat{\bs{u}}_{t|k}\in\bb{R}^m$ are the estimates of $\bs{x}_t$ and $\bs{u}_t$ given $\bs{Y}^{k}_1$ with associated covariances $\bs{P}_{t|k}^{\bs{x}}\in\bb{R}^{n\times n}$ and $\bs{P}_{t|k}^{\bs{u}}\in\bb{R}^{m\times m}$, respectively, for any $t$ and $k$. Also, $\bs{P}_{t|k}^{\bs{xu}}\in\bb{R}^{n\times m}$ is the cross-covariance of $\bs{x}_t$ and $\bs{u}_t$ given $\bs{Y}_1^k$.
\end{itemize}
Since we consider the non-sparse input case, we also make the following strong assumption:
\begin{itemize}[leftmargin=0.8cm]
    \item [\textbf{A4:}] The input matrices $\bs{D}_k$ have full column rank.
\end{itemize}

Under Assumptions A1-A4, the MAP estimates $\hat{\bs{x}}_{k\mid K},\hat{\bs{u}}_{k\mid K}$ of the states and inputs are computed using the joint distribution of all the states and inputs $\lc\bs{x}_k,\bs{u}_k\rc_{k=1}^K$ given all observations $\bs{Y}_{1}^K$, i.e.,
\begin{align}
\lc \hat{\bs{x}}_{k\mid K},\hat{\bs{u}}_{k\mid K}\rc_{k=1}^K&=\underset{\substack{\bs{x}_{k},\bs{u}_{k}\\k=1,\ldots,K}}{\arg \max}\; p\lb \lc\bs{x}_{k},\bs{u}_{k}\rc_{k=1}^K\mid \bs{Y}^K_1\rb\\
&=\underset{\substack{\bs{x}_{k},\bs{u}_{k}\\k=1,\ldots,K}}{\arg \max}\; \prod_{k=1}^{K}p(\bs{y}_k\mid\bs{x}_k,\bs{u}_k)\notag\\&\hspace{1.5cm}\times p(\bs{x}_k\mid\bs{x}_{k-1},\bs{u}_{k-1}) p(\bs{u}_k),\label{eq:joint_cost}
\end{align}
where we define $\bs{x}_0=\zero$ and $\bs{u}_0=\zero$ and use Assumption A2. We do not make any assumption on the distribution $p(\bs{x}_1\mid\bs{x}_{0},\bs{u}_{0}) =p(\bs{x}_1)$.  Similarly, in the non-sparse inputs setting, we have no prior knowledge of the distribution $p(\bs{u}_k)$. Hence, we drop the dependence of \eqref{eq:joint_cost} on $p(\bs{x}_1)$ and $p(\bs{u}_k)$, which corresponds to imposing (an improper) uniform prior. 
So, from \eqref{eq:state_eqn} and \eqref{eq:meas_eqn} and Assumption A1, \eqref{eq:joint_cost} reduces to
\begin{multline}
\lc \hat{\bs{x}}_{k\mid K},\hat{\bs{u}}_{k\mid K}\rc_{k=1}^K\!\!=\!\underset{ \substack{\bs{x}_{k},\bs{u}_{k}\\k=1,\ldots,K}}{\arg {\min}}\; \sum_{k=1}^{K}\lV\bs{y}_k-\bs{C}_k\bs{x}_k-\bs{D}_k\bs{u}_k\rV_{\bs{R}_k}^2\\+\sum_{k=1}^{K-1}\lV\bs{x}_{k+1}-\bs{A}_{k}\bs{x}_{k}-\bs{B}_{k}\bs{u}_{k}\rV_{\bs{Q}_k}^2. \label{eq:RKS_cost}
\end{multline}
Similar to the traditional Kalman smoothing, the above optimization problem can be solved recursively using the prediction, filtering, and smoothing steps. This algorithm is referred to as the robust Kalman smoothing (RKS), and its pseudocode is given in \Cref{alg:RobustKalman}. An outline of the derivation of the algorithm is presented next.

\begin{algorithm}
	\caption{Robust Kalman Smoothing}
	\setstretch{1.2}
	\begin{algorithmic}[1]
		\REQUIRE $\{\bs{y}_k, \bs{A}_k,\bs{B}_k,\bs{C}_k,\bs{D}_k,\bs{Q}_k,\bs{R}_k\}_{k=1}^K$
		\STATEx \hspace{-0.5cm}\textbf{Initialization:} $\hat{\bs{x}}_{0|0}=\zero,\hat{\bs{u}}_{0|0}=\zero$, and $\bs{P}^{\bs{\xi}}_{0\mid 0}$ and $\bs{Q}_0=\zero$
		\STATE $\tilde{\bs{A}}_k=\begin{bmatrix}
\bs{A}_k & \bs{B}_k\end{bmatrix}$ and $\bs{T}= \begin{bmatrix}\bs{I} & \bs{0}\end{bmatrix}^{\T}\in\bb{R}^{(n+m)\times n}$
		\FOR{$k=1,2,\ldots,K$}
		\STATEx \#\emph{Prediction:}
		\STATE $\hat{\bs{x}}_{k\mid k-1}=\tilde{\bs{A}}_{k-1}\hat{\bs{\xi}}_{k-1\mid k-1}$
		\STATE $\bs{P}^{\bs{x}}_{k\mid k-1}=\tilde{\bs{A}}_{k-1}
\bs{P}^{\bs{\xi}}_{k-1\mid k-1}\tilde{\bs{A}}_{k-1}^{\T}+\bs{Q}_{k-1}$
		\STATEx \#\emph{Filtering:}
		\STATE Compute matrices $\bs{J}_k$ and $\bs{L}_k$ using \eqref{eq:J_defn} and \eqref{eq:L_defn}.
		\STATE Compute the gain matrix $\bs{G}_k$ using \eqref{eq:G_defn}.
		\STATE $
		\hat{\bs{\xi}}_{k\mid k}=\bs{T}
		\hat{\bs{x}}_{k\mid k-1}+\bs{G}_k\lb \bs{y}_k-\bs{C}_k\hat{\bs{x}}_{k\mid k-1}\rb$
        \STATE $\bs{P}^{\bs{\xi}}_{k\mid k}\!
=\!\lb\bs{T}\!-\!\bs{G}_k\bs{C}_k\rb\!\!\bs{P}^{\bs{x}}_{k\mid k-1}\!\lb\bs{T}\!-\!\bs{G}_k\bs{C}_k\rb\!^{\T}\!+\bs{G}_k\bs{R}_k\bs{G}_k^{\T}$
\STATE Compute $\hat{\bs{x}}_{k\mid k}$ and $\bs{P}_{k\mid k}^{\bs{x}}$  using \eqref{eq:xi_defn} from $\hat{\bs{\xi}}_{k\mid k}$ and $\bs{P}_{k\mid k}^{\bs{\xi}}$
		\ENDFOR
		\STATEx \#\emph{Smoothing:}
		\FOR{$k=K-1,K-2,\ldots,1$}
		\STATE $\bs{K}_k=\bs{P}_{k\mid k}^{\bs{\xi}}\tilde{\bs{A}}_k^{\T}\lb\bs{P}_{k+1\mid k}^{\bs{x}}\rb^{-1}$
		\STATE $\bs{P}_{k\mid K}^{\bs{\xi}}=\bs{P}_{k\mid k}^{\bs{\xi}}+\bs{K}_k\lb\bs{P}_{k+1\mid K}^{\bs{x}}-\bs{P}_{k+1\mid k}^{\bs{x}}\rb\bs{K}_k^{\T}$
        \STATE $\hat{\bs{\xi}}_{k\mid K}=\hat{\bs{\xi}}_{k\mid k}+\bs{K}_k\lb\hat{\bs{x}}_{k+1\mid K}-\tilde{\bs{A}}_k\hat{\bs{\xi}}_{k\mid k}\rb$
        \STATE Compute $\hat{\bs{x}}_{k\mid K}$ and $\bs{P}_{k\mid K}^{\bs{x}}$ using \eqref{eq:xi_defn} from $\hat{\bs{\xi}}_{k\mid K}$ and $\bs{P}_{k\mid K}^{\bs{\xi}}$
		\ENDFOR
		\ENSURE $\{\hat{\bs{x}}_{k \mid K}\}_{k=1}^K$ and $\{\hat{\bs{u}}_{k \mid K}\}_{k=1}^K$
	\end{algorithmic}
	\label{alg:RobustKalman}
\end{algorithm}

In the prediction step, we compute the mean and covariance of $\bs{x}_{k}$ and $\bs{u}_{k}$ given measurements $\bs{Y}_1^{k-1}$. 
From Assumptions A1 and A3 and \eqref{eq:state_eqn}, it is straightforward to derive the Kalman filter-type prediction steps given by Steps 3 and 4 of \Cref{alg:RobustKalman}. 

In the filtering step, we determine the statistics of the joint Gaussian posterior distribution of $\bs{x}_{k}$ and $\bs{u}_{k}$ given $\bs{Y}_1^k$. For brevity, we define 
\begin{equation}\label{eq:xi_defn}
    \bs{\xi}_k = \begin{bmatrix}\bs{x}_{k}^{\T} & \bs{u}_{k}^{\T}\end{bmatrix}^{\T}.
\end{equation}
Then, $p\lb\bs{\xi}_t|\bs{Y}^{k}_1\rb\sim\mcal{N}\lb\hat{\bs{\xi}}_{t|k},
    \bs{P}_{t|k}^{\bs{\xi}}\rb$, where $\hat{\bs{\xi}}_{t|k}$ and $\bs{P}_{t|k}^{\bs{\xi}}$ are the statistics of the distribution in Assumption A3. To compute the MAP estimate of the posterior distribution,  the filtering step maximizes $p(\bs{\xi}_{k} \mid \bs{Y}_1^{k})=p(\bs{x}_{k},\bs{u}_{k} \mid \bs{Y}_1^{k})$. Since $p(\bs{y}_{k} \mid \bs{Y}_1^{k-1})$ is independent of $\bs{\xi}_k$  and from Assumption A2, we derive
\begin{equation}
\hat{\bs{\xi}}_{k\mid k}\!=\!\underset{\bs{\xi}_k: \bs{\xi}_k^{\T}= [\bs{x}_{k}^{\T}  \bs{u}_{k}^{\T}]}{\arg \max}\;p\lb\bs{y}_{k} \mid \bs{x}_{k},\bs{u}_{k}\rb p\lb\bs{x}_{k} \mid \bs{Y}_1^{k-1}\rb p\lb\bs{u}_{k}\rb,\label{eq:RKS_filter_inter1}
\end{equation}
Using Assumptions A1 and A3, the above optimization simplifies as
\begin{equation}
\underset{\bs{\xi}_k: \bs{\xi}_k^{\T}= [\bs{x}_{k}^{\T} \; \bs{u}_{k}^{\T}]}{\arg \min }\!\! \mathcal{L}=\lV\bs{y}_k\!-\!\bs{C}_k\bs{x}_k\!-\!\bs{D}_k\!\bs{u}_k\rV_{\bs{R}_k}^2\!\!+\lV\bs{x}_k\!-\!\hat{\bs{x}}_{k\mid k-1}\!\rV_{\bs{P}_{k\mid k-1}^{\bs{x}}}^2\!\!\!\!\!.\label{eq:xu_min_filtering}
\end{equation}
We note that \eqref{eq:xu_min_filtering} does not include any term corresponding to $p(\bs{u}_k)$ because we do not assume any prior knowledge of this distribution. Similar to the derivation of the Kalman filter, we set the gradient of the cost function with respect to $\bs{x}_k$ and $\bs{u}_k$ to zero to obtain the MAP estimates. Step 7 of \Cref{alg:RobustKalman} summarizes the derived filter updates. Also, we define the gain matrix in the update equations as
\begin{equation}
\bs{G}_k\triangleq\begin{bmatrix}
(\bs{I}-\bs{L}_k\bs{D}_k\bs{J}_k\bs{C}_k)^{-1}\bs{L}_k(\bs{I}-\bs{D}_k\bs{J}_k)\\
(\bs{I}-\bs{J}_k\bs{C}_k\bs{L}_k\bs{D}_k)^{-1}\bs{J}_k(\bs{I}-\bs{C}_k\bs{L}_k)
\end{bmatrix}\in\bb{R}^{(n+m)\times p},\label{eq:G_defn}
\end{equation}
where matrices $\bs{J}_k\in\bb{R}^{m\times p}$ and $\bs{L}_k\in\bb{R}^{n\times p}$ are
\begin{align}
\bs{J}_k&\triangleq\lb\bs{D}_k^{\T}\bs{R}_k^{-1}\bs{D}_k\rb^{-1}\bs{D}_k^{\T}\bs{R}_k^{-1}\label{eq:J_defn}\\
\bs{L}_k&\triangleq\bs{P}_{k\mid k-1}^{\bs{x}}\bs{C}_k^{\T}\lb\bs{R}_k+\bs{C}_k\bs{P}_{k\mid k-1}^{\bs{x}}\bs{C}_k^{\T}\rb^{-1}.\label{eq:L_defn}
\end{align}
Here, we use Assumption A4 to ensure the invertibility of matrix $\bs{D}_k^{\T}\bs{R}_k^{-1}\bs{D}_k$. 
Further, we substitute $\hat{\bs{\xi}}_{k\mid k}$ from Step 7 of \Cref{alg:RobustKalman} in the covariance computation using
\begin{equation}\label{eq:cov_exp}
\bs{P}_{k|k}^{\bs{\xi}} =    \mathbb{E}\lc \lb
\bs{\xi}_{k}-\hat{\bs{\xi}}_{k\mid k} \rb\lb
\bs{\xi}_{k}-\hat{\bs{\xi}}_{k\mid k}\rb^\T\rc,
\end{equation}
to derive Step 8 of \Cref{alg:RobustKalman}. Here, we use the facts that $\bs{J}_k\bs{D}_k=\bs{I}$ and $
\bs{G}_k\bs{D}_k=\begin{bmatrix}
\zero&\bs{I}
\end{bmatrix}^{\T}$. 

Finally, in the smoothing step, to estimate $\hat{\bs{\xi}}_{k\mid K}$ and $\bs{P}_{k\mid K}^{\bs{\xi}}$, we start with the joint posterior distribution of $\bs{\xi}_{k+1}$ and $\bs{\xi}_{k}$ given $\bs{Y}_1^K$. Using Assumption A2, we have
\begin{align}
\log p\lb\bs{\xi}_{k+1}, \bs{\xi}_{k} \mid\bs{Y}_1^{K}\rb\notag\\
&\hspace{-3.15cm}=\log p\lb\bs{x}_{k+1} \mid \bs{x}_{k}, \bs{u}_{k}\rb+\log p\lb\bs{x}_{k}, \bs{u}_{k} \mid\bs{Y}_1^{k}\rb\notag\\
&\hspace{-2.2cm}-\log p\lb\bs{x}_{k+1} \mid\bs{Y}_1^{k}\rb+\log p\lb\bs{x}_{k+1}, \bs{u}_{k+1} \mid\bs{Y}_1^{K}\rb\\
&\hspace{-3.15cm}\propto-\frac{1}{2} \bs{\xi}_{k+1}^{\T}\!\lc\! \bs{T}\ls\bs{Q}_k^{-1}\!-\!\lb\bs{P}_{k+1\mid k}^{\bs{x}}\rb^{-1}\rs\bs{T}^{\T}\!+\!\lb\bs{P}_{k+1\mid K}^{\bs{\xi}}\rb^{-1}\!\rc \! \bs{\xi}_{k+1} \notag\\
&\hspace{-2.3cm}+\bs{\xi}_{k+1}^{\T}\bs{T}\bs{Q}_k^{-1}\tilde{\bs{A}}_k \bs{\xi}_k+\bs{\xi}_{k}^{\T}\lb\bs{P}_{k\mid k}^{\bs{\xi}}\rb^{-1} \hat{\bs{\xi}}_{k\mid k}\notag\\
&\hspace{-2cm}-\frac{1}{2} \bs{\xi}_{k}^{\T}\ls\tilde{\bs{A}}_k^{\T} \bs{Q}_k^{-1} \tilde{\bs{A}}_k+\lb\bs{P}_{k\mid k}^{\bs{\xi}}\rb^{-1}\rs \bs{\xi}_k,\label{eq:smoothing_opt}
\end{align}
where $\propto$ denotes the equivalence up to an additive constant independent of $\bs{\xi}_{k}$ and $\bs{\xi}_{k+1}$, and $\tilde{\bs{A}}_k$ and $\bs{T}$ are defined in Step 1 of \Cref{alg:RobustKalman}. Now, \eqref{eq:smoothing_opt} takes the same form as that of the joint posterior distribution in the classical Kalman smoothing derivation (see~\cite[Equation (11)]{byron2004derivation}). Following a mathematical proof similar to that in \cite{byron2004derivation}, we arrive at the smoothing equations given by Steps 12-15 of \Cref{alg:RobustKalman}, see Appendix-I for details. We also extend the RKS algorithm to measurements of the form $\bs{y}_k=\bs{C}_k\bs{x}_k+\bs{v}_k$, and the algorithm along with it's derivation is given in Appendix-II.

In summary, this section presented the optimal estimation of states and inputs when there is no prior information on the inputs. We note that the estimator relies on Assumption A4, which holds only when $p\geq m$. To overcome this limitation, we next derive sparsity aware-estimation algorithms that can work in the $p<m$ regime, where the RKS algorithm fails because Assumption A4 does not hold. Since the input is known to be sparse, we encode this information into the estimation model via suitable priors on the inputs. Based on different sparsity-promoting priors, we develop two approaches, (a) regularized RKS and (b) Bayesian RKS, presented next.

\section{Regularized Robust Kalman Smoothing}
\label{sec:reg_RKS}
Inspired by the convex optimization-based sparse signal recovery algorithms~\cite{foucart2013math,wipf2010iterative}, we use the following prior  to induce sparsity:
\begin{equation}\label{eq:convex_prior}
    p\lb \bs{u}_{k}\rb=\prod_{i=1}^{m}\chi\exp\ls-\frac{\tau_k}{2} \lv \bs{u}_k(i)\rv^l\rs,
\end{equation}
where $\chi$ is the normalizing constant and $\tau_k,l>0$ are  known distribution parameters. Here, $\tau_k$ controls the sparsity of the control inputs, i.e., a large value of $\tau_k$ leads to sparser solutions. This tuning parameter is often chosen by cross-validation. The choice of $l$ determines the properties of the optimization problem in \eqref{eq:joint_cost}, leading to different variants of estimators as given below. 

\subsection{$\ell_1$-regularized Robust Kalman Smoothing}
\label{sec:l1RKS}
The most popular choice is $l=1$, which is motivated by the $\ell_1$ norm-based Laplacian prior leading to the $\ell_1$-regularized RKS algorithm. Thus, problem \eqref{eq:RKS_cost} of the RKS algorithm changes to
\begin{multline}
\lc \hat{\bs{x}}_{k\mid K},\hat{\bs{u}}_{k\mid K}\rc_{k=1}^K\!\!=\!\underset{ \substack{\bs{x}_{k},\bs{u}_{k}\\k=1,\ldots,K}}{\arg {\min}}\; \sum_{k=1}^{K}\lV\bs{y}_k-\bs{C}_k\bs{x}_k-\bs{D}_k\bs{u}_k\rV_{\bs{R}_k}^2\\+\sum_{k=1}^{K-1}\lV\bs{x}_{k+1}-\bs{A}_{k}\bs{x}_{k}-\bs{B}_{k}\bs{u}_{k}\rV_{\bs{Q}_k}^2+\sum_{k=1}^{K}\tau_k\lV\bs{u}_k\rV_1.
\label{eq:RKS_cost_mod}
\end{multline}
Due to the Laplacian prior, Assumption A3 does not generally hold, but our derivation assumes A3 for tractability. 

Unlike \eqref{eq:RKS_cost}, the new convex optimization problem in  \eqref{eq:RKS_cost_mod} does not admit a closed-form solution. Therefore, we use the alternating direction method of multipliers (ADMM) algorithm. ADMM decomposes the convex optimization problem in \eqref{eq:RKS_cost_mod} into simpler optimization problems. For this, we reformulate \eqref{eq:RKS_cost_mod} to an equivalent optimization problem using auxiliary variables $\lc\bs{t}_k\in\bb{R}^m\rc_{k=1}^{K}$ and define the  augmented Lagrangian as follows:
\begin{multline}
\mcal{L}\lb\lc\bs{x}_k,\bs{u}_k,\bs{t}_k,\bs{\bs{\lambda}}_k\rc_{k=1}^K\rb = \sum_{k=1}^K\lV\bs{y}_k-\bs{C}_k\bs{x}_k-\bs{D}_k\bs{u}_k\rV_{\bs{R}_k}^2\\ +\sum_{k=1}^{K-1}\lV\bs{x}_{k+1}-\bs{A}_{k}\bs{x}_{k}-\bs{B}_{k}\bs{u}_{k}\rV_{\bs{Q}_k}^2\\
+\sum_{k=1}^{K}\ls \tau_k\lV\bs{t}_k\rV_1
+\bs{\lambda}_k^{\T}\lb\bs{t}_k-\bs{u}_k\rb+c\lV\bs{t}_k-\bs{u}_k\rV^2\rs,\label{eq:xu_l1_eqv2}
\end{multline}
where $\lc\bs{\lambda}_k\in\bb{R}^m\rc_{k=1}^{K}$ are the the Lagrangian multipliers that arise from the $K$ constraints $\bs{t}_k=\bs{u}_k$. Also, $c>0$ is a scalar. ADMM is an iterative algorithm, and the $r$th iteration updates are
\begin{align}
\lc\bs{x}_k^{(r)},\bs{u}_k^{(r)}\rc_{k=1}^K\!\!\!&=\underset{\{\bs{x}_{k},\bs{u}_{k}\}_{k=1}^K}{\arg \min } \mcal{L}\!\lb\!\lc\!\bs{x}_k\!,\bs{u}_k,\bs{t}_k^{(r-1)}\!,\bs{\bs{\lambda}}_k^{(r-1)}\!\rc_{k=1}^K\rb \label{eq:xu_l1_min}\\
\bs{t}_k^{(r)}&=\arg \min _{\bs{t}_k} \mcal{L}\!\lb\!\lc\bs{x}_k^{(r)}\!,\bs{u}_k^{(r)}\!,\bs{t}_k,\bs{\bs{\lambda}}_k^{(r-1)}\rc_{k=1}^K\!\rb \label{eq:t_l1_min}\\
\bs{\lambda}_k^{(r)}&=\bs{\lambda}_k^{(r-1)}+c\lb\bs{u}_{k}^{(r)}-\bs{t}_k^{(r)}\rb,\label{eq:lam_l1_min}
\end{align}
for $k=1,2,\ldots,K$. Further, \eqref{eq:xu_l1_min} can be further simplified as 
\begin{multline}
\lc\bs{x}_k^{(r)},\bs{u}_k^{(r)}\rc_{k=1}^K=\underset{ \substack{\bs{x}_{k},\bs{u}_{k}\\k=1,\ldots,K}}{\arg \min }\sum_{k=1}^K\lV\bs{y}_k^{\ell_1}-\tilde{\bs{C}}_k\bs{x}_k-\tilde{\bs{D}}_k\bs{u}_k\rV_{\bs{R}_k^{\ell_1}}^2\\
+\sum_{k=1}^{K-1}\lV\bs{x}_{k+1}-\bs{A}_{k}\bs{x}_{k}-\bs{B}_{k}\bs{u}_{k}\rV_{\bs{Q}_k}^2.\label{eq:xu_l1_simplified}
\end{multline}
Here, we define the new matrices as
\begin{align}
    \bs{y}_k^{\ell_1}& =\begin{bmatrix}
\bs{y}_k\\\bs{t}_k^{(r-1)}-c^{-1}\bs{\lambda}_k^{(r-1)}
\end{bmatrix} &
\bs{R}_k^{\ell_1} &=\begin{bmatrix}
\bs{R}_k & 0\\0 & c^{-1}\bs{I}
\end{bmatrix}\label{eq:yR_defn}\\
\tilde{\bs{C}}_k&=\begin{bmatrix}
\bs{C}_k^{\T}&\bs{0}
\end{bmatrix}^{\T}
 &\tilde{\bs{D}}_k&=\begin{bmatrix}
\bs{D}_k^{\T}&\bs{I}
\end{bmatrix}^{\T}.\label{eq:CD_defn}
\end{align}
Since \eqref{eq:xu_l1_simplified} takes the same form as \eqref{eq:RKS_cost}, its solution is given by the RKS algorithm with a modified measurement model. Further, from \eqref{eq:t_l1_min}, the auxiliary variable $\bs{t}_k$ update is given by
\begin{align}
\bs{t}_k^{(r)}&=\arg \min _{\bs{t}_k}\;\tau_k\lV\bs{t}_k\rV_1+c\lV\bs{t}_k-c^{-1}\bs{\lambda}_k^{(r-1)}-\bs{u}_{k}^{(r)}\rV^2\\
&= S_{c^{-1}\tau_k}\lb\bs{u}_{k}^{(r)}+c^{-1}\bs{\lambda}_k^{(r-1)}\rb,
\end{align}
where the last step follows from the solution to the LASSO problem. Also, $S(\cdot)$ is the entry-wise soft thresholding function given by \begin{equation}
 S_b(a)\triangleq\operatorname{Sgn}(a)\max\lc |a|-b,0\rc.\label{eq:S_defn}   
\end{equation}
 The overall $\ell_1$-regularized RKS algorithm is given in \Cref{alg:l1_RobustKalman}.
\begin{algorithm}[ht]
	\caption{$\ell_1$-regularized Robust Kalman Smoothing}
	\setstretch{1.2}
	\begin{algorithmic}[1]
		\REQUIRE $\{\bs{y}_k, \bs{A}_k,\bs{B}_k,\bs{C}_k,\bs{D}_k,\bs{Q}_k,\bs{R}_k\}_{k=1}^K$
			\STATEx \hspace{-0.5cm}\textbf{Parameters:} $c$, $r_{\max}$, and $\tau_k$ for $k=1,2,\ldots,K$
		\STATEx \hspace{-0.5cm}\textbf{Initialization:}  $\bs{t}_{k}^{(0)}=\bs{\lambda}_{k}^{(0)}=\zero$, for $k=1,2,\ldots,K$
		\STATE Compute $ \bs{R}_k^{\ell_1} $, $\tilde{\bs{C}}_k$, $\tilde{\bs{D}}_k$ using \eqref{eq:yR_defn} and \eqref{eq:CD_defn}
		\FOR{$r=1,2,\ldots,r_{\max}$}
		\STATE Compute $\bs{y}_k^{\ell_1}$ using \eqref{eq:yR_defn}
        \STATE \parbox[t]{219pt}{Compute $\bs{x}_{k}^{(r)}$ and  $\bs{u}_{k}^{(r)}$ via  \Cref{alg:RobustKalman} replacing $\bs{y}_k$, $\bs{C}_k$, $\bs{D}_k$, and $
\bs{R}_k $ with $\bs{y}_k^{\ell_1}$, $\tilde{\bs{C}}_k$, $\tilde{\bs{D}}_k$, and $
\bs{R}_k^{\ell_1} $ , respectively}
        \FOR{$k=1,2,\ldots,K$}
        \STATE Using \eqref{eq:S_defn}, $\bs{t}_k^{(r)}=S_{c^{-1}\tau_k}\lb\bs{u}_{k}^{(r)}+c^{-1}\bs{\lambda}_k^{(r-1)}\rb$
		\STATE $\bs{\lambda}_k^{(r)}=\bs{\lambda}_k^{(r-1)}+c\lb\bs{u}_{k}^{(r)}-\bs{t}_k^{(r)}\rb$
        \ENDFOR
		\ENDFOR
		\ENSURE $\lc\bs{x}_{k}^{(r)}\rc_{k=1}^K$ and $\lc\bs{u}_{k}^{(r)}\rc_{k=1}^K$
	\end{algorithmic}
	\label{alg:l1_RobustKalman}
\end{algorithm}
\begin{algorithm}[t]
	\caption{Reweighted $\ell_2$-regularized Robust Kalman Smoothing}
	\setstretch{1.2}
	\begin{algorithmic}[1]
		\REQUIRE $\{\bs{y}_k, \bs{A}_k,\bs{B}_k,\bs{C}_k,\bs{D}_k,\bs{Q}_k,\bs{R}_k\}_{k=1}^K$
		\STATEx \hspace{-0.5cm}\textbf{Parameters:} $l$, $r_{\max}$, $\kappa$, and $\tau_k$ for $k=1,2,\ldots,K$
		\STATEx \hspace{-0.5cm}\textbf{Initialization:} $\bs{u}_{k}^{(0)}=\bs{1}$, for $k=1,2,\ldots,K$
		\FOR{$r=1,2,\ldots,r_{\max}$}
            \STATE Determine $\mathcal{I}_k$ for each $\bs{u}_k^{(r-1)}$ and compute $\tilde{\bs{u}}_k^{(r-1)}$, $\tilde{\bs{W}}_k^{(r)}$ using \eqref{eq: W_tilde comp}
		\STATE \parbox[t]{219pt}{Compute $\bs{y}_k^{\ell_2}$, $\bs{B}_k^{\ell_2}$, $\bs{C}_k^{\ell_2}$, $\bs{D}_k^{\ell_2}$, $\bs{R}_k^{\ell_2}$ using \eqref{eq:yR_defn_mod}, and \eqref{eq: B,C,D defn}}
        \STATE \parbox[t]{219pt}{Compute $\bs{x}_{k}^{(r)}$ and  $\tilde{\bs{u}}_{k}^{(r)}$ via  \Cref{alg:RobustKalman} replacing $\bs{y}_k$, $\bs{B}_k$, $\bs{C}_k$, $\bs{D}_k$, and $
\bs{R}_k $ with $\bs{y}_k^{\ell_2}$, $\bs{B}_k^{\ell_2}$, $\bs{C}_k^{\ell_2}$, $\bs{D}_k^{\ell_2}$, and $\bs{R}_k^{\ell_2} $}  respectively
        \STATE Sparsify $\tilde{\bs{u}}_{k}^{(r)}$ to get $\bs{u}_{k}^{(r)}$
		\ENDFOR
		\ENSURE $\lc\bs{x}_{k}^{(r)}\rc_{k=1}^K$ and $\lc\bs{u}_{k}^{(r)}\rc_{k=1}^K$
	\end{algorithmic}
	\label{alg:l2_RobustKalman}
\end{algorithm}
\subsection{Reweighed $\ell_2$-regularized Robust Kalman Smoothing}
\label{sec:reweightedl2}
When $0<l<2$, an alternative estimation algorithm, similar in spirit to the iterative reweighting-based sparse recovery~\cite{wipf2010iterative}, can be derived. In this case, the optimization problem \eqref{eq:RKS_cost} changes to
\begin{multline}
\lc \hat{\bs{x}}_{k\mid K},\hat{\bs{u}}_{k\mid K}\rc_{k=1}^K\!\!=\!\underset{ \substack{\bs{x}_{k},\bs{u}_{k}\\k=1,\ldots,K}}{\arg {\min}}\; \sum_{k=1}^{K}\lV\bs{y}_k-\bs{C}_k\bs{x}_k-\bs{D}_k\bs{u}_k\rV_{\bs{R}_k}^2\\ \hspace{-0.2cm}+\sum_{k=1}^{K-1}\lV\bs{x}_{k+1}-\bs{A}_{k}\bs{x}_{k}-\bs{B}_{k}\bs{u}_{k}\rV_{\bs{Q}_k}^2+\sum_{k=1}^{K}\tau_k\sum_{i=1}^m\lv\bs{u}_k(i)\rv^l.\label{eq:RKS_cost_mod2}
\end{multline}
Further, we note that for any $u\in\bb{R}$, the function $\lv u\rv^l=\lb\lv u\rv^2\rb^{l/2}$ is a concave function of $\lv u\rv^2$. Therefore, the function is bounded above by its first-order Taylor approximation, and for any $u\in\bb{R}$,
\begin{equation}
    \lb\lv \bs{u}_k(i)\rv^2\rb^{l/2} \leq \lb\lv u\rv^2\rb^{l/2} +\frac{l}{2}\lb\lv u\rv^2\rb^{l/2-1} (\lv \bs{u}_k(i)\rv^2-\lv u\rv^2).
\end{equation}
With $u=\bs{u}_{k}^{(r-1)}(i)$, we get
\begin{multline}
  \lb\lv \bs{u}_k(i)\rv^2\rb^{l/2} \leq  \frac{l}{2}\lv \bs{u}_{k}^{(r-1)}(i)\rv^{-(2-l)} \lv \bs{u}_k(i)\rv^2\\+\lb1-\frac{l}{2}\rb \lv \bs{u}_{k}^{(r-1)}(i)\rv^{l},\label{eq:concave}
\end{multline}
Using the above bound, we can iteratively solve \eqref{eq:RKS_cost_mod2} by optimizing the upper bounded cost function in each iteration. 
So, we compute 
\begin{multline}
\lc \bs{x}_{k}^{(r)},\bs{u}_{k}^{(r)}\rc_{k=1}^K\!\!=\!\underset{ \substack{\bs{x}_{k},\bs{u}_{k}\\k=1,\ldots,K}}{\arg {\min}}\; \sum_{k=1}^{K}\lV\bs{y}_k-\bs{C}_k\bs{x}_k-\bs{D}_k\bs{u}_k\rV_{\bs{R}_k}^2\\+\sum_{k=1}^{K-1}\lV\bs{x}_{k+1}-\bs{A}_{k}\bs{x}_{k}-\bs{B}_{k}\bs{u}_{k}\rV_{\bs{Q}_k}^2+\frac{l}{2}\sum_{k=1}^{K}\tau_k\lV\bs{u}_k\rV_{\bs{W}_k^{(r)}}^2.\label{eq:xu_min_mod2}
\end{multline}
Here, we define the diagonal weight matrix $\bs{W}_k^{(r)}\in\bb{R}^{m\times m}$ as 
\begin{equation}
    \bs{W}_k^{(r)}=\operatorname{diag}\lc
    \lv \bs{u}_{k}^{(r-1)}\rv
    \rc^{(2-l)},\label{eq:W_defn}
\end{equation}
with $\lv\cdot\rv$ representing the element-wise operation. When some entries of $\bs{u}_{k}^{(r-1)}$ become close to zeros, $\bs{W}_k^{(r)}$ can potentially become non-invertible. To avoid numerical instabilities, when any $\bs{u}_k^{(r-1)}(i)$ falls below some small threshold $\kappa$ (e.g., $\kappa = 10^{-6}$) in magnitude, those entries and the corresponding dictionary columns are pruned from the model. The modified optimization problem in \eqref{eq:xu_min_mod2} is rewritten as 
\begin{multline}
\lc \bs{x}_{k}^{(r)},\tilde{\bs{u}}_{k}^{(r)}\rc_{k=1}^K\!\!=
\underset{ \substack{\bs{x}_{k},\tilde{\bs{u}}_{k}\\k=1,\ldots,K}}{\arg \min }\sum_{k=1}^K\lV\bs{y}_k^{\ell_2}-\bs{C}_k^{\ell_2}\bs{x}_k-\bs{D}_k^{\ell_2}\tilde{\bs{u}}_k\rV_{\bs{R}_k^{\ell_2}}^2\\
+\sum_{k=1}^{K-1}\lV\bs{x}_{k+1}-\bs{A}_{k}\bs{x}_{k}-\bs{B}_{k}^{\ell_2}\tilde{\bs{u}}_{k}\rV_{\bs{Q}_k}^2,\label{eq:RKS_cost_mod3}
\end{multline}
\begin{align}
\text{where\quad} \bs{y}_k^{\ell_2}& =\begin{bmatrix}
\bs{y}_k\\\bs{0}_{\vert\mathcal{I}_k\vert\times 1}
\end{bmatrix}, &
\bs{R}_k^{\ell_2} &=\begin{bmatrix}
\bs{R}_k & 0\\0 & \frac{2}{\tau_kl}\tilde{\bs{W}}_k^{(r)}
\end{bmatrix},\label{eq:yR_defn_mod}
\end{align}
\begin{align}
\bs{B}_k^{\ell_2}\!=\!\lb\bs{B}_k\rb_{\mathcal{I}_k}, \bs{C}_k^{\ell_2}\!=\!\ls\bs{C}_k^{\T},\bs{0}_{n\times \vert\mathcal{I}_k\vert}\rs^{\T}\!\!, \bs{D}_k^{\ell_2}\!=\!\ls\lb\bs{D}_k\rb_{\mathcal{I}_k}^{\T},\bs{I}_{\vert\mathcal{I}_k\vert}\rs^{\T}\!\!,\label{eq: B,C,D defn}
\end{align}\vspace{-0.2cm}
\begin{align}
\text{and\quad} \tilde{\bs{W}}_k^{(r)}=\operatorname{diag}\lc
\lv \tilde{\bs{u}}_{k}^{(r-1)}\rv\rc^{(2-l)},\label{eq: W_tilde comp}
\end{align}
where $\tilde{\bs{u}}_{k}^{(r-1)}$ is the vector we get after removing the entries of $\bs{u}_{k}^{(r-1)}$ whose absolute values are less than $\kappa$. $\mathcal{I}_k$ is the set of indices corresponding to the entries of $\bs{u}_{k}^{(r-1)}$ which have absolute value more than $\kappa$. $\tilde{\bs{u}}_k=\lb\bs{u}_k\rb_{\mathcal{I}_k}$. We construct $\bs{u}_k^{(r)}$ by sparsifying $\tilde{\bs{u}}_k^{(r)}$, i.e., by inserting zeros at locations indexed by $\mathcal{I}_k^{c}$. We note that \eqref{eq:RKS_cost_mod3} takes the same form as \eqref{eq:RKS_cost_mod}. Further, the  modified measurement matrix $\tilde{\bs{D}}_k^{\ell_2}$ satisfies Assumption A4. Thus, \eqref{eq:RKS_cost_mod3} can be solved using the RKS algorithm in \Cref{alg:RobustKalman}. The pseudocode for the overall reweighed $\ell_2$-regularized RKS is summarized in \Cref{alg:l2_RobustKalman}.
\vspace{-0.2cm}
\subsection{Extension to Jointly Sparse Inputs}\label{sec:l1_joint_sparse}
We can extend $\ell_1$-regularized RKS to estimate jointly sparse inputs by modifying \eqref{eq:RKS_cost_mod} as follows:
\begin{multline}
\lc \hat{\bs{x}}_{k\mid K},\hat{\bs{u}}_{k\mid K}\rc_{k=1}^K\!\!=\!\underset{ \substack{\bs{x}_{k},\bs{u}_{k}\\k=1,\ldots,K}}{\arg {\min}}\; \sum_{k=1}^{K}\lV\bs{y}_k-\bs{C}_k\bs{x}_k-\bs{D}_k\bs{u}_k\rV_{\bs{R}_k}^2\\+\sum_{k=1}^{K-1}\lV\bs{x}_{k+1}-\bs{A}_{k}\bs{x}_{k}-\bs{B}_{k}\bs{u}_{k}\rV_{\bs{Q}_k}^2+\tau \sum_{i=1}^{m} \sqrt{\sum_{k=1}^{K}\bs{u}_k^2(i)}.\label{eq:xu_l1_filter_mod}
\end{multline} 
where regularizer $\tau \sum_{i=1}^{m} \sqrt{\sum_{k=1}^{K}\bs{u}_k^2(i)}$ is inspired by the LASSO type regularization~\cite{angelosante2009lasso}. As the parameter $\tau>0$ increases, the size of the common support output by the algorithm decreases, making the inputs sparser. Now following the same approach as in \Cref{sec:l1RKS}, we introduce auxiliary variables $\bs{t}_k$'s to reformulate \eqref{eq:xu_l1_filter_mod} and solve it using ADMM.
\begin{multline}
\lc \hat{\bs{x}}_{k\mid K},\hat{\bs{u}}_{k\mid K}\rc_{k=1}^K\!\!=\!\underset{ \substack{\bs{x}_{k},\bs{u}_{k}\\k=1,\ldots,K}}{\arg {\min}}\; \sum_{k=1}^{K}\lV\bs{y}_k-\bs{C}_k\bs{x}_k-\bs{D}_k\bs{u}_k\rV_{\bs{R}_k}^2\\+\sum_{k=1}^{K-1}\lV\bs{x}_{k+1}-\bs{A}_{k}\bs{x}_{k}-\bs{B}_{k}\bs{u}_{k}\rV_{\bs{Q}_k}^2+\tau \sum_{i=1}^{m} \sqrt{\sum_{k=1}^{K}t_k^2(i)}
\\+\sum_{k=1}^{K}\bs{\lambda}_k^{\T}\lb\bs{t}_k-\bs{u}_k\rb+c\sum_{k=1}^{K}\lV\bs{t}_k-\bs{u}_k\rV^2
\label{eq:xu_l1_eqv2_mod},
\end{multline}	
where $\lc\bs{\lambda}_k\in\bb{R}^m\rc_{k=1}^{K}$ are the Lagrangian multipliers that arise from the equality constraints $\bs{t}_k=\bs{u}_k$ and $c>0$ is a positive scalar. Since \eqref{eq:xu_l1_eqv2_mod} is identical to \eqref{eq:xu_l1_eqv2} except for the terms in $\bs{t}_k$, our modified version of $\ell_1$-regularized RKS is identical to \Cref{alg:l1_RobustKalman} except for Step 8 which changes as follows~\cite{angelosante2009lasso}:
\begin{multline}
\bs{t}^{(r)}(i,:)= \frac{\bs{u}^{(r)}(i,:)+c^{-1} \bs{\lambda}^{(r-1)}(i,:)}{\lV\bs{u}^{(r)}(i,:)+c^{-1} \bs{\lambda}^{(r-1)}(i,:)\rV}\\
S_{c^{-1} \tau}\lb\lV\bs{u}^{(r)}(i,:)+c^{-1} \bs{\lambda}^{(r-1)}(i,:)\rV\rb,
\end{multline}
where $S$ is defined in \eqref{eq:S_defn}, and $\bs{t}^{(r)}(i,:)$, $\bs{u}^{(r)}(i,:)$, $\bs{\lambda}^{(r-1)}(i,:)$ follow the definition, 
\begin{equation}\label{eq:rearrange_block}
    \bs{a}(i,:)\triangleq\begin{bmatrix}
\bs{a}_1(i)&\bs{a}_2(i)&\ldots& \bs{a}_K(i)
\end{bmatrix}^{\T}\in\bb{R}^K,
\end{equation} for any set $\{\bs{a}_k\in\bb{R}^m\}_{k=1}^K$ and $i=1,2,\ldots,K$. This modified version is referred to as \emph{group $\ell_1$-regularized RKS}.\vspace{-0.1cm}

\section{Bayesian Robust Kalman Smoothing}\label{sec:Bayesian}
In this section, we present an alternative approach, called Bayesian RKS, to estimate the states and sparse inputs $\{\bs{x}_k, \bs{u}_k\}_{k=1}^K$ using measurements $\{\bs{y}_k\}_{k=1}^K$ in \eqref{eq:meas_eqn}. Unlike the regularized RKS that assumes the knowledge of the parameters of the sparse input prior, the Bayesian approach uses hierarchical priors which account for the uncertainty in the prior distribution. Specifically, we use a Gaussian prior\footnote{The underlying true sparse inputs  need not be generated according to the assumed priors in \eqref{eq:convex_prior}, \eqref{eq:sbl_prior}. The choice of prior is motivated by the fact that they result in sparse posterior estimates, as desired.} on the inputs to promote sparsity:
\begin{equation}
p(\bs{u}_k ; \bs{\gamma}_k)=\prod_{i=1}^{m}\frac{1}{\sqrt{2 \pi \gamma_k(i)}} \exp \lb-\frac{\bs{u}_k(i)^{2}}{2 \bs{\gamma}_k(i)}\rb,\label{eq:sbl_prior}
\end{equation}
where $\bs{\gamma}_k\in\bb{R}^{m}$ is the unknown hyperparameter of the distribution. The Bayesian RKS learns the hyperparameters from the measurements, making it different from the regularized RKS. We present two Bayesian RKS algorithms, namely, SBL-RKS and VB-RKS. 

\subsection{Sparse Bayesian Learning-based Robust Kalman Smoothing}
\label{sec:SBL_RKS}
\vspace{-0.1cm}
In the SBL framework, we first compute the ML estimate $\hat{\bs{\gamma}}_{k}^{\mathrm{ML}}$ of the hyperparameter as
\begin{equation}
\hat{\bs{\gamma}}_{k}^{\mathrm{ML}}=\underset{\bs{\gamma} \in \mathbb{R}_{+}^{m \times 1}}{\arg \max}\; p\lb\bs{Y}^K_1 ; \bs{\gamma}\rb.\label{eq:gamma_ML}
\end{equation}
 Using the estimate $\hat{\bs{\gamma}}_{k}^{\mathrm{ML}}$, we can estimate the states and inputs using the Kalman filtering and smoothing algorithm. For this, we note that \eqref{eq:state_eqn} and \eqref{eq:meas_eqn} are equivalent to the following:
\begin{equation}
\bs{\xi}_{k+1}=\begin{bmatrix}\tilde{\bs{A}}_k\\ \bs{0} \end{bmatrix}\bs{\xi}_k+\begin{bmatrix}\bs{w}_k\\\bs{z}_k\end{bmatrix} \hspace{0.2cm}\text{and}\hspace{0.2cm}
\bs{y}_{k}=\begin{bmatrix}\bs{C}_k & \bs{D}_k\end{bmatrix}\bs{\xi}_k+\bs{v}_{k},\label{eq:eqn_mod}
\end{equation}
where $\bs{\xi}_k$ is defined in \eqref{eq:xi_defn} and  $\bs{z}_k=\bs{u}_{k+1}$ is an auxiliary variable. Further, from Assumption A1 and \eqref{eq:sbl_prior}, we obtain 
\begin{equation}
    \begin{bmatrix}\bs{w}_k\\\bs{z}_k\end{bmatrix} \sim \mcal{N}\lb\bs{0},\begin{bmatrix}\bs{Q}_{k} & \bs{0} \\ \bs{0} & \operatorname{Diag}\lc\bs{\gamma}_{k+1}\rc\end{bmatrix}\rb,\label{eq:noise_mod}
\end{equation}
Hence, estimating the states and inputs is equivalent to estimating $\lc\bs{\xi}_k\rc_{k=1}^K$ using $\bs{Y}_1^K$ via the standard Kalman filtering and smoothing due to the Gaussian assumptions.

\begin{algorithm}[t]
	\caption{RKS with Sparse Bayesian Learning}
	\begin{algorithmic}[1]
		\REQUIRE $\{\bs{y}_k, \bs{A}_k,\bs{B}_k,\bs{C}_k,\bs{D}_k,\bs{Q}_k,\bs{R}_k\}_{k=1}^K$
		\STATEx \hspace{-0.5cm}\textbf{Parameters:} $r_{\max}$, $\epsilon_{\mathrm{thres}}$
		\STATEx \hspace{-0.5cm}\textbf{Initialization:} $\bs{\gamma}_k^{(0)}=\bs{1}$ for $k=1,2,\ldots,K$, $r=1$, $\epsilon=2\epsilon_{\mathrm{thres}}$
		\STATE $\bar{\bs{A}}_k=\begin{bmatrix}
\bs{A}_k & \bs{B}_k\\
\zero & \zero\end{bmatrix}\in\bb{R}^{(n+m)\times (n+m)}$, $\bar{\bs{C}}_k=\begin{bmatrix}
\bs{C}_k & \bs{D}_k\end{bmatrix}$
	\WHILE {($r<r_{max}$) and ($\epsilon>\epsilon_{\mathrm{thres}}$)}
		\STATEx \#\emph{E-Step:}
		\STATE $\hat{\bs{\xi}}_{0\mid 0}=\bs{0}$, $\bs{P}_{0\mid 0}^{\bs{\xi}}=\bs{I}$
		\FOR{$k=1,2,\ldots,K$}
  	\STATE $\bar{\bs{Q}}_{k-1}=\begin{bmatrix}\bs{Q}_{k-1} & \bs{0} \\ \bs{0} & \operatorname{Diag}\lc\bs{\gamma}_{k}\rc\end{bmatrix}$
		\STATEx \#\emph{Prediction:}
		\STATE $\hat{\bs{\xi}}_{k\mid k-1}=\bar{\bs{A}}_{k-1}\hat{\bs{\xi}}_{k-1\mid k-1}$
		\STATE $\bs{P}^{\bs{\xi}}_{k\mid k-1}=\bar{\bs{A}}_{k-1}
\bs{P}^{\bs{\xi}}_{k-1\mid k-1}\bar{\bs{A}}_{k-1}^{\T}+\bar{\bs{Q}}_{k-1}$
		\STATEx \#\emph{Filtering:}
		\STATE $\bs{G}_{k}=\bs{P}_{k \mid k-1}^{\bs{\xi}} \bar{\bs{C}}_k^{\T}\lb\bs{R}_k+\bar{\bs{C}}_k \bs{P}_{k \mid k-1}^{\bs{\xi}} \bar{\bs{C}}_k^{\T}\rb^{-1}$
		\STATE $\hat{\bs{\xi}}_{k \mid k}=\hat{\bs{\xi}}_{k \mid k-1}+\bs{G}_{k}\lb\bs{y}_k-\bar{\bs{C}}_k \hat{\bs{\xi}}_{k \mid k-1}\rb$
		\STATE $\bs{P}_{k \mid k}^{\bs{\xi}}=\lb\bs{I}-\bs{G}_{k}\bar{\bs{C}}_k\rb \bs{P}_{k \mid k-1}^{\bs{\xi}}$
		\ENDFOR
		\STATEx \#\emph{Smoothing:}
		\FOR{$k=K-1,K-2,\ldots,1$}
		\STATE $\bs{K}_k=\bs{P}_{k\mid k}^{\bs{\xi}}\bar{\bs{A}}_{k}^{\T}\lb\bs{P}_{k+1\mid k}^{\bs{\xi}}\rb^{-1}$
		\STATE $\bs{P}_{k\mid K}^{\bs{\xi}}=\bs{P}_{k\mid k}^{\bs{\xi}}+\bs{K}_k\lb\bs{P}_{k+1\mid K}^{\bs{\xi}}-\bs{P}_{k+1\mid k}^{\bs{\xi}}\rb\bs{K}_k^{\T}$
        \STATE $\hat{\bs{\xi}}_{k\mid K}=\hat{\bs{\xi}}_{k\mid k}+\bs{K}_k\lb\hat{\bs{\xi}}_{k+1\mid K}-\bar{\bs{A}}_k\hat{\bs{\xi}}_{k\mid k}\rb$
        \STATE $\bs{P}_{k+1,k\mid K}^{\bs{\xi}}=\begin{bmatrix}\bs{P}_{k+1\mid K}^{\bs{x}} & \bs{P}_{k+1\mid K}^{\bs{xu}}\end{bmatrix}^{\T}\bs{K}_k^{\T}$
        \STATE Compute $\hat{\bs{u}}_{k\mid K}$ and $\bs{P}_{k\mid K}^{\bs{u}}$ using \eqref{eq:xi_defn} from $\hat{\bs{\xi}}_{k\mid K}$ and $\bs{P}_{k\mid K}^{\bs{\xi}}$
		\ENDFOR
		\STATEx \#\emph{M-step:}
        \STATE $\bs{\gamma}_k^{(r)}=\operatorname{Diag}\lc\hat{\bs{u}}_{k\mid K}\hat{\bs{u}}_{k\mid K}^{\T}+\bs{P}^{\bs{u}}_{k\mid K}\rc$, for $k=1,2,\ldots,K$
		\ENDWHILE
		\STATE Compute $\lc \hat{\bs{x}}_{k\mid K},\hat{\bs{u}}_{k\mid K}\rc_{k=1}^K$ using \eqref{eq:xi_defn} from $\hat{\bs{\xi}}_{k\mid K}$
		\ENSURE $\{\hat{\bs{x}}_{k \mid K}\}_{k=1}^K$ and $\{\hat{\bs{u}}_{k \mid K}\}_{k=1}^K$
	\end{algorithmic}
	\label{alg:SBL_RKS}
\end{algorithm}
The optimization problem in \eqref{eq:gamma_ML} does not admit a closed-form solution, so we employ the EM algorithm to solve it. The EM algorithm is an iterative method with the expectation (E) and maximization (M) steps. 
In the $r$th iteration, the E-step computes the expected log-likelihood function $\mcal{Q}^{(r)}$ of $\lc\bs{\gamma}_k\rc_{k=1}^K$ with respect to the current distribution of the available data $\bs{Y}_1^K$ and the hidden data $\lc\bs{\xi}_k\rc_{k=1}^K$  given the estimate $\bs{\gamma}_k^{(r-1)}$ of the hyperparameter $\bs{\gamma}_k$ obtained in the previous iteration. The M-step maximizes the expected log-likelihood to obtain the new estimate of $\bs{\gamma}_k$. Further, from the state space model in \eqref{eq:eqn_mod}, the  distribution of the data is given by 
\begin{equation}
p\lb \bs{Y}^K_1,\lc\bs{\xi}_k\rc_{k=1}^K; \lc\bs{\gamma}_k\rc_{k=1}^K \rb=\prod_{k=1}^{K}p(\bs{y}_k\vert\bs{\xi}_k)p\lb\bs{\xi}_k\mid\bs{\xi}_{k-1}; \bs{\gamma}_k\rb,\label{eq:joint_dist}
\end{equation}
where $\bs{\xi}_0=\zero$.
Thus, the E-step is given by 
\begin{multline}
    \mcal{Q}^{(r)}\lb \lc\bs{\gamma}_k\rc_{k=1}^K \rb 
    = \sum_{k=1}^K\bb{E}_{\bs{\xi}_k,\bs{\xi}_{k-1}\mid \bs{Y}^K_1 ; \bs{\gamma}_k^{(r-1)}}\Big\{\log p(\bs{y}_k\vert\bs{\xi}_k)\\
    \times \ld p\lb\bs{x}_k\mid\bs{\xi}_{k-1}\rb p\lb\bs{u}_k; \bs{\gamma}_k\rb\rc.
\end{multline}
From the above relation, the M-step that maximizes $\mcal{Q}^{(r)}$ with respect to $\lc\bs{\gamma}_k\rc_{k=1}^K$ is separable, and ignoring the terms independent of $ \bs{\gamma}_k$, the M-step reduces to 
\begin{equation}
    \bs{\gamma}_k^{(r+1)} = \underset{\bs{\gamma} }{\arg \max}\; \bb{E}_{\bs{u}_k\mid \bs{Y}^K_1 ; \bs{\gamma}_k^{(r-1)}}\lc p\lb\bs{u}_k; \bs{\gamma}\rb\rc.
\end{equation}
Using \eqref{eq:sbl_prior}, we derive the M-step as
\begin{align}
\hspace{-0.21cm}\bs{\gamma}_k^{(r+1)}&\hspace{-0.1cm}=\hspace{-0.3cm}\underset{\bs{\gamma}:\bs{\Gamma}=\operatorname{Diag}\lc\bs{\gamma}\rc}{\arg \min } \hspace{-0.2cm}\log |\bs{\Gamma}|+ \operatorname{Tr}\lc\bs{\Gamma}^{-1}(\hat{\bs{u}}_{k\mid K}\hat{\bs{u}}_{k\mid K}^{\T}+\bs{P}^{\bs{u}}_{k\mid K})\rc\\
&=\operatorname{Diag}\lc\hat{\bs{u}}_{k\mid K}\hat{\bs{u}}_{k\mid K}^{\T}+\bs{P}^{\bs{u}}_{k\mid K}\rc.
\end{align}
Further, $\hat{\bs{u}}_{k \mid K}$ and $\bs{P}_{k \mid K}^{\bs{u}}$ can be computed by applying Kalman filtering and smoothing on the modified state space model in \eqref{eq:eqn_mod}. The overall SBL-RKS algorithm is summarized in \Cref{alg:SBL_RKS}. 

\noindent\emph{Remark:} When all the inputs are jointly sparse, we can use a common prior $\bs{u}_k\sim\mcal{N}\lb\bs{0},\operatorname{Diag}\lc\bs{\gamma}\rc\rb$, i.e., we have $\bs{\gamma}_k=\bs{\gamma}$ for $k=1,2,\ldots,K$. Then, the SBL-RKS for joint sparse input recovery is identical to \Cref{alg:SBL_RKS} except for Steps 3 and 19. The modified noise variance in Step 3 changes as follows:
\begin{equation}
    \bar{\bs{Q}}_{k}=\begin{bmatrix}\bs{Q}_{k} & \bs{0} \\ \bs{0} & \operatorname{Diag}\lc\bs{\gamma}\rc\end{bmatrix}, k=1,2,\ldots,K.
\end{equation}
Similarly, the M-step in Step 19 is modified as
\begin{equation}
\bs{\gamma}^{(r+1)}=\frac{1}{K}\sum_{k=1}^{K}\operatorname{Diag}\lc\hat{\bs{u}}_{k\mid K}\hat{\bs{u}}_{k\mid K}^{\T}+\bs{P}^{\bs{u}}_{k\vert K}\rc.
\end{equation}
The modified SBL-RKS for jointly sparse inputs is referred to as \emph{multiple measurement vector SBL-RKS (MSBL-RKS)}.

We also extend the SBL-RKS algorithm for measurements of the form $\bs{y}_k=\bs{C}_k\bs{x}_k+\bs{v}_k$ and the algorithm along with it's derivation is given in Appendix-II.C.
\vspace{-0.2cm}
\subsection{Variational Bayesian Robust Kalman Smoothing}\label{sec:VB_RKS}
\vspace{-0.1cm}
In the variational Bayesian approach, we employ a two-stage hierarchical prior. Specifically, we assume $\bs{\beta}_k(i)\sim \operatorname{Gamma}(a,b)$ where the precision hyperparameter $\bs{\beta}_k(i) = 1/\bs{\gamma}_k(i)$ in \eqref{eq:sbl_prior}, and $\operatorname{Gamma}(a,b)$ is the Gamma distribution with shape parameter $a>0$ and rate parameter $b>0$, i.e.,
\begin{equation}\label{eq.beta_distri}
     p(\bs{\beta}_k)=\prod_{i=1}^m \Gamma^{-1}(a)\;b^a(\bs{\beta}_k(i))^{a-1}\operatorname{exp}(-b\bs{\beta}_k(i)).
\end{equation}
VB-RKS estimates the unknowns $\mcal{Z} = \lc\bs{X}_1^K, \bs{U}_1^K, \{\bs{\beta}_k\}_{k=1}^K \rc$ as the mean of their posterior distribution. However, the posterior distribution computation is intractable, and we approximate it using a family of factorized 
 distributions:
 \begin{equation}
     p(\mcal{Z}\vert \bs{Y}_1^K)\approx q(\mcal{Z}) = \prod_{k=1}^K q^{\bs{x}}_k(\bs{x}_k) q^{\bs{u}}_k(\bs{u}_k) q^{\bs{\beta}}_k(\bs{\beta}_k),
 \end{equation}
 where $q^{\bs{x}}_k(\cdot), q^{\bs{u}}_k(\cdot)$, and $ q^{\bs{\beta}}_k(\cdot)$ are the marginal distributions of the latent variables $\bs{x}_k,\bs{u}_k$, and $\bs{\beta}_k$, respectively.

The optimal marginal distribution that minimizes the Kullback-Leibler divergence between the true and factorized posteriors can be obtained using the majorization-maximization procedure~\cite{nguyen2023depth} and is given by 
\begin{equation}
    \ln q^{\bs{x}}_k(\bs{x}_k)\propto\mathbb{E}_{q(\mcal{Z}\setminus\bs{x}_k)}\lc\ln p\lb \mcal{Z},\bs{Y}_1^K\rb \rc ,
\end{equation} where $ \propto$ denotes the equivalence up to an additive constant and the expectation is with respect to all the latent variables except $\bs{x}_k$. Further, we recall that
\begin{equation}
    p(\mcal{Z},\bs{Y}_1^K) \! = \!\prod_{k=1}^K p(\bs{y}_k|\bs{x}_k,\bs{u}_k)p(\bs{x}_{k}|\bs{x}_{k-1},\bs{u}_{k-1})
 p(\bs{u}_k|\bs{\beta}_k) p(\bs{\beta}_k),
\end{equation}
where $p(\bs{u}_k|\bs{\beta}_k)$ and $ p(\bs{\beta}_k)$ are given by \eqref{eq:sbl_prior} with $\bs{\beta}_k(i) = 1/\bs{\gamma}_k(i)$, and \eqref{eq.beta_distri}, respectively. Consequently, we arrive at
\begin{multline}
\ln q^{\bs{x}}_k(\bs{x}_k)\propto \Vert\bs{y}_k-\bs{C}_k\bs{x}_k-\bs{D}_k\langle\bs{u}_k\rangle\Vert_{\bs{R}_k}^2\\
+\Vert\langle\bs{x}_{k+1}\rangle-\bs{A}_{k}\bs{x}_k-\bs{B}_{k}\langle\bs{u}_{k}\rangle\Vert_{\bs{Q}_{k}}^2\\
+\Vert\bs{x}_k-\bs{A}_{k-1}\langle\bs{x}_{k-1}\rangle-\bs{B}_{k-1}\langle\bs{u}_{k-1}\rangle\Vert_{\bs{Q}_{k-1}}^2,
\end{multline}
where $\langle\cdot\rangle$ denotes the mean of a random variable following the marginal distribution derived from $q(\cdot)$. Hence, the marginal distribution $q^{\bs{x}}_k(\bs{x}_k)$ is Gaussian. Its mean can be computed by setting the gradient with respect to $\bs{x}_k$ to 0, leading to
\begin{multline}
\langle\bs{x}_k\rangle=
\bs{P}^{\bs{x}}_k\ls \bs{C}_k^{\T}\bs{R}_k^{-1}\bs{y}_k+\bs{Q}_{k-1}^{-1}\bs{B}_{k-1}\langle\bs{u}_{k-1}\rangle\rd\\
-\lb \bs{C}_k^{\T}\bs{R}_k^{-1}\bs{D}_k+\bs{A}_k^{\T}\bs{Q}^{-1}_k\bs{B}_k\rb \langle\bs{u}_k\rangle\\
\ld+\bs{Q}_{k-1}^{-1}\bs{A}_{k-1}\langle\bs{x}_{k-1}\rangle+\bs{A}_k^{\T}\bs{Q}_{k}^{-1}\langle\bs{x}_{k+1}\rangle\rs,\label{eq:VB_x_update}
\end{multline}
where we define
\begin{equation}
\bs{P}^{\bs{x}}_k=\lb \bs{C}_k^{\T}\bs{R}_k^{-1}\bs{C}_k+\bs{Q}_{k-1}^{-1}+\bs{A}_k^{\T}\bs{Q}_k^{-1}\bs{A}_k\rb ^{-1}.\label{eq:VB_Px_update} 
\end{equation}
Similarly, the marginal distribution of $\bs{u}_k$ is computed as
\begin{align}
\ln q^{\bs{u}}_k(\bs{u}_k)&\propto \mathbb{E}_{q\lb \mathcal{Z}\setminus\bs{u}_k\rb }\lc\ln p\lb \mathcal{Z},\bs{Y}_1^K\rb \rc\\
&\propto\Vert\bs{y}_k-\bs{C}_k\langle\bs{x}_k\rangle-\bs{D}_k\bs{u}_k\Vert_{\bs{R}_k}^2\notag\\
&\hspace{0.35cm}+\Vert\bs{x}_{k+1}\!-\!\bs{A}_{k}\langle\bs{x}_k\rangle\!-\!\bs{B}_{k}\bs{u}_{k}\Vert_{\bs{Q}_{k}}^2
+ \sum_{i=1}^{m}\bs{u}_k(i)^{2}\!\langle\bs{\beta}_k(i)\rangle\!.
\end{align}
 The mean of the Gaussian distribution $q^{\bs{u}}_k(\bs{x}_k)$ is
 \begin{multline}
 \langle\bs{u}_k\rangle=\bs{P}^{\bs{u}}_k\ls\bs{D}^{\T}_k\bs{R}_k^{-1}\bs{y}_k-\lb \bs{D}^{\T}_k\bs{R}_k^{-1}\bs{C}_k+\bs{B}_k^{\T}\bs{Q}^{-1}_{k}\bs{A}_k\rb\rd\\\times \langle\bs{x}_k\rangle
 \ld+\bs{B}_k^{\T}\bs{Q}_k^{-1}\langle\bs{x}_{k+1}\rangle\rs,\label{eq:VB_u_update}
 \end{multline}
 where the matrix $\bs{P}^{\bs{u}}_k$ is
 \begin{equation}
\bs{P}^{\bs{u}}_k=\lb \bs{D}_k^{\T}\bs{R}_k^{-1}\bs{D}_k+\bs{B}_k^{\T}\bs{Q}_k^{-1}\bs{B}_k+\langle\operatorname{diag}\lc \bs{\beta}_k \rc\rangle\rb ^{-1}.
\end{equation}
We use $\boldsymbol{x}_{K+1}=\boldsymbol{0}$ for $k=K$ and $\boldsymbol{x}_{0}=\boldsymbol{0}$ for $k=1$ in \eqref{eq:VB_x_update} and \eqref{eq:VB_u_update}.
Finally, $q(\bs{\beta}_k)=\prod_{i=1}^m q(\bs{\beta}_k(i))$ is a Gamma distribution with mean
\begin{equation}
\langle \bs{\beta}_k(i)\rangle = \frac{a+0.5}{b+0.5\langle\bs{u}_k^2(i)\rangle}= \frac{a+0.5}{b+0.5\ls \langle\bs{u}_k(i)\rangle^2 + \bs{P}^{\bs{u}}_k(i,i)\rs}.\label{eq:VB_RKS_beta_update}
\end{equation}
Using \eqref{eq:VB_x_update}, \eqref{eq:VB_u_update}, and \eqref{eq:VB_RKS_beta_update}, the marginal distribution parameters are iteratively updated until convergence to obtain the approximate posterior distribution. The pseudocode is summarized in \Cref{alg:vb_RobustKalman}. 

The VB-RKS relies on the majorization-maximization procedure, it is guaranteed to converge to a stationary point of the posterior log-likelihood function. Also, when all the inputs are jointly sparse, similar to SBL-RKS, we use a common prior $\bs{u}_k\sim\mathcal{N}\lb \bs{0},\operatorname{Diag}\{\bs{\beta}\}\rb $, i.e., $\bs{\beta}_k=\bs{\beta}$ for $k=1,2,\ldots,K$. 
In that case, the VB-RKS for joint sparse input recovery
is identical to \Cref{alg:vb_RobustKalman} except that \eqref{eq:VB_RKS_beta_update} in Step 6 changes as follows. We refer this algorithm to as ultiple measurement vector VB-RKS (MVB-RKS).
\begin{equation}
\langle \bs{\beta}(i)\rangle = \frac{a+0.5}{b+\frac{0.5}{K}\sum_{k=1}^K\langle\bs{u}_k^2(i)\rangle}
\end{equation}
\noindent\emph{Remark:} A special case of our problem is the conventional KF problem when $\bs{u}_k=\bs{0}$ $\forall k$. The update steps consist of only two equations \eqref{eq:VB_x_update} and \eqref{eq:VB_Px_update} with $\bs{B}_k=\bs{0}$, $\bs{D}_k=\bs{0}$. The state update equation is given by,
\begin{align}
\langle\bs{x}_k\rangle=\bs{P}^{\bs{x}}_k\Big[\bs{C}_k^{\T}\bs{R}_k^{-1}\bs{y}_k+\bs{Q}_{k-1}^{-1}\bs{A}_{k-1}\langle\bs{x}_{k-1}\rangle+\bs{A}_k^{\T}\bs{Q}_{k}^{-1}\langle\bs{x}_{k+1}\rangle\Big]
\end{align}\vspace{-0.2cm}
\begin{algorithm}[ht]
	\caption{Variational Bayesian RKS}
	\setstretch{1.2}
	\begin{algorithmic}[1]
		\REQUIRE $\{\bs{y}_k, \bs{A}_k,\bs{B}_k,\bs{C}_k,\bs{D}_k,\bs{Q}_k,\bs{R}_k\}_{k=1}^K$
		\STATEx \hspace{-0.5cm}\textbf{Parameters:} $r_{\max}$ and $\tilde{r}_{\max}$
		\STATEx \hspace{-0.5cm}\textbf{Initialization:} $\langle\bs{x}_{k}\rangle=\bs{0}$, $\langle\bs{u}_{k}\rangle=\bs{0}$, $\langle\bs{\beta}_{k}\rangle=\bs{1}$ for $k=1,2,\ldots,K$
		\FOR{$r=1,2,\ldots,r_{\max}$}
        \FOR{$\tilde{r}=1,2,\ldots,\tilde{r}_{\max}$}
        \STATE Compute $\bs{x}_{k}^{(r,\tilde{r})}=\langle\bs{x}_{k}\rangle$ using \eqref{eq:VB_x_update} for $k=1,\ldots,K$ 
        \STATE Compute $\bs{u}_{k}^{(r,\tilde{r})}=\langle\bs{u}_{k}\rangle$ using \eqref{eq:VB_u_update} for $k=1,\ldots,K$ 
		\ENDFOR
        \STATE Compute $\bs{\beta}_k^{(r)}=\langle\bs{\beta}_{k}\rangle$ using \eqref{eq:VB_RKS_beta_update} for $k=1,\ldots,K$ 
        \ENDFOR
		\ENSURE $\left\{\bs{x}_{k}^{(r,\tilde{r})}\right\}_{k=1}^K$ and $\left\{\bs{u}_{k}^{(r,\tilde{r})}\right\}_{k=1}^K$
	\end{algorithmic}
	\label{alg:vb_RobustKalman}
\end{algorithm}
\vspace{-0.2cm}
\subsection{Comparison of Regularized RKS and Bayesian-RKS}\label{sec:comparison}
We next compare the two approaches discussed above to get insights on choosing between them. 
\subsubsection{Parameter Tuning}
Both regularization and Bayesian RKS approaches make appropriate statistical assumptions (like \eqref{eq:convex_prior} and \eqref{eq:sbl_prior}) on the solution and apply estimation techniques to identify the desired sparse solution. However, they use different solution approaches. The regularized RKS assumes the knowledge of its parameters (such as $c, \tau_k$) and uses type I ML estimation to compute the states and inputs. In contrast, Bayesian-RKS uses type II ML estimation, i.e., a  hierarchical Bayesian framework, and learns the hyperparameters from the measurements. Consequently, Bayesian-RKS does not require hand-tuning of any sensitive parameters. The performance of the algorithm does not drastically change with the small changes in its parameters  $r_{\max}$ (and  $\tilde{r}_{\max}$), which is set to be a large value. However, parameter $\tau_k$ of the regularized RKS determines the sparsity of the solution and is typically chosen using a trial-and-error method. This parameter calibration is a drawback of all regularized sparse recovery methods compared to the Bayesian RKS methods. 

\subsubsection{Complexity Comparisons}\label{sec:complexity}
All four algorithms, $\ell_1$-regularized RKS, reweighed $\ell_2$-regularized RKS, SBL-RKS, and VB-RKS are iterative. Each iteration of all the algorithms has time complexity $\mathcal{O}(K(n^3+m^3+p^3))$ for the versions with and without the joint sparsity assumption. The computational complexity of regularized RKS is dominated by the matrix inversion in the RKS algorithm (Steps 5 and 12 of \Cref{alg:RobustKalman}) used by both regularized RKS algorithms. Similarly, the complexity of SBL-RKS is dominated by the matrix inversions in Steps 8 and 13 of \Cref{alg:SBL_RKS}. Since the sparsity-driven algorithms consider low-dimensional measurements where $m\geq p$, the time complexity reduces to $\mathcal{O}(K(n^3+m^3))$. For comparison, we consider the state-of-the-art $\ell_1$ minimization-based algorithm, referred to as basis pursuit (BP)-RKS (group BP-RKS for the joint support case). We have derived BP-RKS and group BP-RKS by extending the algorithm in~\cite{sefati2015linear} to handle noise and joint support recovery. BP-RKS is a non-iterative algorithm whose complexity scales as $\mathcal{O}(K^{\frac{7}{2}}m^{\frac{3}{2}}p^2+K(n^3+p^3))$ due to the convex programming optimization using the interior point method. So, our algorithms have low complexity order when the number of iterations is small.

Further, we have observed from our simulation results that the other algorithms require a larger number of iterations and a longer time to converge than SBL-RKS (see \Cref{table:runtime}). In particular, VB-RKS takes a longer run time owing to the large number of iterations required for convergence, although the asymptotic per iteration complexity of VB-RKS is the same as others. Also, unlike the standard sparse recovery problems in the compressed sensing literature where the SBL-based approach is typically slower than regularization-based methods, SBL-RKS is surprisingly faster than the regularized RKS algorithms. In our case, SBL-RKS relies on the conventional Kalman filtering and smoothing algorithms (Steps 5 to 18 of \Cref{alg:SBL_RKS}) due to the underlying Gaussian assumption. On the contrary, the regularized RKS uses the RKS algorithm (\Cref{alg:RobustKalman}), which has more matrix inversions and multiplications compared to the conventional Kalman filtering and smoothing. Hence, although the asymptotic computational complexity is the same for all the algorithms, SBL-RKS is more computationally efficient than the regularized RKS in practice. 

The auxiliary space complexity of all the algorithms is $\mathcal{O}(p^2+K(n^2+m^2))$. Here, the term $K(n^2+m^2)$ arises because we store $\bs{P}_{k \mid k}^{\bs{\xi}}\in\bb{R}^{(n+m)\times (n+m)}$ for $k=1,2,\ldots,K$ in each iteration. In the low-dimensional measurement regime where $m\geq p$, the space complexity of all the algorithms is $\mathcal{O}(K(n^2+m^2))$. 

In summary, we prefer SBL-RKS over the regularized RKS and VB-RKS algorithms due to its low complexity, hand-tuning-free approach, and superior recovery accuracy (see \Cref{sec:simulation} for the empirical results).
\section{Simulation Results}
\label{sec:simulation}

\begin{figure*}[!htbp]
	\centering
	\begin{subfigure}[t]{0.33\textwidth}
		\centering
		\includegraphics[width=\textwidth]{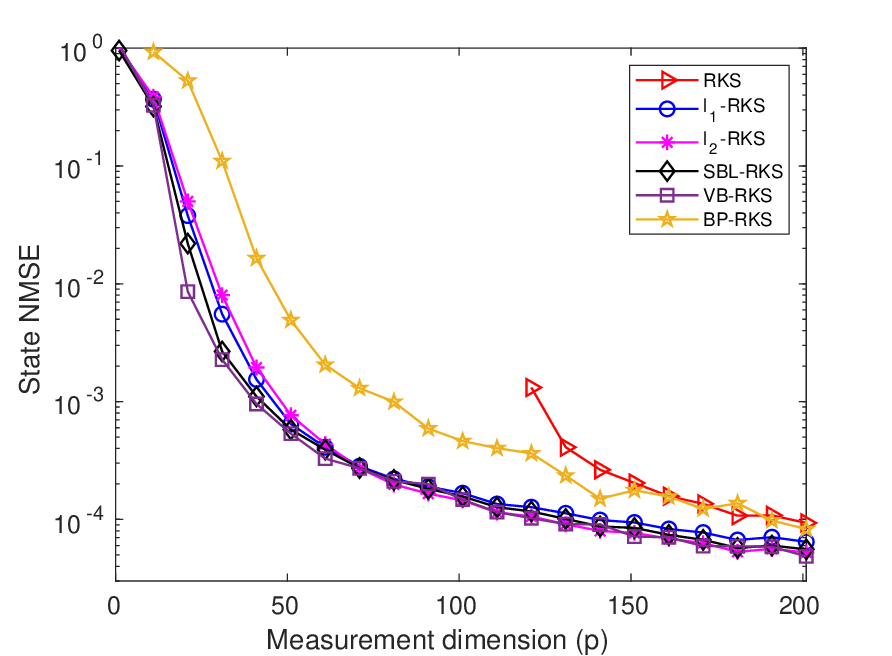}
		\caption{NMSE in state estimation}
		\label{fig:NMSE_state_varying}
	\end{subfigure}
	\begin{subfigure}[t]{0.33\textwidth}
		\centering
		\includegraphics[width=\textwidth]{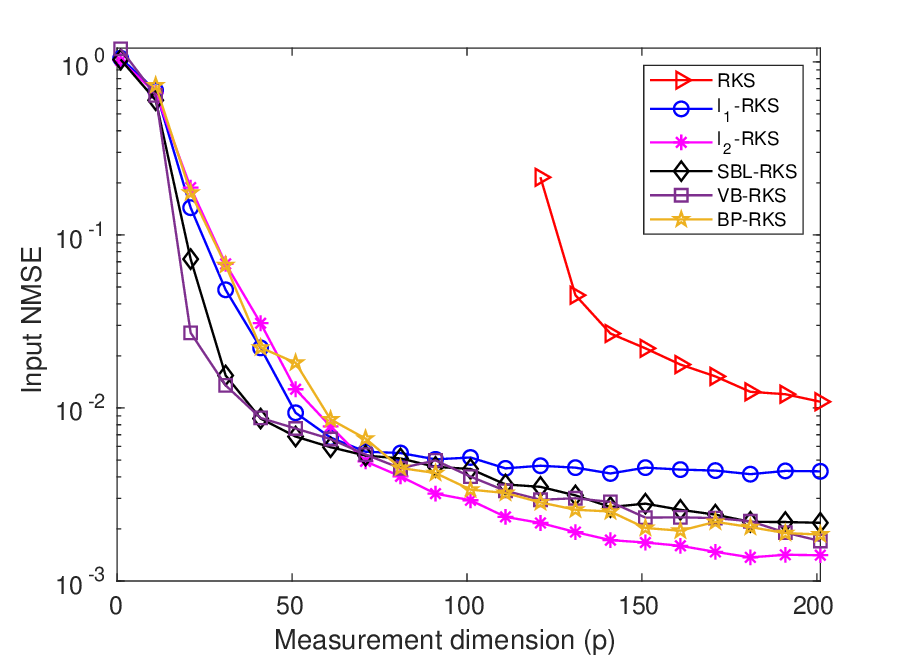}
		\caption{NMSE in input estimation}
		\label{fig:NMSE_input_varying}
	\end{subfigure}
	\begin{subfigure}[t]{0.33\textwidth}
		\centering
		\includegraphics[width=\textwidth]{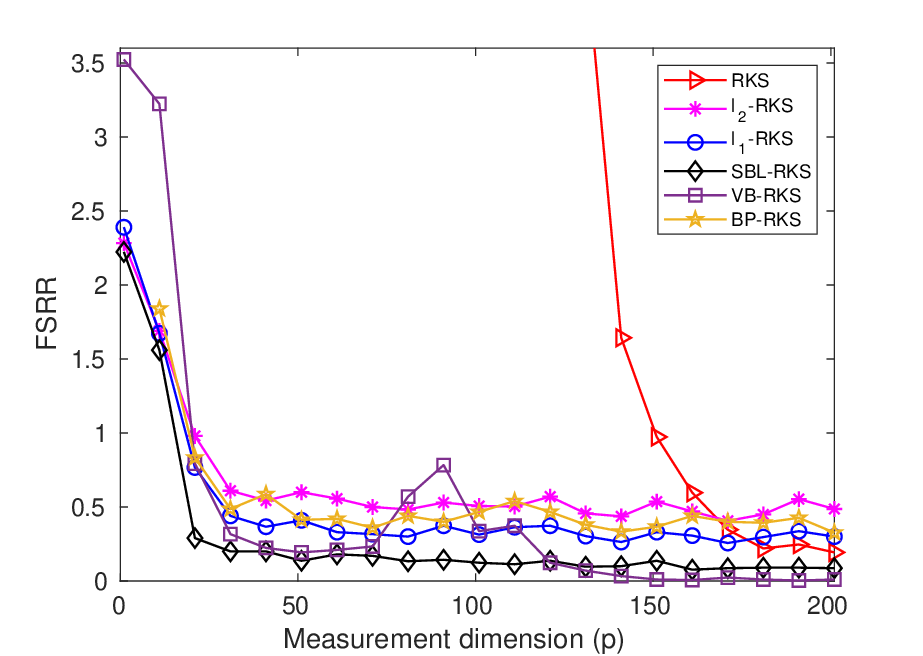}
		\caption{Percentage of false support recovery}
		\label{fsr_varying}
	\end{subfigure}
	\caption{Performance comparison of our sparse recovery algorithms and RKS as a function of measurement dimension $p$ when the support of control inputs are time-varying support with $n=30$, $m=100$, $K=30$, $s=5$, and SNR $=20$~dB.}
	\label{fig:varying}
\end{figure*}
\begin{figure*}[!htbp]
	\centering
	\begin{subfigure}[t]{0.32\textwidth}
		\centering
		\includegraphics[width=\textwidth]{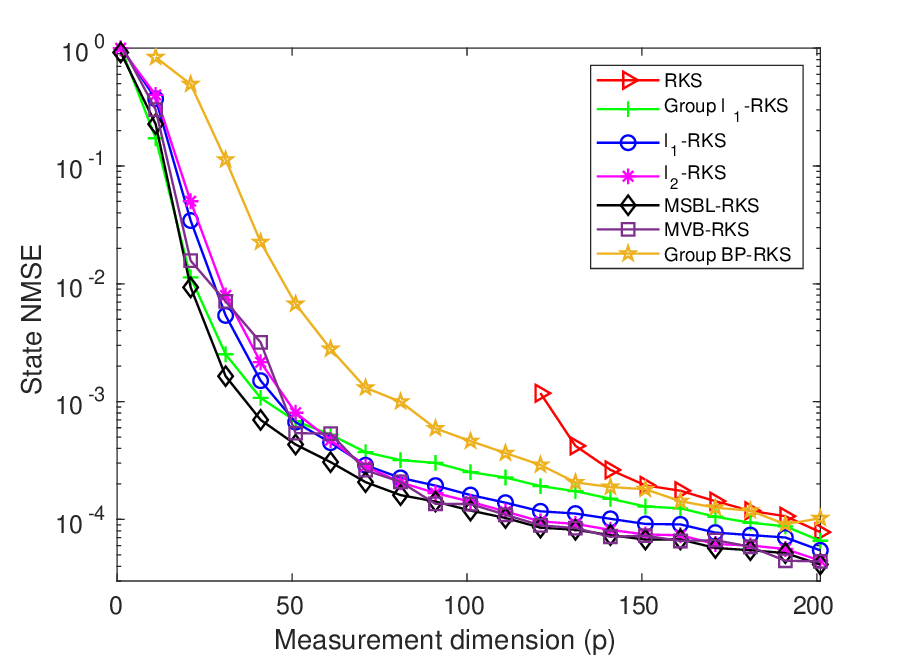}
		\caption{NMSE in state estimation}
		\label{fig:NMSE_state_invariant}
	\end{subfigure}
	\begin{subfigure}[t]{0.32\textwidth}
		\centering
		\includegraphics[width=\textwidth]{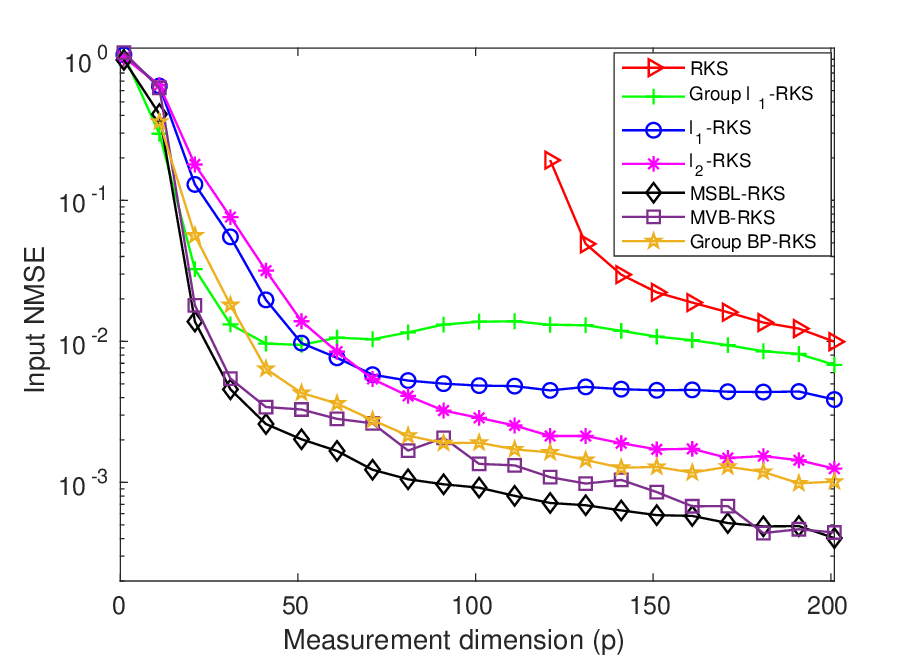}
		\caption{NMSE in input estimation}
		\label{fig:NMSE_input_invariant}
	\end{subfigure}
	\begin{subfigure}[t]{0.32\textwidth}
		\centering
		\includegraphics[width=\textwidth]{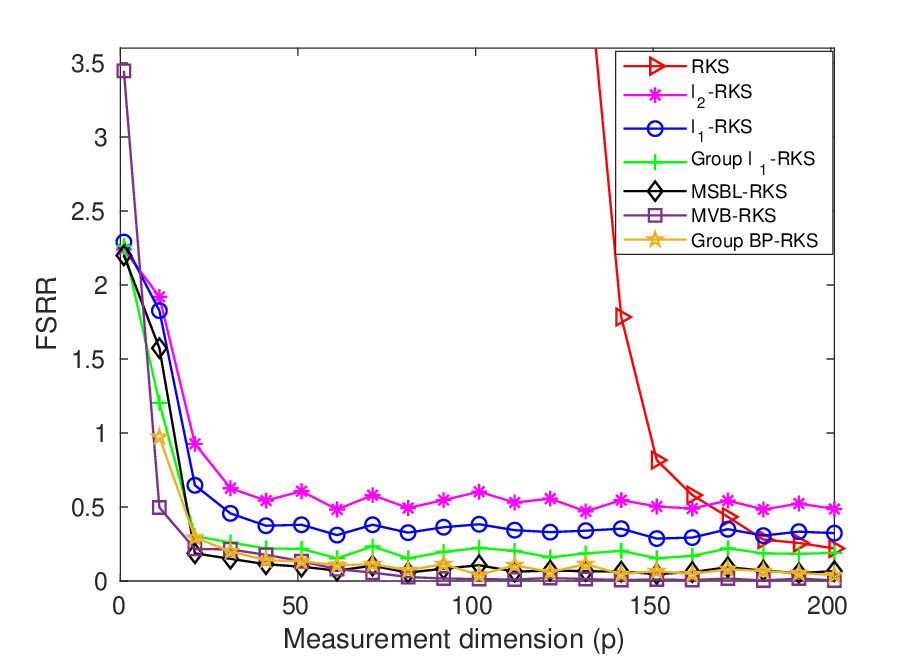}
		\caption{Percentage false support recovery rate}
		\label{fsr_invariant}
	\end{subfigure}
	\caption{Performance comparison of our sparse recovery algorithms and RKS as a function of measurement dimension $p$ when the control inputs are jointly sparse with $n=30$, $m=100$, $K=30$, $s=5$, and SNR  $=20$~dB.}
	\label{fig:invariant}
\end{figure*}
\begin{figure*}[!htb]
\minipage{0.3\textwidth}
\includegraphics[width=\textwidth]{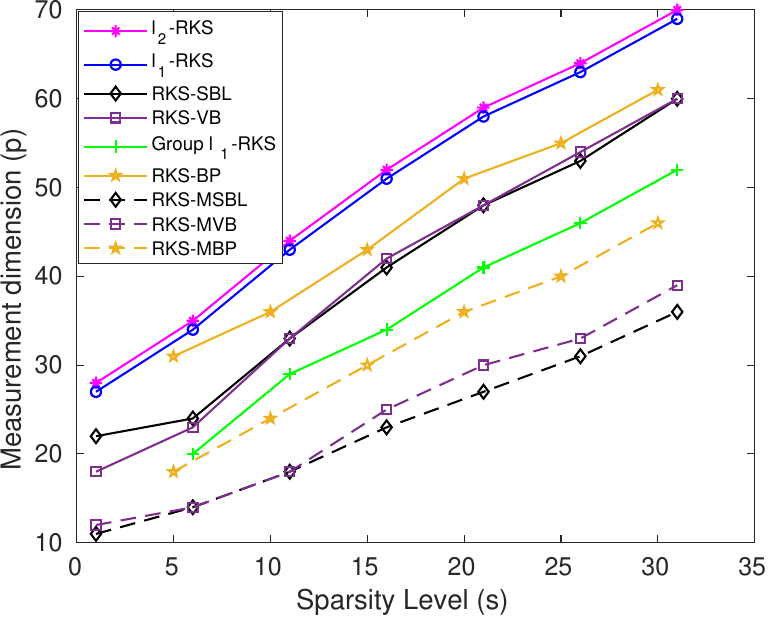}
  \caption{Phase transition diagram for our sparse recovery algorithms with $n=30$, $m=100$, $K=30$, and SNR $=20$~dB.}
	\label{fig:phasedia}
\endminipage
\hfill
\minipage{0.66\textwidth}
\centering
 \begin{subfigure}[t]{0.48\textwidth}
		\centering
		\includegraphics[width=\textwidth]{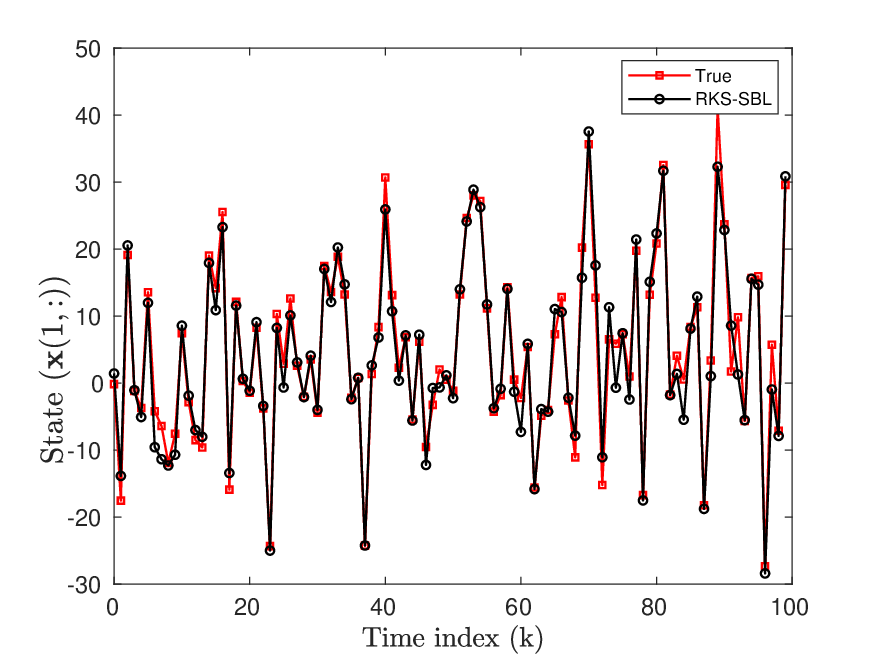}
		\caption{Time tracking of state}
		\label{fig:time_state}
	\end{subfigure}
 	\begin{subfigure}[t]{0.48\textwidth}
		\centering
		\includegraphics[width=\textwidth]{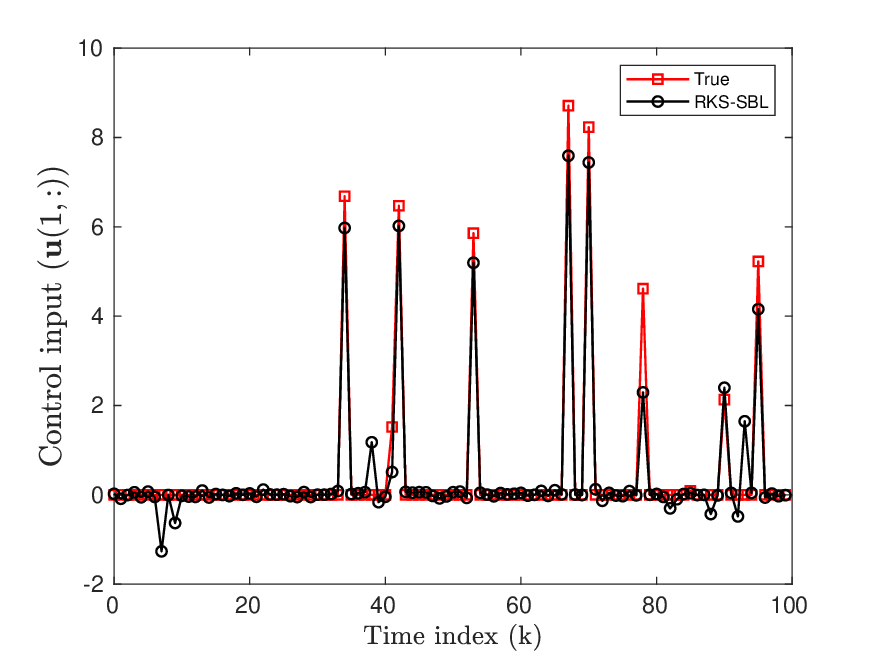}
		\caption{Time tracking of input}
		\label{fig:time_inp}
	\end{subfigure}
	\caption{Time domain tracking performance of SBL-RKS with $n=30$, $m=100$, $p=20$, $K=100$, $s=5$, and SNR $=20$~dB.}
	\label{fig:time_tracking}
\endminipage
\end{figure*}
\begin{figure}[!htbp]
	\centering
	\begin{subfigure}[t]{0.45\textwidth}
  \includegraphics[width=\textwidth]{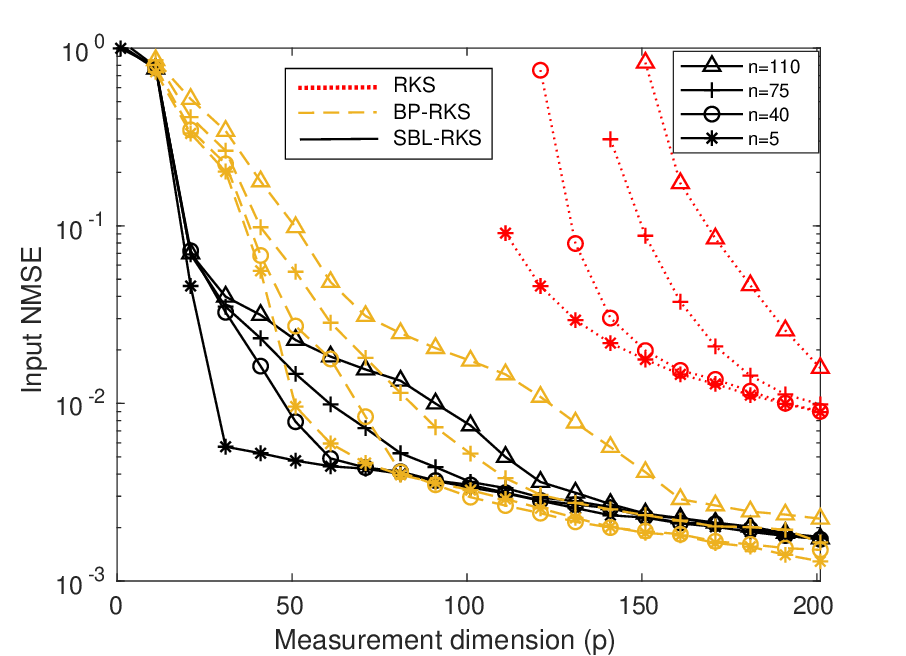}
  \caption{Variation w.r.t. state dimension; SNR $=20$~dB.}
 			\label{fig:dimension}
\end{subfigure}

	\begin{subfigure}[t]{0.45\textwidth}
  \includegraphics[width=\textwidth]{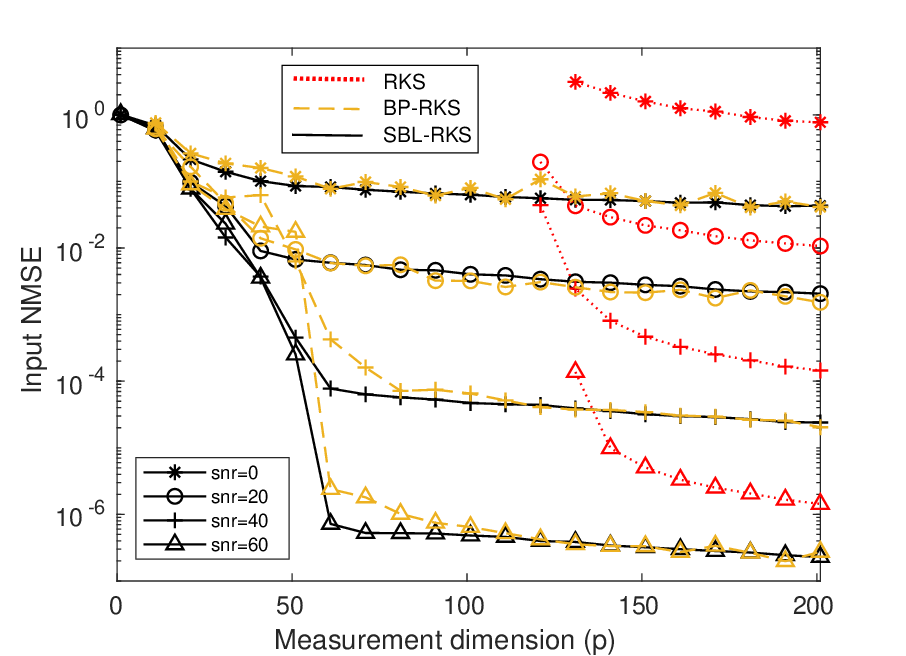}
  \caption{Variation w.r.t. SNR; $n=30$.}
			\label{fig:SNR}
   \end{subfigure}
   \caption{NMSE in input estimation of SBL-RKS and RKS as a function of measurement dimension when the control inputs have time-varying support, with  $m=100$, $K=30$, and $s=5$.}\label{fig:dim_snrVsp}
\end{figure}

In this section, we present the empirical results that demonstrate the superior performance of the algorithms that exploit sparsity. Our setting is as follows. We choose the state dimension $n=30$, the input dimension $m=100$, the output dimension $p=20$, and the number of time steps $K=30$. The sparsity level of the input is $s=5$, and the locations of $s$ nonzero entries are chosen uniformly at random from the set $\{1,2,\ldots,m\}$. Further, the nonzero entries are drawn independently from a normal distribution $\mathcal{N}(0,\sigma_u^2)$ with $\sigma_u=5$. For the time-varying support case, we choose different support for each time instant $k$, and for the jointly sparse case, we use the same support for all values of $k$. We set the system matrices, $\bs{A}_k=\bs{A}$, $\bs{B}_k=\bs{B}$, $\bs{D}_k=\bs{C}$, and $\bs{D}_k=\bs{D}$, for all values of $k=1,2,\ldots,K$. The entries of the matrices $\bs{A},\bs{B},\bs{C}$, and $\bs{D}$ and the initial state $\bs{x}_1$ are independently drawn from the standard normal distribution. Also, the process noise covariance $\bs{Q}$ and the measurement noise covariance $\bs{R}$ are the identity matrix and $\sigma_v^2\eye,$ respectively. Finally, $\sigma_v$ is computed from the measurement SNR via the relation $\mathrm{SNR} =s\sigma_u^2/\sigma_v^2$. 

For the above setting, we compare the performance of our algorithms: $\ell_1$-regularized RKS, reweighed $\ell_2$-regularized RKS, SBL-RKS and {VB-RKS} for the time-varying support and jointly sparse cases. We also consider two benchmark approaches: basis pursuit (BP)-RKS and group BP-RKS, which are adapted from the algorithm in~\cite{sefati2015linear} (see Appendix-III for details), and the RKS algorithm (\Cref{alg:RobustKalman}). The following metrics are used for comparison: normalized mean squared error (NMSE) in the state and input estimation, false support recovery rate (FSRR) for input estimation, and run time. Here, FSRR is defined as the ratio of the Hamming distance between the true and estimated support vectors and the length of the support vectors. For this computation, we define the estimated support vector as a binary vector with ones corresponding to the entries greater than 0.8$\sigma_u$ and zeros elsewhere. Thus, the FSRR is the sum of the false alarm and missed detection rates of the support estimation. The results are summarized in \Cref{fig:varying,fig:invariant,fig:time_tracking,fig:dim_snrVsp,fig:phasedia} and \Cref{table:runtime}. Since the RKS algorithm requires a large $p$ to estimate the states and input, in most cases, we plot the performance of the algorithms as a function of $p$, the measurement dimension. We make the following observations:

\begin{table}[t]
\caption{Run time comparison of our algorithms when $n=30$, $p=20$,  $m=100$, $K=30$, $s=5$ and SNR is 20 dB}
\centering
	\begin{tabular}{| c |c | c|} 
		\hline
		\bf Support & \bf Algorithm & \bf Runtime \\ 
		\hline
		\hline
		\multirow{5}{*}{Time varying}&
        BP-RKS & 73.94~s\\
        \cline{2-3}
		&$\ell_1$-regularized RKS & 33.76~s\\ 
		\cline{2-3}
		&Reweighed $\ell_2$-regularized RKS & 44.5~s\\ 
		\cline{2-3}
	&	SBL-RKS & 14.5~s\\ 
        \cline{2-3}
	    &VB-RKS & $\sim$5~min\\
		\hline
		\hline
		\multirow{4}{*}{\begin{tabular}{c}
		     Jointly sparse
		\end{tabular}}
        & Group BP-RKS & 52~s\\
        \cline{2-3}
		& Group $\ell_1$-regularized RKS & 34.6~s\\
        \cline{2-3}
        &MSBL-RKS & 13.5~s\\
         \cline{2-3}
        &MVB-RKS & $\sim$5~min\\
		\hline
	\end{tabular}
\label{table:runtime}
\end{table}

\subsubsection{Comparison with RKS}
From \Cref{fig:varying,fig:invariant}, we infer that the conventional Kalman filtering algorithm, RKS, has poor NMSE performance compared to the sparsity-driven approaches. This observation underscores the importance of exploiting sparsity, as it helps the algorithms to achieve better NMSE. RKS relies on Assumption A4, and hence, it requires $p>m$. Naturally, the algorithm fails in the low-dimensional measurement regime. Also, the NMSE of RKS is comparable to that of the sparsity-driven algorithms only when $p>m$, but the latter outperform RKS even in that regime. 

\subsubsection{Comparison of Sparsity-driven Algorithms} \Cref{fig:varying,fig:invariant} and \Cref{table:runtime} show that the SBL-RKS and VB-RKS outperform BP-RKS and regularized RKS in terms of NMSE in both states and input estimation, FSRR, and run time, in line with our arguments in \Cref{sec:complexity}. VB-RKS shows similar performances as SBL-RKS except for runtime which is much higher for VB-RKS. As noted earlier, VB-RKS's longer run time is due to the large number of iterations it takes for VB-RKS to converge. The NMSE in state estimation is comparable for all five sparsity-driven algorithms, but there is a slight difference in input estimation performance. For the time-varying support case, the reweighted $\ell_2$-regularized RKS slightly outperforms the other schemes in terms of input NMSE, but SBL-RKS and VB-RKS have the best FSRR. Moreover, SBL-RKS has better run time compared to the other algorithms, making it a better choice. For the joint sparsity case, MSBL-RKS and MVB-RKS have the best NMSE and FSRR values, but MSBL-RKS has the shortest run time. 

\subsubsection{Time-varying Support and Joint Sparsity}
\Cref{fig:varying,fig:invariant} and \Cref{table:runtime} show the different algorithms' performances for the time-varying support and joint sparsity cases. Comparing \Cref{fig:varying,fig:invariant}, we see that the trends in NMSE and FSRR are the same for both cases. Nonetheless, for the same number of measurements $p$, the NMSE in input estimation and FSRR is better for the joint sparsity-aware algorithms: group BP-RKS, group $\ell_1$-regularized RKS, MSBL-RKS, and MVB-RKS. This behavior is because of the additional joint sparse structure exploited by these algorithms. Further, this difference is explicitly shown using the phase transition diagram in \Cref{fig:phasedia}, which plots the minimum value of the number of measurements $p$ required for $90\%$ recovery accuracy (i.e., successful recovery of the sparse signals in $90 \%$ of the random experiments). Here, a sparse vector is said to be successfully recovered if the normalized mean square error between the original signal and the recovered signal is below 0.05. Clearly, the Bayesian RKS algorithms require the least number of measurements followed by group BP-RKS and group $\ell_1$-regularized RKS. The joint sparsity-aware algorithms are followed by Bayesian-RKS, BP-RKS, and regularized RKS. The regularized RKS algorithms have similar phase transition curves, but SBL-RKS and VB-RKS require fewer measurements than the regularized RKS. Finally, \Cref{table:runtime} indicates that joint sparsity-aware algorithms have a shorter run time than their counterparts for time-varying support case because they have fewer parameters to estimate. 

\subsubsection{Time domain tracking of states and inputs}
The time domain state and input tracking performance of SBL-RKS is shown in \Cref{fig:time_tracking} for visual comparison. Here, the estimate of one entry in the state and input vectors across time obtained using SBL-RKS are depicted in \Cref{fig:time_state} and in \Cref{fig:time_inp}, respectively. It is clear from \Cref{fig:time_inp} that SBL-RKS is able to recover the spikes in the entries of input vectors accurately. Also, \Cref{fig:time_state} shows that SBL-RKS tracks the state tightly. We observe similar results for all the other sparsity-aware algorithms, and hence, we omit plotting them to avoid clutter.

\subsubsection{Dependence on State Dimension and SNR}
\Cref{fig:dimension} shows the NMSE in spase input recovery when the state dimension is varied from $n=5$ to $n=110$. We observe that as the state dimension increases, we need more measurements to achieve the same NMSE. However, when the state dimension is large, all the curves  converge to the same NMSE, with the RKS curves converging to a much larger NMSE compared to SBL-RKS and BP-RKS. We omit the curves corresponding to the regularized RKS and VB-RKS, but they follow a similar trend. Further, \Cref{fig:SNR} plots NMSE in the input estimation when SNR varies from $0$~dB to $80$~dB  for the RKS, BP-RKS, and SBL-RKS algorithms. NMSE decreases with an increase in SNR for a given number of measurements, as expected. Also, for the same number of measurements, SBL-RKS and BP-RKS give a lower NMSE than RKS. \Cref{fig:dimension,fig:SNR} together evidence that there is a minimum number of measurements required for joint state and input recovery, beyond which the process and measurement noise levels determine the NMSE.
\section{Conclusion}
\label{eq:sec:comclusion}
In this paper, we studied the joint estimation of states and sparse inputs as an observer design problem in an LDS. We developed novel algorithms for the joint estimation using sparsity-promoting priors on the inputs. We used a regularization-based MAP estimation approach and a Gaussian prior-based hierarchical Bayesian learning approach. We empirically showed that exploiting sparsity enhances the estimation performance, and the hierarchical Bayesian learning has better recovery performance than all the other algorithms. We also extended our approaches for the jointly sparse input case and demonstrated the efficacy of exploiting additional structures with sparsity. The theoretical guarantees of our algorithms and the fundamental limits of sparse recovery algorithms for an LDS are interesting directions for future work. Future work can also extend our algorithms to handle system identification or estimation of the system matrices.

\appendices

\section{The Kalman Filtering and Smoothing Steps} \label{app:Filtering proof}
\subsection{Prediction}
In the prediction step, we compute the MAP estimate of $\bs{x}_{k}$ and $\bs{u}_{k}$ given measurements up to time $k-1$. The posterior pdf $p\left(\bs{x}_{k} \mid \bs{Y}_1^{k-1}\right)$ can be expressed as~\cite{fang2013simultaneous}
\begin{align}
&p\left(\bs{x}_{k} \mid \bs{Y}_1^{k-1}\right)=\nonumber\\
&\iint\hspace{-1mm}p\left(\bs{x}_{k} \mid \bs{x}_{k-1},\bs{u}_{k-1}\right) p\left(\bs{x}_{k-1},\bs{u}_{k-1} \mid \bs{Y}_1^{k-1}\right) \mathrm{~d}\bs{x}_{k-1} \mathrm{d}\bs{u}_{k-1}. \label{pred}
\end{align}
We make use of the following well known result to compute the right hand side of the expression in \eqref{pred}.
\begin{lemma}\label{lemma1}
	If
	$p\left(\boldsymbol{\beta} \mid \boldsymbol{\alpha}\right) = \mathcal{N}\left(\bs{A}\boldsymbol{\alpha}+\bs{c}, \boldsymbol{\Sigma}_{\boldsymbol{\beta}\mid \boldsymbol{\alpha}}\right)$ and\\
	$p\left(\boldsymbol{\alpha}\right) = \mathcal{N}\left(\boldsymbol{\mu}_{\boldsymbol{\alpha}}, \boldsymbol{\Sigma}_{\boldsymbol{\alpha}}\right)$
	then $p\left(\boldsymbol{\beta}\right)=\int p\left(\boldsymbol{\beta} \mid \boldsymbol{\alpha}\right)p\left(\boldsymbol{\alpha}\right)\mathrm{d}\boldsymbol{\alpha} = \mathcal{N}\left(\boldsymbol{\mu}_{\boldsymbol{\beta}}, \boldsymbol{\Sigma}_{\boldsymbol{\beta}}\right)$ where
	\begin{gather}
 \boldsymbol{\mu}_{\boldsymbol{\beta}}=\bs{A}\boldsymbol{\mu}_{\boldsymbol{\alpha}}+\bs{c}\\
	\boldsymbol{\Sigma}_{\boldsymbol{\beta}}=\boldsymbol{\Sigma}_{\boldsymbol{\beta}\mid \boldsymbol{\alpha}}+\bs{A}\boldsymbol{\Sigma}_{\boldsymbol{\alpha}}\bs{A}^{\T}.
	\end{gather}
\end{lemma}
In \eqref{pred}, the first pdf under the integral is $p\left(\bs{x}_{k} \mid \bs{x}_{k-1}, \bs{u}_{k-1}\right) = \mathcal{N}\left(\bs{A}_{k-1}\bs{x}_{k-1}+\bs{B}_{k-1}\bs{u}_{k-1}, \bs{Q}_{k-1}\right)$ due to the markovian nature of the dynamics and $p\left(\bs{x}_{k} \mid \bs{Y}_1^{k-1}\right)$ is also Gaussian due to  \eqref{gaussian_approx1}. Now, applying Lemma 1, we can compute the prediction of $\bs{x}_k$ as its posterior mean, as follows:
\begin{align}
\hat{\bs{x}}_{k\mid k-1}=\underset{\bs{x}_k}{\arg \max}\hspace{2mm}p\left(\bs{x}_{k}\mid \bs{Y}_1^{k-1}\right).
\end{align}
Hence, we get the update equations in Steps 3 and~4 in \Cref{alg:RobustKalman}. 
\subsection{Filtering Steps} \label{app:KFS_filtering}
To solve the minimization problem in \eqref{eq:xu_min_filtering}, we differentiate $\mathcal{L}$ with respect to $\bs{x}_k$ and $\bs{u}_k$ and set it to $0$ to get
\begin{align}
\nabla_{\bs{x}_k}\mathcal{L}&=-2\bs{C}_k^{\T}\bs{R}_k^{-1}(\bs{y}_k-\bs{C}_k\bs{x}_k-\bs{D}_k\bs{u}_k)\nonumber\\
&\hspace{2cm}+2(\bs{P}_{k\mid k-1}^{\bs{x}})^{-1}(\bs{x}_k-\hat{\bs{x}}_{k\mid k-1})=0\label{gradL_x}\\
\hat{\bs{x}}_{k\mid k}&=\big((\bs{P}_{k\mid k-1}^{\bs{x}})^{-1}+\bs{C}_k^{\T}\bs{R}_k^{-1}\bs{C}_k\big)^{-1}\big((\bs{P}_{k\mid k-1}^{\bs{x}})^{-1}\hat{\bs{x}}_{k\mid k-1}\nonumber\\
&\hspace{3cm}+\bs{C}_k^{\T}\bs{R}_k^{-1}(\bs{y}_k-\bs{D}_k\hat{\bs{u}}_{k\mid k})\big).\label{x_hat_thm1}
\end{align}
We use the matrix inversion Lemma on the first factor of \eqref{x_hat_thm1} to get
\begin{gather}
\big((\bs{P}_{k\mid k-1}^{\bs{x}})^{-1}+\bs{C}_k^{\T}\bs{R}_k^{-1}\bs{C}_k\big)^{-1}=\Big(\bs{I}_n-\bs{L}_k\bs{C}_k\Big)\bs{P}^{\bs{x}}_{k\mid k-1}\label{P_L_rel}\\
\text{where \,\,} \bs{L}_k\triangleq\bs{P}_{k\mid k-1}^{\bs{x}}\bs{C}_k^{\T}\big(\bs{R}_k+\bs{C}_k\bs{P}_{k\mid k-1}^{\bs{x}}\bs{C}_k^{\T}\big)^{-1}.\label{identity}
\end{gather}
Substituting \eqref{P_L_rel} in \eqref{x_hat_thm1}, we get,
\begin{align}
\hat{\bs{x}}_{k\mid k} 
&=\Big(\bs{I}_n-\bs{L}_k\bs{C}_k\Big)\hat{\bs{x}}_{k\mid k-1}\nonumber\\&\hspace{0.5cm}+\Big(\bs{I}_n-\bs{L}_k\bs{C}_k\Big)\bs{P}^{\bs{x}}_{k\mid k-1}\bs{C}_k^{\T}\bs{R}_k^{-1}(\bs{y}_k-\bs{D}_k\hat{\bs{u}}_{k\mid k}).\label{x_hat_thm1_int}
\end{align}
The term pre-multiplied with $\bs{y}_k-\bs{D}_k\hat{\bs{u}}_{k\mid k}$ can be simplified as 
\begin{align}
\Big(\bs{I}_n-\bs{L}_k\bs{C}_k\Big)&\bs{P}^{\bs{x}}_{k\mid k-1}\bs{C}_k^{\T}\bs{R}_k^{-1}\nonumber\\
&\hspace{-2cm}=\bs{P}^{\bs{x}}_{k\mid k-1}\bs{C}_k^{\T}\bs{R}_k^{-1}-\bs{L}_k\big(\bs{R}_k+\bs{C}_k\bs{P}^{\bs{x}}_{k\mid k-1}\bs{C}_k^{\T}-\bs{R}_k\big)\bs{R}_k^{-1}\nonumber\\
&\hspace{-2cm}=\bs{P}^{\bs{x}}_{k\mid k-1}\bs{C}_k^{\T}\bs{R}_k^{-1}-\bs{L}_k\big(\bs{R}_k+\bs{C}_k\bs{P}^{\bs{x}}_{k\mid k-1}\bs{C}_k^{\T}\big)\bs{R}_k^{-1}+\bs{L}_k\nonumber\\
&\hspace{-2cm}=\bs{P}^{\bs{x}}_{k\mid k-1}\bs{C}_k^{\T}\bs{R}_k^{-1}-\bs{P}^{\bs{x}}_{k\mid k-1}\bs{C}_k^{\T}\bs{R}_k^{-1}+\bs{L}_k =\bs{L}_k.\label{identity4_thm1}
\end{align}
Using \eqref{identity4_thm1} in \eqref{x_hat_thm1_int}, we get,
\begin{align}
\hat{\bs{x}}_{k\mid k}= \Big(\bs{I}_n-\bs{L}_k\bs{C}_k\Big)\hat{\bs{x}}_{k\mid k-1}+\bs{L}_k(\bs{y}_k-\bs{D}_k\hat{\bs{u}}_{k\mid k}).\label{x_hat_thm1_2}
\end{align}
Next, taking gradient with respect to $\bs{u}_k$,
\begin{align}
\nabla_{\bs{u}_k}\mathcal{L}&=-2\bs{D}_k^{\T}\bs{R}_k^{-1}(\bs{y}_k-\bs{C}_k\bs{x}_k-\bs{D}_k\bs{u}_k)=0\label{gradL_u}\\
\hat{\bs{u}}_{k\mid k}&=\big(\bs{D}_k^{\T}\bs{R}_k^{-1}\bs{D}_k\big)^{-1}\bs{D}_k^{\T}\bs{R}_k^{-1}(\bs{y}_k-\bs{C}_k\hat{\bs{x}}_{k\mid k})\\
&=\bs{J}_k(\bs{y}_k-\bs{C}_k\hat{\bs{x}}_{k\mid k})\label{u_hat_thm1}
\end{align}
where $\bs{J}_k$ is given in \eqref{eq:J_defn}. Now, \eqref{x_hat_thm1_2} and \eqref{u_hat_thm1} are coupled equations. By substituting \eqref{u_hat_thm1} in \eqref{x_hat_thm1_2} and expressing $\hat{\bs{x}}_{k\mid k}$ in terms of $\hat{\bs{x}}_{k\mid k-1}$ and $\bs{y}_k$ we get,
\begin{align}
\hat{\bs{x}}_{k\mid k}&=(\bs{I}_n-\bs{L}_k\bs{D}_k\bs{J}_k\bs{C}_k)^{-1}\Big[(\bs{I}_n-\bs{L}_k\bs{C}_k)\hat{\bs{x}}_{k\mid k-1}
\nonumber\\&\hspace{4cm}+\bs{L}_k(\bs{I}_p-\bs{D}_k\bs{J}_k)\bs{y}_k\Big]\nonumber\\
&=\bs{E}_k\hat{\bs{x}}_{k\mid k-1}+\bs{F}_k\bs{y}_k,\label{x_hat_thm3_EF}
\end{align}
where
\begin{align}
\bs{E}_k\triangleq(\bs{I}_n-\bs{L}_k\bs{D}_k\bs{J}_k\bs{C}_k)^{-1}(\bs{I}_n-\bs{L}_k\bs{C}_k)\\
\bs{F}_k\triangleq(\bs{I}_n-\bs{L}_k\bs{D}_k\bs{J}_k\bs{C}_k)^{-1}\bs{L}_k(\bs{I}_p-\bs{D}_k\bs{J}_k).
\end{align}
Similarly, 
\begin{align}
\hat{\bs{u}}_{k\mid k}&=-(\bs{I}_m-\bs{J}_k\bs{C}_k\bs{L}_k\bs{D}_k)^{-1}\bs{J}_k\Big[\bs{C}_k(\bs{I}_n-\bs{L}_k\bs{C}_k)\hat{\bs{x}}_{k\mid k-1}\nonumber\\
&\hspace{4cm}-(\bs{I}_p-\bs{C}_k\bs{L}_k)\bs{y}_k\Big]\\
&=\bs{G}^{'}_k\hat{\bs{x}}_{k\mid k-1}+\bs{H}_k\bs{y}_k, \label{u_hat_thm1_GH}
\end{align}
where
\begin{align}
\bs{G}^{'}_k&\triangleq-(\bs{I}_m-\bs{J}_k\bs{C}_k\bs{L}_k\bs{D}_k)^{-1}\bs{J}_k\bs{C}_k(\bs{I}_n-\bs{L}_k\bs{C}_k)\\
\bs{H}_k&\triangleq(\bs{I}_m-\bs{J}_k\bs{C}_k\bs{L}_k\bs{D}_k)^{-1}\bs{J}_k(\bs{I}_p-\bs{C}_k\bs{L}_k).
\end{align}
The following identities help us simplify subsequent expressions.
\begin{align}
&\bs{I}_n-\bs{F}_k\bs{C}_k\nonumber\\
&=\bs{I}_n-\hspace{-1mm}(\bs{I}_n-\bs{L}_k\bs{D}_k\bs{J}_k\bs{C}_k)^{-1}(\bs{L}_k\bs{C}_k-\bs{L}_k\bs{D}_k\bs{J}_k\bs{C}_k)\nonumber\\
&=\bs{I}_n+(\bs{I}_n-\bs{L}_k\bs{D}_k\bs{J}_k\bs{C}_k)^{-1}(\bs{I}_n-\bs{L}_k\bs{C}_k)-\bs{I}_n\nonumber\\
&=\bs{E}_k\\
\bs{H}_k\bs{C}_k&=(\bs{I}_m-\bs{J}_k\bs{C}_k\bs{L}_k\bs{D}_k)^{-1}(\bs{J}_k\bs{C}_k-\bs{J}_k\bs{C}_k\bs{L}_k\bs{C}_k)\nonumber\\
&=-\bs{G}^{'}_k
\end{align}
Using the above two identities, we can rewrite the expressions for $\hat{\bs{x}}_{k\mid k}$ and $\hat{\bs{u}}_{k\mid k}$, which yield Step 7 of \Cref{alg:RobustKalman}.

The update errors corresponding to $\bs{x}_k$ and $\bs{u}_k$ are  as follows:
\begin{align}
\bs{e}^{\bs{x}}_{k\mid k}&=\bs{x}_k-\hat{\bs{x}}_{k\mid k}=\bs{x}_k-(\bs{I}_n-\bs{F}_k\bs{C}_k)\hat{\bs{x}}_{k\mid k-1}-\bs{F}_k\bs{y}_k\nonumber\\
&=\bs{x}_k-(\bs{I}_n-\bs{F}_k\bs{C}_k)\hat{\bs{x}}_{k\mid k-1}-\bs{F}_k(\bs{C}_k\bs{x}_k+\bs{D}_k\bs{u}_k+\bs{v}_k)\nonumber\\
&=(\bs{I}_n-\bs{F}_k\bs{C}_k)\bs{e}^{\bs{x}}_{k\mid k-1}-\bs{F}_k\bs{D}_k\bs{u}_k-\bs{F}_k\bs{v}_k\label{ex}
\end{align}
and 
\begin{align}
\bs{e}^{\bs{u}}_{k\mid k}&=\bs{u}_k-\hat{\bs{u}}_{k\mid k}=\bs{u}_k-\bs{H}_k(\bs{y}_k-\bs{C}_k\hat{\bs{x}}_{k\mid k-1})\nonumber\\
&=-\bs{H}_k\bs{C}_k\bs{e}^{\bs{x}}_{k\mid k-1}+(\bs{I}_m-\bs{H}_k\bs{D}_k)\bs{u}_k-\bs{H}_k\bs{v}_k\label{ey}.
\end{align}
Using the identity $\bs{J}_k\bs{D}_k=\bs{I}_m$, it is straightforward to verify that:
\begin{align}
\bs{F}_k\bs{D}_k&=(\bs{I}_n-\bs{L}_k\bs{D}_k\bs{J}_k\bs{C}_k)^{-1}(\bs{L}_k\bs{D}_k-\bs{L}_k\bs{D}_k\bs{J}_k\bs{D}_k)=\bs{0}\label{identity1}\\
\bs{H}_k\bs{D}_k&=(\bs{I}_m-\bs{J}_k\bs{C}_k\bs{L}_k\bs{D}_k)^{-1}(\bs{J}_k\bs{D}_k-\bs{J}_k\bs{C}_k\bs{L}_k\bs{D}_k)=\bs{I}_m.\label{identity2}
\end{align}
Applying identities \eqref{identity1}, \eqref{identity2} in \eqref{ex}, \eqref{ey} respectively, we eliminate the dependence of the estimation error on $\bs{u}_k$, and the update error at the $k^{\text{th}}$ step depends on update error at the ${(k-1)}^{\text{th}}$ step and the measurement noise at the $k^{\text{th}}$ step:
\begin{align}
\bs{e}^{\bs{x}}_{k\mid k}&=(\bs{I}_n-\bs{F}_k\bs{C}_k)\bs{e}^{\bs{x}}_{k\mid k-1}-\bs{F}_k\bs{v}_k\\
\bs{e}^{\bs{u}}_{k\mid k}&=-\bs{H}_k\bs{C}_k\bs{e}^{\bs{x}}_{k\mid k-1}-\bs{H}_k\bs{v}_k.
\end{align}
It is easy to verify that $\bs{e}^{\bs{x}}_{k\mid k}$ and $\bs{e}^{\bs{u}}_{k\mid k}$ have zero mean and hence $\hat{\bs{x}}_{k\mid k}$ and $\hat{\bs{u}}_{k\mid k}$ are unbiased estimates of $\bs{x}_k$ and $\bs{u}_k$.
Now, the error covariances corresponding to the state $\bs{x}_k$ and input $\bs{u}_k$ can be found using $\bs{P}^{\bs{x}}_{k\mid k}=\mathbb{E}\left(\bs{e}^{\bs{x}}_{k\mid k}(\bs{e}^{\bs{x}}_{k\mid k})^{\T}\right)$, $\bs{P}^{\bs{u}}_{k\mid k}=\mathbb{E}\left(\bs{e}^{\bs{u}}_{k\mid k}(\bs{e}^{\bs{u}}_{k\mid k})^{\T}\right)$ and $\bs{P}^{\bs{xu}}_{k\mid k}=\mathbb{E}\left(\bs{e}^{\bs{x}}_{k\mid k}(\bs{e}^{\bs{u}}_{k\mid k})^{\T}\right)$, which yields Step~8 of \Cref{alg:RobustKalman}.
\subsection{Smoothing steps}\label{sec:smoothing_proof}
We develop the smoothing updates for the case of unknown inputs along the lines of the Kalman smoothing in~\cite{byron2004derivation}.
According to Assumption \textbf{A3}, the smoothed posterior distribution $p\left(\bs{x}_{k},\bs{u}_{k} \mid\bs{Y}_1^{K}\right)$ is Gaussian with mean $\hat{\bs{x}}_{k\mid K},\hat{\bs{u}}_{k\mid K}$ and covariance $\bs{P}_{k\mid K}^{\boldsymbol{\xi}}$ as derived below. To this end, we first derive the covariance of the joint posterior distribution $p\left(\boldsymbol{\xi}_{k+1}, \boldsymbol{\xi}_{k} \mid\bs{Y}_1^{K}\right)$, denoted by $\bs{P}_{k+1,k\mid K}^{\boldsymbol{\xi}}$, as follows.
\begin{align}
p\left(\boldsymbol{\xi}_{k+1}, \boldsymbol{\xi}_{k} \mid\bs{Y}_1^{K}\right)&=p\left(\boldsymbol{\xi}_{k} \mid \boldsymbol{\xi}_{k+1},\bs{Y}_1^{K}\right)p\left(\boldsymbol{\xi}_{k+1} \mid\bs{Y}_1^{K}\right)\label{smooth1}\\
&\hspace{-3cm}=\frac{p\left(\boldsymbol{\xi}_{k+1} \mid \boldsymbol{\xi}_{k}\right)p\left(\boldsymbol{\xi}_{k} \mid\bs{Y}_1^{k}\right)}{p\left(\boldsymbol{\xi}_{k+1} \mid\bs{Y}_1^{k}\right)} p\left(\boldsymbol{\xi}_{k+1} \mid\bs{Y}_1^{K}\right)\label{smooth2}\\
&\hspace{-3cm}=\frac{p\left(\bs{x}_{k+1} \mid \bs{x}_{k},\bs{u}_{k}\right)p\left(\bs{u}_{k+1}\right)p\left(\boldsymbol{\xi}_{k} \mid\bs{Y}_1^{k}\right)}{p\left(\bs{x}_{k+1} \mid\bs{Y}_1^{k}\right)p\left(\bs{u}_{k+1}\right)} p\left(\boldsymbol{\xi}_{k+1} \mid\bs{Y}_1^{K}\right)\label{smooth3}\\
&\hspace{-3cm}=\frac{p\left(\bs{x}_{k+1} \mid \bs{x}_{k},\bs{u}_{k}\right)p\left(\boldsymbol{\xi}_{k} \mid\bs{Y}_1^{k}\right)}{p\left(\bs{x}_{k+1} \mid\bs{Y}_1^{k}\right)} p\left(\boldsymbol{\xi}_{k+1} \mid\bs{Y}_1^{K}\right)\label{joint_pdf_thm3}
\end{align}
where \eqref{smooth2} follows since,  given $\boldsymbol{\xi}_{k+1}$,  $\boldsymbol{\xi}_k$ is independent of $\{\bs{y}_{k+1}, \ldots, \bs{y}_K\}$, and \eqref{smooth3} follows since $\bs{u}_{k+1}$ is independent of $\bs{x}_k$, $\bs{x}_{k+1}$, $\bs{u}_k$ and all past observations $\{\bs{y}_1,\ldots,\bs{y}_k\}$.

Taking the logarithm of each term in \eqref{joint_pdf_thm3}, we have
\begin{align}
\log p\left(\bs{x}_{k+1} \mid \bs{x}_{k},\bs{u}_{k}\right)&=-\frac{1}{2} \Vert\bs{x}_{k+1}-\bs{A}_k\bs{x}_k-\bs{B}_k\bs{u}_k\Vert_{\bs{Q}_k}^2\\
&=-\frac{1}{2} \Vert\bs{T}\boldsymbol{\xi}_{k+1}-\tilde{\bs{A}}_k\boldsymbol{\xi}_k\Vert_{\bs{Q}_k}^2\label{pdf1}
\end{align}
where we used $\bs{x}_{k+1}=\bs{T}\boldsymbol{\xi}_{k+1}$ with $\bs{T}\triangleq \left[\bs{I}_n,\bs{0}_{n\times m}\right]$ and $\tilde{\bs{A}}_k=\left[\bs{A}_k,\bs{B}_k\right]$. Similarly, we can show that
\begin{align}
\log p\left(\bs{x}_{k+1}\mid\bs{Y}_1^{k}\right)&=-\frac{1}{2} \Vert\bs{x}_{k+1}-\hat{\bs{x}}_{k+1\mid k}\Vert_{\bs{P}_{k+1\mid k}^{\bs{x}}}^2\\
&\hspace{-0.5cm}=-\frac{1}{2} \Vert\bs{T}\left(\boldsymbol{\xi}_{k+1}-\hat{\boldsymbol{\xi}}_{k+1\mid k}\right)\Vert_{\bs{P}_{k+1\mid k}^{\bs{x}}}^2\label{pdf2}
\end{align}
\begin{align}
\log p\left(\boldsymbol{\xi}_{k} \mid\bs{Y}_1^{k}\right)&=-\frac{1}{2} \Vert\boldsymbol{\xi}_k-\hat{\boldsymbol{\xi}}_{k\mid k}\Vert_{\bs{P}_{k\mid k}^{\boldsymbol{\xi}}}^2\label{pdf3}\\
\log p\left(\boldsymbol{\xi}_{k+1} \mid\bs{Y}_1^{K}\right)&=-\frac{1}{2} \Vert\boldsymbol{\xi}_{k+1}-\hat{\boldsymbol{\xi}}_{k+1\mid K}\Vert_{\bs{P}_{k+1\mid K}^{\boldsymbol{\xi}}}^2\label{pdf4}
\end{align}
Now using \eqref{pdf1}, \eqref{pdf2}, \eqref{pdf3}, \eqref{pdf4} and clubbing the like quadratic terms together, we get 
\begin{align}
&\log p\big(\boldsymbol{\xi}_{k+1}, \boldsymbol{\xi}_{k} \mid\bs{Y}_1^{K}\big)\nonumber\\
&=-\frac{1}{2} \boldsymbol{\xi}_{k+1}^{\T}\Big(\bs{T}^{\T}\bs{Q}_k^{-1}\bs{T}-\bs{T}^{\T}\left(\bs{P}_{k+1\mid k}^{\bs{x}}\right)^{-1}\bs{T}\nonumber\\
&\hspace{5cm}+\left(\bs{P}_{k+1\mid K}^{\boldsymbol{\xi}}\right)^{-1}\Big) \boldsymbol{\xi}_{k+1} \nonumber\\
&-\frac{1}{2} \boldsymbol{\xi}_{k+1}^{\T}\left(-\bs{T}^{\T}\bs{Q}_k^{-1}\tilde{\bs{A}}_k\right) \boldsymbol{\xi}_k-\frac{1}{2} \boldsymbol{\xi}_k^{\T}\left(-\tilde{\bs{A}}_k^{\T} \bs{Q}_k^{-1}\bs{T}\right) \boldsymbol{\xi}_{k+1}\nonumber\\
&-\frac{1}{2} \boldsymbol{\xi}_{k}^{\T}\left(\tilde{\bs{A}}_k^{\T} \bs{Q}_k^{-1} \tilde{\bs{A}}_k+\left(\bs{P}_{k\mid k}^{\boldsymbol{\xi}}\right)^{-1}\right) \boldsymbol{\xi}_k+\boldsymbol{\xi}_{k}^{\T}\left(\left(\bs{P}_{k\mid k}^{\boldsymbol{\xi}}\right)^{-1} \hat{\boldsymbol{\xi}}_{k\mid k}\right)\nonumber\\
&\hspace{0cm}+ \text{linear and constant terms.}
\end{align}
We now follow the proof technique for deriving the smoothing updates in \cite{byron2004derivation}, and we use same notation $\bs{S}_{11}$, $\bs{S}_{12}$, $\bs{S}_{21}$, $\bs{S}_{22}$ and $\bs{F}_{11}$ as given in the reference which indicate same variables in our context. We can find $\bs{S}_{22}^{-1}$ in our case as follows:
\begin{align}
\bs{S}_{22}^{-1}&=\left(\tilde{\bs{A}}_k^{\T}\bs{Q}_k^{-1}\tilde{\bs{A}}_k+\left(\bs{P}_{k\mid k}^{\boldsymbol{\xi}}\right)^{-1}\right)^{-1}\nonumber\\
&=\bs{P}_{k\mid k}^{\boldsymbol{\xi}}-\bs{P}_{k\mid k}^{\boldsymbol{\xi}}\tilde{\bs{A}}_k^{\T}\left(\bs{Q}_k+\tilde{\bs{A}}_k\bs{P}_{k\mid k}^{\boldsymbol{\xi}}\tilde{\bs{A}}_k^{\T}\right)^{-1}\tilde{\bs{A}}_k\bs{P}_{k\mid k}^{\boldsymbol{\xi}}\nonumber\\
&=\bs{P}_{k\mid k}^{\boldsymbol{\xi}}-\bs{P}_{k\mid k}^{\boldsymbol{\xi}}\tilde{\bs{A}}_k^{\T}\left(\bs{P}_{k+1\mid k}^{\bs{x}}\right)^{-1}\tilde{\bs{A}}_k\bs{P}_{k\mid k}^{\boldsymbol{\xi}}\nonumber\\
&=\bs{P}_{k\mid k}^{\boldsymbol{\xi}}-\bs{K}_k\bs{P}_{k+1\mid k}^{\bs{x}}\bs{K}_k^{\T}
\end{align}
where $\bs{K}_k\triangleq\bs{P}_{k\mid k}^{\boldsymbol{\xi}}\tilde{\bs{A}}_k^{\T}\left(\bs{P}_{k+1\mid k}^{\bs{x}}\right)^{-1}.$
Similarly, $\bs{S}_{21}$ is given by
 $\bs{S}_{21}=-\tilde{\bs{A}}_k^{\T}\bs{Q}_k^{-1}\bs{T}.$
Therefore, $\bs{S}_{22}^{-1}\bs{S}_{21}$ is
\begin{align}
\bs{S}_{22}^{-1}\bs{S}_{21}&=-\Big(\bs{P}_{k\mid k}^{\boldsymbol{\xi}}-\bs{P}_{k\mid k}^{\boldsymbol{\xi}}\tilde{\bs{A}}_k^{\T}\left(\bs{Q}_k+\tilde{\bs{A}}_k\bs{P}_{k\mid k}^{\boldsymbol{\xi}}\tilde{\bs{A}}_k^{\T}\right)^{-1}\nonumber\\
&\hspace{3cm}\tilde{\bs{A}}_k\bs{P}_{k\mid k}^{\boldsymbol{\xi}}\Big)\tilde{\bs{A}}_k^{\T}\bs{Q}_k^{-1}\bs{T}\nonumber\\
&\hspace{-1cm}=-\bs{P}_{k\mid k}^{\boldsymbol{\xi}}\tilde{\bs{A}}_k^{\T}\left(\bs{I}_n-\left(\bs{Q}_k+\tilde{\bs{A}}_k\bs{P}_{k\mid k}^{\boldsymbol{\xi}}\tilde{\bs{A}}_k^{\T}\right)^{-1}\right.\nonumber\\
&\hspace{3cm}\left.\tilde{\bs{A}}_k\bs{P}_{k\mid k}^{\boldsymbol{\xi}}\tilde{\bs{A}}_k^{\T}\right)\bs{Q}_k^{-1}\bs{T}\nonumber\\
&\hspace{-1cm}=-\bs{P}_{k\mid k}^{\boldsymbol{\xi}}\tilde{\bs{A}}_k^{\T}\left(\bs{Q}_k+\tilde{\bs{A}}_k\bs{P}_{k\mid k}^{\boldsymbol{\xi}}\tilde{\bs{A}}_k^{\T}\right)^{-1}\bs{T}\nonumber\\
&\hspace{-1cm}=-\bs{P}_{k\mid k}^{\boldsymbol{\xi}}\tilde{\bs{A}}_k^{\T}\left(\bs{P}_{k+1\mid k}^{\bs{x}}\right)^{-1}\bs{T}=-\bs{K}_k\bs{T}.
\end{align}
Hence, the covariance update is given by (see (19) of \cite{byron2004derivation}),
\begin{align}
&\bs{P}_{k\mid K}^{\boldsymbol{\xi}}=\bs{S}_{22}^{-1}+\bs{S}_{22}^{-1}\bs{S}_{21}\bs{F}_{11}^{-1}\bs{S}_{12}\bs{S}_{22}^{-1}\nonumber\\
&=\left(\bs{P}_{k\mid k}^{\boldsymbol{\xi}}-\bs{K}_k\bs{P}_{k+1\mid k}^{\bs{x}}\bs{K}_k^{\T}\right)+\left(-\bs{K}_k\bs{T}\right)\bs{P}_{k+1\mid K}^{\boldsymbol{\xi}}\left(-\bs{K}_k\bs{T}\right)^{\T}\nonumber\\
&=\bs{P}_{k\mid k}^{\boldsymbol{\xi}}+\bs{K}_k\left(\bs{P}_{k+1\mid K}^{\bs{x}}-\bs{P}_{k+1\mid k}^{\bs{x}}\right)\bs{K}_k^{\T}\label{eq:P_xi_k_K}
\end{align}
where we used $\bs{T}\bs{P}_{k+1\mid K}^{\boldsymbol{\xi}}\bs{T}^{\T}=\bs{P}_{k+1\mid K}^{\bs{x}}$ in the last step.

Next, we can compute $\bs{P}_{k+1,k\mid K}^{\boldsymbol{\xi}}$ as (see (20) in \cite{byron2004derivation}) 
\begin{align}
\bs{P}_{k+1,k\mid K}^{\boldsymbol{\xi}}&=-\bs{F}_{11}^{-1}\bs{S}_{12}\bs{S}_{22}^{-1}=-\bs{P}_{k+1\mid K}^{\boldsymbol{\xi}}\left(-\bs{K}_k\bs{T}\right)^{\T}\nonumber\\
&=\begin{bmatrix}
\bs{P}_{k+1\mid K}^{\bs{x}} & \bs{P}_{k+1\mid K}^{\bs{xu}}\\
\left(\bs{P}_{k+1\mid K}^{\bs{xu}}\right)^{\T} & \bs{P}_{k+1\mid K}^{\bs{u}}
\end{bmatrix}
\begin{bmatrix}
\bs{I}_n\\\bs{0}_{m\times n}
\end{bmatrix}\bs{K}_k^{\T}\nonumber\\
&=\begin{bmatrix}
\bs{P}_{k+1\mid K}^{\bs{x}}\bs{K}_k^{\T}\\
\left(\bs{P}_{k+1\mid K}^{\bs{xu}}\right)^{\T}\bs{K}_k^{\T}
\end{bmatrix}
\end{align}
We can find the smoothed posterior estimate using the following relation, (see (23) of \cite{byron2004derivation})
\begin{align}
&\bs{S}_{21}\hat{\boldsymbol{\xi}}_{k+1\mid K}+\bs{S}_{22}\hat{\boldsymbol{\xi}}_{k\mid K}=\left(\bs{P}^{\boldsymbol{\xi}}_{k\mid k}\right)^{-1}\hat{\boldsymbol{\xi}}_{k\mid k}\\
\implies\hat{\boldsymbol{\xi}}_{k\mid K}&=-\bs{S}_{22}^{-1}\bs{S}_{21}\hat{\boldsymbol{\xi}}_{k+1\mid K}+\bs{S}_{22}^{-1}\left(\bs{P}^{\boldsymbol{\xi}}_{k\mid k}\right)^{-1}\hat{\boldsymbol{\xi}}_{k\mid k}\\
&=-\left(-\bs{K}_k\bs{T}\right)\hat{\boldsymbol{\xi}}_{k+1\mid K}\nonumber\\
&\hspace{-1.5cm}+\left(\bs{P}_{k\mid k}^{\boldsymbol{\xi}}-\bs{P}_{k\mid k}^{\boldsymbol{\xi}}\tilde{\bs{A}}_k^{\T}\left(\bs{P}_{k+1\mid k}^{\bs{x}}\right)^{-1}\tilde{\bs{A}}_k\bs{P}_{k\mid k}^{\boldsymbol{\xi}}\right)\hspace{-1mm}\left(\bs{P}_{k\mid k}^{\boldsymbol{\xi}}\right)^{-1}\hat{\boldsymbol{\xi}}_{k\mid k}\nonumber\\
&\hspace{-1.5cm}=\bs{K}_k\hat{\bs{x}}_{k+1\mid K}+\left(\bs{I}_{n+m}-\bs{K}_k\tilde{\bs{A}}_k\right)\hat{\boldsymbol{\xi}}_{k\mid k}.\label{eq:xi_k_K}
\end{align}
Equations \eqref{eq:P_xi_k_K} and \eqref{eq:xi_k_K} yield Step~13 and 14 of \Cref{alg:RobustKalman}, respectively. This completes proof of all the steps in the algorithm.
\section{RKS Algorithm for Noisy Compressive State Measurements}\label{sec_thm3}
In this section, we consider that the measurements are a linear function of the state, and that there is no direct feed-through of the input vector, unlike in \eqref{eq:meas_eqn}. That is, the observations are given by 
\begin{align}
\bs{y}_k=\bs{C}_k\bs{x}_k+\bs{v}_k. \label{eq:meas_thm3}
\end{align}
With measurements of this form, we can only estimate the inputs up to $\bs{u}_{k-1}$ at time step $k$. When $p\ge m$, i.e., the number of measurements, $p$, is at least equal to the dimension of the control inputs, $m$, the filtering updates are given by following expressions:
\begin{align}
\hat{\bs{x}}_{k\mid k}&=(\bs{I}_p-\bs{F}_k\bs{C}_k)\bs{A}_{k-1}\hat{\bs{x}}_{k-1\mid k-1}+\bs{F}_k\bs{y}_k\label{x_update_thm3}\\
\hat{\bs{u}}_{k-1\mid k} &=\bs{M}_k(\bs{y}_k-\bs{C}_k\bs{A}_{k-1}\hat{\bs{x}}_{k-1\mid k-1})
\end{align}
In the above, we define
\begin{align}
\bs{F}_k&\triangleq(\bs{I}_n-(\bs{I}_n-\bs{L}_k\bs{C}_k)\bs{B}_{k-1}\bs{J}_k)^{-1}\bs{L}_k\label{eq:F_k}\\
\bs{M}_k&\triangleq(\bs{I}_m-\bs{J}_k\big(\bs{I}_n-\bs{L}_k\bs{C}_k\big)\bs{B}_{k-1})^{-1}\bs{J}_k\bs{L}_k,\label{eq:M_k}
\end{align}
\begin{align}
\text{where \,\,}
\bs{P}^{*}_{k\mid k-1}&\triangleq \bs{A}_{k-1}\bs{P}_{k-1\mid k-1}^{\bs{x}}\bs{A}_{k-1}^{\T}+\bs{Q}_{k-1}, \label{P_star}\\
\bs{J}_k\triangleq\big(\bs{B}_{k-1}^{\T}&\bs{P}^{*}_{k\mid k-1}\bs{B}_{k-1}\big)^{-1}\bs{B}_{k-1}^{\T}(\bs{P}^{*}_{k\mid k-1})^{-1},\label{J_k_thm3}\\
\bs{L}_k&\triangleq\bs{P}_{k\mid k-1}^{*}\bs{C}_k^{\T}\big(\bs{R}_k+\bs{C}_k\bs{P}_{k\mid k-1}^{*}\bs{C}_k^{\T}\big)^{-1}.\label{Lk_thm3}
\end{align}
The covariance updates are given by,
\begin{align}
\bs{P}^{\bs{x}}_{k\mid k}
&=(\bs{I}_n-\bs{F}_k\bs{C}_k)\bs{P}^{*}_{k\mid k-1}(\bs{I}_n-\bs{F}_k\bs{C}_k)^{\T}+\bs{F}_k\bs{R}_k\bs{F}_k^{\T}\label{Px_thm3}\\
\bs{P}^{\bs{u}}_{k-1\mid k}
&=\bs{M}_k\bs{C}_k\bs{P}^{*}_{k\mid k-1}\bs{C}_k^{\T}\bs{M}_k^{\T}+\bs{M}_k\bs{R}_k\bs{M}_k^{\T},\\
\bs{P}^{\bs{x}\bs{u}}_{k,k-1\mid k}&=-(\bs{I}_n-\bs{F}_k\bs{C}_k)\bs{A}_{k-1}\bs{P}^{\bs{x}}_{k-1\mid k-1}\bs{A}_{k-1}^{\T}\bs{C}_k^{\T}\bs{M}_k^{\T}\nonumber\\
&\hspace{0.5cm}-(\bs{I}_n-\bs{F}_k\bs{C}_k)\bs{Q}_{k-1}\bs{C}_k^{\T}\bs{M}_k^{\T}+\bs{F}_k^{\T}\bs{R}_k\bs{M}_k^{\T}.\label{Pxu_thm3}
\end{align}
We prove the above steps in the sequel.

\begin{proof}
Let us denote ${\boldsymbol{\xi}}_k\triangleq [\bs{x}_k^{\T},\bs{u}_{k-1}^{\T}]^{\T}$. We make the following assumption similar to \Cref{sec:RKS}, i.e. for any $t$ and $k$,
\begin{multline}
p\left(\bs{x}_{k},\bs{u}_{k-1}\mid \bs{Y}_1^{t}\right)\sim \\\mathcal{N}\left(\left[\begin{array}{c}\hat{\bs{x}}_{k\mid t} \\ \hat{\bs{u}}_{k-1\mid t}\end{array}\right],\left[\begin{array}{cc}\bs{P}_{k\mid t}^{\bs{x}}& \bs{P}_{k,k-1\mid t}^{\bs{xu}} \\ \left(\bs{P}_{k,k-1\mid t}^{ \bs{xu}}\right)^{\T} & \bs{P}_{k-1\mid t}^{\bs{u}}\end{array}\right]\right).\label{gaussian_approx2_thm3}
\end{multline}
\subsection{Filtering Steps}
The filtering problem in the $k^{\text{th}}$ time step involves estimating two posterior distributions $p\left(\bs{x}_{k} \mid \bs{Y}_1^{k}\right)$ and $p\left(\bs{x}_{k},\bs{u}_{k-1}\mid \bs{Y}_1^{k}\right)$. These two posteriors  are coupled by integral equations and estimating them for the $k^{\text{th}}$ step will help us find the posteriors for $(k+1)^{\text{th}}$ time, which gives rise to the forward filtering updates.

\noindent\textbf{Step 1:} We find $\hat{\bs{x}}_{k\mid k}$ as the mode of $p\left(\bs{x}_{k} \mid \bs{Y}_1^{k}\right)$, where
\begin{align}
p\left(\bs{x}_{k} \mid \bs{Y}_1^{k}\right)=
\int p\left(\bs{x}_{k},\bs{u}_{k-1} \mid \bs{Y}_1^{k}\right)\mathrm{d}\bs{u}_{k-1}.
\end{align}
\textbf{Step 2:} The posterior distribution $p\left(\bs{x}_{k},\bs{u}_{k-1}\mid \bs{Y}_1^{k}\right)$ inside the integral is factored as 
\begin{align}
p\left(\bs{x}_{k},\bs{u}_{k-1}\mid \bs{Y}_1^{k}\right)=\frac{p\left(\bs{y}_{k} \mid \bs{x}_{k}\right) p\left(\bs{x}_{k},\bs{u}_{k-1} \mid \bs{Y}_1^{k-1}\right)}{p\left(\bs{y}_{k} \mid \bs{Y}_1^{k-1}\right)}\label{filtering_thm3}
\end{align}
The $p\left(\bs{x}_{k},\bs{u}_{k-1} \mid \bs{Y}_1^{k-1}\right)$ term in \eqref{filtering_thm3} can be expressed as
\begin{align}
&p\left(\bs{x}_{k},\bs{u}_{k-1} \mid \bs{Y}_1^{k-1}\right)=\nonumber\\
&\int\hspace{-1mm}p\left(\bs{x}_{k} \mid \bs{x}_{k-1},\bs{u}_{k-1}\right) p\left(\bs{x}_{k-1},\bs{u}_{k-1} \mid \bs{Y}_1^{k-1}\right) \mathrm{~d}\bs{x}_{k-1}\label{prediction1_them3}\\
&=p\left(\bs{u}_{k-1}\right)\int\hspace{-1mm}p\left(\bs{x}_{k} \mid \bs{x}_{k-1},\bs{u}_{k-1}\right) p\left(\bs{x}_{k-1} \mid \bs{Y}_1^{k-1}\right) \mathrm{~d}\bs{x}_{k-1}\label{prediction_them3}
\end{align}
To get \eqref{prediction_them3} from \eqref{prediction1_them3}, we have used the fact that $\bs{u}_{k-1}$ is independent of the state $\bs{x}_{k-1}$ and the measurements $\{\bs{y}_1,\ldots,\bs{y}_{k-1}\}$. We can find
$p\left(\bs{x}_{k-1} \mid \bs{Y}_1^{k-1}\right) \mathrm{~d}\bs{x}_{k-1}$ inside the integral by marginalizing the joint posterior  $p\left(\bs{x}_{k-1},\bs{u}_{k-2} \mid \bs{Y}_1^{k-1}\right) \mathrm{~d}\bs{x}_{k-1}$, which is in turn estimated from the $(k-1)^{\text{th}}$ update step.

The pdf terms inside the integral are Gaussian distributed and are given as follows:
\begin{align}
p\left(\bs{x}_{k} \mid \bs{x}_{k-1},\bs{u}_{k-1}\right)&= \mathcal{N}\left(\bs{A}_{k-1}\bs{x}_{k-1}+\bs{B}_{k-1}\bs{u}_{k-1}, \bs{Q}_{k-1}\right)\\
p\left(\bs{x}_{k-1} \mid \bs{Y}_1^{k-1}\right)&= \mathcal{N}\left(\hat{\bs{x}}_{k-1\mid k-1}, \bs{P}_{k-1\mid k-1}^{\bs{x}}\right).
\end{align}
The integral in \eqref{prediction_them3} can be evaluated by applying \cref{lemma1}:
\begin{align}
\int\hspace{-1mm}p\left(\bs{x}_{k} \mid \bs{x}_{k-1},\bs{u}_{k-1}\right) p\left(\bs{x}_{k-1} \mid \bs{Y}_1^{k-1}\right) \mathrm{~d}\bs{x}_{k-1}\nonumber\\ 
= \mathcal{N}\left(\bs{A}_{k-1}\hat{\bs{x}}_{k-1\mid k-1}+\bs{B}_{k-1}\bs{u}_{k-1}, \bs{P}^{*}_{k\mid k-1}\right)
\end{align}
where $\bs{P}^{*}_{k\mid k-1}$ is as defined in \eqref{P_star}. Hence, \eqref{filtering_thm3} can be expressed as
\begin{align}
&p\left(\bs{x}_{k},\bs{u}_{k-1}\mid \bs{Y}_1^{k}\right)\nonumber\\
&\propto p\left(\bs{y}_{k} \mid \bs{x}_{k}\right) \int\hspace{-1mm}p\left(\bs{x}_{k} \mid \bs{x}_{k-1},\bs{u}_{k-1}\right) p\left(\bs{x}_{k-1} \mid \bs{Y}_1^{k-1}\right) \mathrm{~d}\bs{x}_{k-1}\\
&\propto \mathcal{N}\left(\bs{C}_{k}\bs{x}_{k}, \bs{R}_{k}\right)\times\mathcal{N}\left(\bs{A}_{k-1}\hat{\bs{x}}_{k-1\mid k-1}+\bs{B}_{k-1}\bs{u}_{k-1}, \bs{P}^{*}_{k\mid k-1}\right).
\end{align}
Hence, the MAP estimate of the filtering step can be posed as following minimization problem:
\begin{align}
\Big\{\hat{\bs{x}}_{k\mid k},\hat{\bs{u}}_{k-1\mid k}\Big\}&=\underset{\bs{x}_k,\bs{u}_{k-1}}{\arg \max}\hspace{2mm}p\left(\bs{x}_{k},\bs{u}_{k-1} \mid \bs{Y}_1^{k}\right)\\
&\hspace{-2cm}=\underset{\bs{x}_k,\bs{u}_{k-1}}{\arg \min }\hspace{1mm} \Vert\bs{y}_k-\bs{C}_k\bs{x}_k\Vert_{\bs{R}_k}^2\nonumber\\&\hspace{-2cm}+\Vert\bs{x}_k-\bs{A}_{k-1}\hat{\bs{x}}_{k-1\mid k-1}-\bs{B}_{k-1}\bs{u}_{k-1}\Vert_{(\bs{P}_{k\mid k-1}^{*})}^2.\label{xu_min_filtering_thm3}
\end{align}
Now, following similar steps as in the Appendix~\ref{app:KFS_filtering}, we get
\begin{align}
&\bs{e}^{\bs{x}}_{k\mid k}=(\bs{I}_n-\bs{F}_k\bs{C}_k)\bs{A}_{k-1}\bs{e}^{\bs{x}}_{k-1\mid k-1} +(\bs{I}_n-\bs{F}_k\bs{C}_k)\bs{B}_{k-1}\bs{u}_{k-1}\nonumber\\
&\hspace{2cm}+(\bs{I}_n-\bs{F}_k\bs{C}_k)\bs{w}_{k-1}-\bs{F}_k\bs{v}_k\label{ex_thm3}\\
&\bs{e}^{\bs{u}}_{k-1\mid k}=-\bs{M}_k\bs{C}_k\bs{A}_{k-1}\bs{e}^{\bs{x}}_{k-1\mid k-1}+(\bs{I}_m-\bs{M}_k\bs{C}_k\bs{B}_{k-1})\bs{u}_{k-1}\nonumber\\
&\hspace{3cm}-\bs{M}_k\bs{v}_k-\bs{M}_k\bs{C}_k\bs{w}_{k-1}.\label{eu_thm3}
\end{align}
Using the identities $(\bs{I}_n-\bs{F}_k\bs{C}_k)\bs{B}_{k-1}=\bs{0}$, $\bs{I}_m-\bs{M}_k\bs{C}_k\bs{B}_{k-1}=\bs{0}$ and $\bs{I}_m-\bs{J}_k\bs{B}_{k-1}=\bs{0}$ in \eqref{ex_thm3} and \eqref{eu_thm3} respectively, we can eliminate the dependence on $\bs{u}_{k-1}$ in the estimation errors. Now, the error covariances corresponding to the state $\bs{x}_k$ and input $\bs{u}_{k-1}$ can be found using $\bs{P}^{\bs{x}}_{k\mid k}=\mathbb{E}\left(\bs{e}^{\bs{x}}_{k\mid k}(\bs{e}^{\bs{x}}_{k\mid k})^{\T}\right)$, $\bs{P}^{\bs{u}}_{k-1\mid k}=\mathbb{E}\left(\bs{e}^{\bs{u}}_{k-1\mid k}(\bs{e}^{\bs{u}}_{k-1\mid k})^{\T}\right)$ and $\bs{P}^{\bs{xu}}_{k,k-1\mid k}=\mathbb{E}\left(\bs{e}^{\bs{x}}_{k\mid k}(\bs{e}^{\bs{u}}_{k-1\mid k})^{\T}\right),$ which results in the expressions provided in \eqref{Px_thm3}-\eqref{Pxu_thm3}.
\end{proof}
\subsection{Smoothing Steps}
In this subsection, we prove the smoothing  steps:
\begin{align}
\hat{\bs{\xi}}_{k\mid K}&=\hat{\bs{\xi}}_{k\mid k}+\bs{K}_k\lb\hat{\bs{B}}_k\hat{\bs{\xi}}_{k+1\mid K}-\hat{\bs{A}}_k\hat{\bs{\xi}}_{k\mid k}\rb\\
\bs{P}_{k\mid K}^{\bs{\xi}}&=\bs{P}_{k\mid k}^{\boldsymbol{\xi}}+\bs{K}_k\left(\hat{\bs{B}}_k\bs{P}_{k+1\mid K}^{\bs{\xi}}\hat{\bs{B}}_k^{\T}-\bs{P}_{k+1\mid k}^{*}\right)\bs{K}_k^{\T}\\
\bs{K}_k&=\bs{P}_{k\mid k}^{\boldsymbol{\xi}}\hat{\bs{A}}_k^{\T}\left(\bs{P}_{k+1\mid k}^{*}\right)^{-1}\\
\text{where } &\hat{\bs{B}}_k\triangleq \left[\bs{I}_n,-\bs{B}_k\right] \text{and }\hat{\bs{A}}_k\triangleq\left[\bs{A}_k,\bs{0}\right].
\end{align}

\begin{proof}
Following similar steps as in \eqref{smooth1}-\eqref{joint_pdf_thm3}, we get
\begin{multline}
p\left(\boldsymbol{\xi}_{k+1}, \boldsymbol{\xi}_{k} \mid\bs{Y}_1^{K}\right)=\\ \frac{p\left(\bs{x}_{k+1} \mid \bs{x}_{k},\bs{u}_{k}\right)p\left(\boldsymbol{\xi}_{k} \mid\bs{Y}_1^{k}\right)}{p\left(\bs{x}_{k+1} \mid\bs{Y}_1^{k}\right)} p\left(\boldsymbol{\xi}_{k+1} \mid\bs{Y}_1^{K}\right),
\end{multline}
\begin{align}
\log p\left(\bs{x}_{k+1} \mid \bs{x}_{k},\bs{u}_{k}\right)&=-\frac{1}{2} \Vert\bs{x}_{k+1}-\bs{A}_k\bs{x}_k-\bs{B}_k\bs{u}_k\Vert_{\bs{Q}_k}^2\nonumber\\
&=-\frac{1}{2} \Vert\hat{\bs{B}}_k\boldsymbol{\xi}_{k+1}-\hat{\bs{A}}_k\boldsymbol{\xi}_k\Vert_{\bs{Q}_k}^2\label{thm3_pdf1}
\end{align}
where $\hat{\bs{B}}_k\triangleq \left[\bs{I}_n,-\bs{B}_k\right]$ and $\hat{\bs{A}}_k\triangleq\left[\bs{A}_k,\bs{0}\right]$. All the other log pdf terms remain the same as in \eqref{pdf2}-\eqref{pdf4}. Thus,
\begin{align}
&\log p\big(\boldsymbol{\xi}_{k+1}, \boldsymbol{\xi}_{k} \mid\bs{Y}_1^{K}\big)\nonumber\\
&=-\frac{1}{2} \boldsymbol{\xi}_{k+1}^{\T}\Big(\hat{\bs{B}}_k^{\T}\bs{Q}_k^{-1}\hat{\bs{B}}_k-\bs{T}^{\T}\left(\bs{P}_{k+1\mid k}^{\bs{x}}\right)^{-1}\bs{T}\nonumber\\
&\hspace{5cm}+\left(\bs{P}_{k+1\mid K}^{\boldsymbol{\xi}}\right)^{-1}\Big) \boldsymbol{\xi}_{k+1} \nonumber\\
&-\frac{1}{2} \boldsymbol{\xi}_{k+1}^{\T}\left(-\hat{\bs{B}}_k^{\T}\bs{Q}_k^{-1}\hat{\bs{A}}_k\right) \boldsymbol{\xi}_k-\frac{1}{2} \boldsymbol{\xi}_k^{\T}\left(-\hat{\bs{A}}_k^{\T} \bs{Q}_k^{-1}\hat{\bs{B}}_k\right) \boldsymbol{\xi}_{k+1}\nonumber\\
&-\frac{1}{2} \boldsymbol{\xi}_{k}^{\T}\left(\hat{\bs{A}}_k^{\T} \bs{Q}_k^{-1} \hat{\bs{A}}_k+\left(\bs{P}_{k\mid k}^{\boldsymbol{\xi}}\right)^{-1}\right) \boldsymbol{\xi}_k+\boldsymbol{\xi}_{k}^{\T}\left(\left(\bs{P}_{k\mid k}^{\boldsymbol{\xi}}\right)^{-1} \hat{\boldsymbol{\xi}}_{k\mid k}\right)\nonumber\\
&\hspace{7cm}+\ldots
\end{align}
The rest of the derivation is similar to that in Appendix \ref{sec:smoothing_proof}, hence we omit them.
\end{proof}
\subsection{Sparse Recovery Algorithms for Noisy Compressive State Measurements}
We can extend the regularized-RKS algorithms for measurements of the form \eqref{eq:meas_thm3} using the filtering and smoothing steps described in this section, similar to the procedure in~\Cref{sec:reg_RKS}. We omit the expressions here for brevity.

For extending the Bayesian-RKS, we need to modify the filtering step by incorporating prior distribution of $\bs{u}_k$. While evaluating \eqref{prediction_them3} we ignored the term $p(\bs{u}_{k-1})$. However, in SBL-RKS, we impose the following prior distribution on the inputs:
\begin{align}
\bs{u}_k\sim \mathcal{N}\left(\bs{0},\bs{P}^{u}_k\right),\,\,\text{where, }\bs{P}^{u}_k\triangleq\operatorname{Diag}\{\bs{\gamma}_k\}.\label{thm3_sbl}
\end{align}
The optimization problem in \eqref{xu_min_filtering_thm3} then modifies as
\begin{align}
\Big\{\hat{\bs{x}}_{k\mid k},\hat{\bs{u}}_{k-1\mid k}\Big\}&=\underset{\bs{x}_k,\bs{u}_k}{\arg \min }\hspace{1mm}\Vert\bs{y}_k-\bs{C}_k\bs{x}_k\Vert_{\bs{R}_k}^2+\Vert\bs{u}_{k-1}\Vert^2_{\bs{P}^{u}_{k-1}}\nonumber\\&\hspace{-2cm}+\Vert\bs{x}_k-\bs{A}_{k-1}\hat{\bs{x}}_{k-1\mid k-1}-\bs{B}_{k-1}\bs{u}_{k-1}\Vert_{\bs{P}_{k\mid k-1}^{*}}^2.
\end{align}
Hence, the expression for $\bs{J}_k$ in \eqref{J_k_thm3} modifies to
\begin{align}
\bs{J}_k\triangleq\big(\bs{B}_{k-1}^{\T}&\bs{P}^{*}_{k\mid k-1}\bs{B}_{k-1}+(\bs{P}^{u}_{k-1})^{-1}\big)^{-1}\bs{B}_{k-1}^{\T}(\bs{P}^{*}_{k\mid k-1})^{-1}
\end{align}
Hence, the \emph{mean} of the update errors in \eqref{ex_thm3} and \eqref{eu_thm3} can be obtained as
\begin{align}
&\mathbb{E}(\bs{e}^{\bs{x}}_{k\mid k})=(\bs{I}_n-\bs{F}_k\bs{C}_k)\bs{A}_{k-1}\mathbb{E}(\bs{e}^{\bs{x}}_{k-1\mid k-1})\\
&\mathbb{E}(\bs{e}^{\bs{u}}_{k-1\mid k})=-\bs{M}_k\bs{C}_k\bs{A}_{k-1}\mathbb{E}(\bs{e}^{\bs{x}}_{k-1\mid k-1}),
\end{align}
where we used $\mathbb{E}(\bs{u}_{k-1})=\bs{0}$ from \eqref{thm3_sbl} even though $(\bs{I}_n-\bs{F}_k\bs{C}_k)\bs{B}_{k-1}\neq \bs{0}$ and $(\bs{I}_m-\bs{M}_k\bs{C}_k\bs{B}_{k-1})\neq \bs{0}$ (since $\bs{J}_k\bs{B}_{k-1}\neq\bs{I}_m$).
Hence $\mathbb{E}(\bs{e}^{\bs{x}}_{1\mid 1})=\bs{0}$ makes all $\mathbb{E}(\bs{e}^{\bs{x}}_{k\mid k})=\bs{0}$ which implies that the estimates are unbiased, i.e., $\mathbb{E}(\hat{\bs{x}}_{k\mid k})=\bs{x}_k~\forall~k$. As a consequence, we can obtain the covariance update as
\begin{align}
\bs{P}^{\bs{\xi}}_{k\mid k}=\bs{Z}_k\begin{bmatrix}
\bs{P}^{\bs{x}}_{k-1\mid k-1} & \bs{0}\\ \bs{0} & \bs{P}^{\bs{u}}_{k-1}
\end{bmatrix}\bs{Z}_k^{\T}+\bs{N}_k\begin{bmatrix}
\bs{Q}_{k-1} & \bs{0}\\ \bs{0} & \bs{R}_{k}
\end{bmatrix}\bs{N}_k^{\T}
\end{align}
\begin{align}
&\hspace{-0.5cm}\text{where  }\bs{Z}_k\triangleq\begin{bmatrix}
(\bs{I}_n-\bs{F}_k\bs{C}_k)\bs{A}_{k-1} & (\bs{I}_n-\bs{F}_k\bs{C}_k)\bs{B}_{k-1}\\
-\bs{M}_k\bs{C}_k\bs{A}_{k-1} & (\bs{I}_m-\bs{M}_k\bs{C}_k\bs{B}_{k-1})
\end{bmatrix}\\
&\text{and }\bs{N}_k\triangleq \begin{bmatrix}
(\bs{I}_n-\bs{F}_k\bs{C}_k) & -\bs{F}_k\\
-\bs{M}_k\bs{C}_k & -\bs{M}_k
\end{bmatrix}.
\end{align}
All the other filtering and smoothing steps remain unchanged. We summarize the SBL based algorithm for the measurement model \eqref{eq:meas_thm3} in \Cref{alg:SBL_RKS_thm3}.
\begin{algorithm}[t]
	\caption{SBL-RKS for Noisy Compressive State Measurements}
	\begin{algorithmic}[1]
		\REQUIRE $\{\bs{y}_k, \bs{A}_k,\bs{B}_k,\bs{C}_k,\bs{D}_k,\bs{Q}_k,\bs{R}_k\}_{k=1}^K$
		\STATEx \hspace{-0.5cm}\textbf{Parameters:} $r_{\max}$
		\STATEx \hspace{-0.5cm}\textbf{Initialization:} $\bs{\gamma}_k^{(0)}=\bs{1}$ for $k=1,2,\ldots,K$, $r=1$
		\STATE $\hat{\bs{A}}_k=\begin{bmatrix}
\bs{A}_k & \bs{0}\end{bmatrix}\in\bb{R}^{n\times (n+m)}$, $\hat{\bs{B}}_k=\begin{bmatrix}
\bs{I} & -\bs{B}_k\end{bmatrix}$
		\FOR{$r=1,2,\ldots,r_{\max}$}
		\STATEx \#\emph{E-Step:}
		\STATE $\hat{\bs{\xi}}_{0\mid 0}=\bs{0}$, $\bs{P}_{0\mid 0}^{\bs{\xi}}=\bs{I}$
\STATEx \#\emph{Filtering:}
\FOR{$k=1,2,\ldots,K$}
\STATE $\bs{P}^{u}_{k-1}=\operatorname{Diag}\lc\bs{\gamma}_{k-1}^{(r-1)}\rc$
\STATE $\bs{P}^{*}_{k\mid k-1}= \bs{A}_{k-1}\bs{P}_{k-1\mid k-1}^{\bs{x}}\bs{A}_{k-1}^{\T}+\bs{Q}_{k-1}$
\STATE \hspace{-5mm}$\bs{J}_k\!\!=\!\!\big(\bs{B}_{k-1}^{\T}\!\bs{P}^{*}_{k\mid k-1}\!\bs{B}_{k-1}\!\!+\!\!(\bs{P}^{u}_{k-1})^{-1}\big)^{-1}\!\!\bs{B}_{k-1}^{\T}(\bs{P}^{*}_{k\mid k-1})^{-1}$
\STATE $\bs{L}_k=\bs{P}_{k\mid k-1}^{*}\bs{C}_k^{\T}\big(\bs{R}_k+\bs{C}_k\bs{P}_{k\mid k-1}^{*}\bs{C}_k^{\T}\big)^{-1}$
\STATE $\bs{F}_k=(\bs{I}_n-(\bs{I}_n-\bs{L}_k\bs{C}_k)\bs{B}_{k-1}\bs{J}_k)^{-1}\bs{L}_k$
\STATE $\bs{M}_k=(\bs{I}_m-\bs{J}_k\big(\bs{I}_n-\bs{L}_k\bs{C}_k\big)\bs{B}_{k-1})^{-1}\bs{J}_k\bs{L}_k$
\STATE $\bs{Z}_k=\begin{bmatrix}
(\bs{I}_n-\bs{F}_k\bs{C}_k)\bs{A}_{k-1} & (\bs{I}_n-\bs{F}_k\bs{C}_k)\bs{B}_{k-1}\\
-\bs{M}_k\bs{C}_k\bs{A}_{k-1} & (\bs{I}_m-\bs{M}_k\bs{C}_k\bs{B}_{k-1})
\end{bmatrix}$
\STATE $\bs{N}_k= \begin{bmatrix}
(\bs{I}_n-\bs{F}_k\bs{C}_k) & -\bs{F}_k\\
-\bs{M}_k\bs{C}_k & -\bs{M}_k
\end{bmatrix}$
\STATE \hspace{-7mm}$\bs{P}^{\bs{\xi}}_{k\mid k}\hspace{-1mm}=\hspace{-1mm}\bs{Z}_k\hspace{-1mm}\begin{bmatrix}
\bs{P}^{\bs{x}}_{k-1\mid k-1} & \hspace{-3mm}\bs{0}\\ \bs{0} &\bs{P}^{u}_{k-1}
\end{bmatrix}\hspace{-1mm}\bs{Z}_k^{\T}\hspace{-1mm}+\hspace{-1mm}\bs{N}_k\hspace{-1mm}\begin{bmatrix}
\bs{Q}_{k-1} & \hspace{-3mm}\bs{0}\\ \bs{0} & \hspace{-3mm}\bs{R}_{k}
\end{bmatrix}\hspace{-1mm}\bs{N}_k^{\T}$
\STATE $\bs{\xi}_{k\mid k}=\begin{bmatrix}
\bs{I}_p-\bs{F}_k\bs{C}_k \\ -\bs{M}_k\bs{C}_k
\end{bmatrix}\bs{A}_{k-1}\hat{\bs{x}}_{k-1\mid k-1}+\begin{bmatrix}\bs{F}_k\\\bs{M}_k\end{bmatrix}\bs{y}_k$
\ENDFOR
		\STATEx \#\emph{Smoothing:}
		\FOR{$k=K-1,K-2,\ldots,1$}
		\STATE $\bs{K}_k=\bs{P}_{k\mid k}^{\boldsymbol{\xi}}\hat{\bs{A}}_k^{\T}\left(\bs{P}_{k+1\mid k}^{*}\right)^{-1}$
		\STATE $\bs{P}_{k\mid K}^{\bs{\xi}}=\bs{P}_{k\mid k}^{\boldsymbol{\xi}}+\bs{K}_k\left(\hat{\bs{B}}_k\bs{P}_{k+1\mid K}^{\bs{\xi}}\hat{\bs{B}}_k^{\T}-\bs{P}_{k+1\mid k}^{*}\right)\bs{K}_k^{\T}$
        \STATE $\hat{\bs{\xi}}_{k\mid K}=\hat{\bs{\xi}}_{k\mid k}+\bs{K}_k\lb\hat{\bs{B}}_k\hat{\bs{\xi}}_{k+1\mid K}-\hat{\bs{A}}_k\hat{\bs{\xi}}_{k\mid k}\rb$
        \STATE $\bs{P}_{k+1,k\mid K}^{\bs{\xi}}=\bs{P}_{k+1\mid K}^{\boldsymbol{\xi}}\hat{\bs{B}}_k^{\T}\bs{K}_k^{\T}$
        \STATE Compute $\hat{\bs{u}}_{k\mid K}$ and $\bs{P}_{k\mid K}^{\bs{u}}$ from $\hat{\bs{\xi}}_{k\mid K}$ and $\bs{P}_{k\mid K}^{\bs{\xi}}$
		\ENDFOR
		\STATEx \#\emph{M-step:}
        \STATE $\bs{\gamma}_k^{(r)}=\operatorname{Diag}\lc\hat{\bs{u}}_{k\mid K}\hat{\bs{u}}_{k\mid K}^{\T}+\bs{P}^{\bs{u}}_{k\mid K}\rc$, for $k=1,2,\ldots,K-1$
		\ENDFOR
		\STATE Compute $\lc \hat{\bs{x}}_{k\mid K},\hat{\bs{u}}_{k-1\mid K}\rc_{k=1}^K$ from $\hat{\bs{\xi}}_{k\mid K}$
		\ENSURE $\{\hat{\bs{x}}_{k \mid K}\}_{k=1}^K$ and $\{\hat{\bs{u}}_{k \mid K}\}_{k=1}^{K-1}$
	\end{algorithmic}
	\label{alg:SBL_RKS_thm3}
\end{algorithm}
\section{Derivation of BP-RKS and Group BP-RKS} \label{app:BP-RKS and GBP-RKS}
In this section, we derive two algorithms, BP-RKS for the time-varying support case, and its extension, group BP-RKS for the joint sparsity case~\cite{sefati2015linear}. To derive BP-RKS, we first consider the problem of estimating the initial state $\bs{x}_{1}$ and inputs $\bs{U}_1^K$ for the linear system defined in \eqref{eq:state_eqn}-\eqref{eq:meas_eqn}, which can be written as 
\begin{equation}
\tilde{\bs{y}}_{K}=\bs{O}_{K}\bs{x}_{1}+\bs{\Gamma}_{K}\tilde{\bs{u}}_{K}+\bs{M}_{K}\tilde{\bs{w}}_{K-1}+\tilde{\bs{v}}_{K},\label{eq:y_vec}
\end{equation}
where $\tilde{\bs{y}}_{K}=\begin{bmatrix}
    \bs{y}_{1}^{\T} & \bs{y}_{2}^{\T} &\ldots& \bs{y}_{K}^{\T}
\end{bmatrix}^{T}\in\mathbb{R}^{Kp}$ denotes the concatenated measurement vector. Likewise, $\tilde{\bs{u}}_{K}\in \mathbb{R}^{Km}, \tilde{\bs{w}}_{K-1}\in\mathbb{R}^{(K-1)n}$, and $ \tilde{\bs{v}}_{K}\in \mathbb{R}^{Kp}$ are obtained by concatenating the inputs, process noise terms and measurement noise terms, respectively. The system matrices, $\bs{O}_{K}\in\mathbb{R}^{Kp\times n}$, $\bs{\Gamma}_{K}\in \mathbb{R}^{Kp\times Km}$ and $\bs{M}_K\in\mathbb{R}^{Kp\times(K-1)n}$ in \eqref{eq:y_vec} are given by
\begin{align}
\bs{O}_{K}&=\begin{bmatrix}
\bs{C}\\
\bs{C}\bs{A}\\
\vdots\\
\bs{C}\bs{A}^{K-1}
\end{bmatrix}\label{eq:Observability_matrx},\;
\bs{M}_K=\begin{bmatrix}\bs{0}&\bs{0}&\cdots&\bs{0}\cr \bs{C}&\bs{0}&\cdots&\bs{0}\cr \vdots&\vdots&\ddots&\vdots\cr \bs{C}\bs{A}^{K-2}& \bs{C}\bs{A}^{K-3}&\cdots&\bs{C}\end{bmatrix}
\\
\bs{\Gamma}_{K}&=\begin{bmatrix}
\bs{D} & \bs{0} & \bs{0} & \cdots & \bs{0}\\
\bs{C}\bs{B} & \bs{D} & \bs{0} & \cdots & \bs{0}\\
\bs{C}\bs{A}\bs{B} & \bs{C}\bs{B} & \bs{D} & \cdots & \bs{0}\\
\vdots & \vdots & \ddots & \\
\bs{C}\bs{A}^{K-2}\bs{B} & \bs{C}\bs{A}^{K-3}\bs{B}& \bs{C}\bs{A}^{K-4}\bs{B} & \cdots & \bs{D}\label{eq:Gamma_matrix}
\end{bmatrix}.
\end{align}
For notational simplicity, we present \eqref{eq:Observability_matrx} and \eqref{eq:Gamma_matrix} for the constant system matrices case, i.e.,  $\bs{A}_k=\bs{A}$, $\bs{B}_k=\bs{B}$, $\bs{D}_k=\bs{C}$, and $\bs{D}_k=\bs{D}$,  $k=1,2,\ldots,K$. However, the extension to time-varying matrices is straightforward. 

Similar to the approach in~\cite{sefati2015linear}, we first eliminate the initial state term in \eqref{eq:y_vec} by multiplying it with $\bs{\Pi}$, the projection matrix onto the orthogonal complement of the column space of the observability matrix $\bs{O}_K$, to get
\begin{equation}
\bs{\Pi} \tilde{\bs{y}}_K=\bs{\Pi}\bs{\Gamma}_{K}\tilde{\bs{u}}_K+\bs{\Pi}\tilde{\bs{n}}_K,\label{eq:y_pi}
\end{equation}
where the noise term $\tilde{\bs{n}}_K = \bs{M}_{K}\tilde{\bs{w}}_{K-1}+\tilde{\bs{v}}_{K}$ and the projection matrix is 
\begin{equation}
\bs{\Pi}=\bs{I}-\bs{O}_{K}(\bs{O}_{K}^{T}\bs{O}_{K})^{-1}\bs{O}_{K}^{T}.\label{eq:Pi_matrix}
\end{equation}
Now we estimate the states and inputs in two steps. We first solve for inputs $\tilde{\bs{u}}_{K}$ using \eqref{eq:y_pi}. Substituting the estimate $\tilde{\bs{u}}^*_{K}$ into \eqref{eq:y_vec}, we then compute the weighted least square estimate $\bs{x}^*_1$  as 
\begin{equation}
\bs{x}^*_{1}=(\bs{O}_{K}^{\T}\bs{Q}_{\tilde{n}}^{-1}\bs{O}_{K})^{-1}\bs{O}_{K}^{\T}\bs{Q}_{\tilde{n}}^{-1}(\tilde{\bs{y}}_{K}-\bs{\Gamma}_{K}\tilde{\bs{u}}^*_{K}).\label{eq:x_1*}
\end{equation}
Here, $\tilde{\bs{Q}}_{K}$ is the covariance of the noise term $\tilde{\bs{n}}$ in \eqref{eq:y_vec} given by 
\begin{equation}\label{eq:tildeQ}
    \tilde{\bs{Q}}_{K}=\bs{M}_K\operatorname{Blkdiag}(\bs{Q}_1,\ldots,\bs{Q}_K)\bs{M}_K^{\T}+\operatorname{Blkdiag}(\bs{R}_1,\ldots,\bs{R}_K),
\end{equation}
with the operator $\operatorname{Blkdiag}(\cdot)$ denoting a block diagonal matrix with its arguments as its block diagonal entries. Having estimated the initial state and inputs, the other states $\bs{X}_2^K$ can be reconstructed using the Kalman smoothing algorithm.

To estimate the sparse inputs $\tilde{\bs{u}}_{K}$ from \eqref{eq:y_pi}, we note that the project step can potentially make $\bs{\Pi}\bs{\Gamma}_{K}$ rank deficient, and we further reduce the system in \eqref{eq:y_pi} to get linearly independent measurements. Denote the singular value decomposition of the matrix $\bs{\Pi}\bs{\Gamma}_{K}$ by
\begin{equation}
\bs{\Pi}\bs{\Gamma}_{K}=\begin{bmatrix}
    \bs{\Psi}_1\\
    \bs{\Psi}_2
\end{bmatrix}^{\mathsf{H}}\begin{bmatrix}
\bs{\Lambda}\in{\bb{R}}^{R\times R} & \bs{0}\\\bs{0} & \bs{0}\end{bmatrix}\begin{bmatrix}\bs{\Phi}_1\\\bs{\Phi}_2\end{bmatrix}=\bs{\Psi}_1^{\mathsf{H}}\bs{\Lambda}\bs{\Phi}_1,\label{eq:Gamma_Pi}
\end{equation}
where $\bs{\Psi}=\begin{bmatrix}
    \bs{\Psi}_1^{\mathsf{H}}&
    \bs{\Psi}_2^{\mathsf{H}}
\end{bmatrix}^{\mathsf{H}}$ is an orthonormal matrix and $R$ is the rank of $\bs{\Pi}\bs{\Gamma}_{K}$. When the system matrices have full rank, we have $R=\min\{Kp-n,Km\}$. From \eqref{eq:y_pi}, we derive 
\begin{equation}
\bs{\Psi}\bs{\Pi} \tilde{\bs{y}}_K = \begin{bmatrix}
 \bs{\Psi}_1\bs{\Pi} \tilde{\bs{y}}_K \\
 \bs{\Psi}_2\bs{\Pi} \tilde{\bs{y}}_K 
\end{bmatrix}=\begin{bmatrix}\bs{\Lambda}\bs{\Phi}_1\\\bs{0}\end{bmatrix}\tilde{\bs{u}}_K+\begin{bmatrix}
 \bs{\Psi}_1\bs{\Pi} \tilde{\bs{n}}_K \\
 \bs{\Psi}_2\bs{\Pi} \tilde{\bs{n}}_K 
\end{bmatrix}.
\end{equation}
Hence, the reduced system of equations is
\begin{equation}\label{eq:reduced_mes}
    \bs{\Psi}_1\bs{\Pi} \tilde{\bs{y}}_K  = \bs{\Psi}_1\bs{\Pi}\bs{\Gamma}_{K}\tilde{\bs{u}}_K+\bs{\Psi}_1\bs{\Pi} \tilde{\bs{n}}_K,
\end{equation}
\begin{algorithm}[ht]
	\caption{Basis Pursuit Robust Kalman Filtering}
	\setstretch{1.2}
	\begin{algorithmic}[1]
		\REQUIRE $\{\bs{y}_k, \bs{A}_k,\bs{B}_k,\bs{C}_k,\bs{D}_k,\bs{Q}_k,\bs{R}_k\}_{k=1}^K$
		\STATE Compute $\tilde{\bs{y}}_K,\bs{O}_K$, $\bs{M}_K$, $\bs{\Gamma}_K$ from \eqref{eq:y_vec}, and $\bs{\Pi}$ using \eqref{eq:Pi_matrix}
        \STATE Determine $\bs{\Phi}_1$ using the singular value decomposition of $\bs{\Pi}\bs{\Gamma}_{K}$ as given in \eqref{eq:Gamma_Pi}
        \STATE Compute $\bar{\bs{Q}}_K$ using \eqref{eq:tildeQ} and \eqref{eq:Q_n_1}
        \STATE Compute $\bar{\bs{y}}_K$ and $\bar{\bs{\Gamma}}_K$ from \eqref{eq:W}
        \STATE Set parameter $\epsilon$ according to \eqref{eq:epsilon}
        \STATE Solve for inputs $\bs{U}_1^K$ using the convex optimization problem \eqref{eq:BPQC}
        \STATE Calculate the initial state estimate $\bs{x}_1$ using \eqref{eq:x_1*}
        \STATE Set $\bs{x}_{1|K}=\bs{x}_{1|1}=\bs{x}_1$
        \STATEx \emph{Kalman Smoother }
                                                \FOR{$k=2,\ldots,K$}
 	\STATEx \#\emph{Prediction:}
		\STATE $\hat{\bs{x}}_{k\mid k-1}=\bs{A}_{k-1}\hat{\bs{x}}_{k-1\mid k-1}{+\bs{B}_k\bs{u}_{k-1}}$
		\STATE $\bs{P}^{\bs{x}}_{k\mid k-1}=\bs{A}_{k-1}
\bs{P}^{\bs{x}}_{k-1\mid k-1}\bs{A}_{k-1}^{\T}+\bs{Q}_{k-1}$
		\STATEx \#\emph{Filtering:}
		\STATE $\bs{G}_{k}=\bs{P}_{k \mid k-1}^{\bs{x}} \bs{C}_k^{\T}\lb\bs{R}_k+\bs{C}_k \bs{P}_{k \mid k-1}^{\bs{x}}\bs{C}_k^{\T}\rb^{-1}$
		\STATE $\hat{\bs{x}}_{k \mid k}=\hat{\bs{x}}_{k \mid k-1}+\bs{G}_{k}\lb\bs{y}_k{-\bs{D}_k\bs{u}_k}-\bs{C}_k \hat{\bs{x}}_{k \mid k-1}\rb$
		\STATE $\bs{P}_{k \mid k}^{\bs{x}}=\lb\bs{I}-\bs{G}_{k}\bs{C}_k\rb \bs{P}_{k \mid k-1}^{\bs{x}}$
		\ENDFOR
		\STATEx \#\emph{Smoothing:}
		\FOR{$k=K-1,K-2,\ldots,2$}
		\STATE $\bs{K}_k=\bs{P}_{k\mid k}^{\bs{x}}\bs{A}_{k}^{\T}\lb\bs{P}_{k+1\mid k}^{\bs{x}}\rb^{-1}$
		\STATE $\bs{P}_{k\mid K}^{\bs{x}}=\bs{P}_{k\mid k}^{\bs{x}}+\bs{K}_k\lb\bs{P}_{k+1\mid K}^{\bs{x}}-\bs{P}_{k+1\mid k}^{\bs{x}}\rb\bs{K}_k^{\T}$
        \STATE\vspace{-8mm}\begin{multline}
        \hat{\bs{x}}_{k\mid K}=\hat{\bs{x}}_{k\mid k}+\bs{K}_k\lb\hat{\bs{x}}_{k+1\mid K}-\bs{A}_k\hat{\bs{x}}_{k\mid k}\rb\\
        {-\bs{P}_{k\mid k}^{\bs{x}}(\bs{I}-\bs{K}_k\bs{A}_k)\bs{A}_k^{\T}\bs{Q}_k^{-1}\bs{B}_k\bs{u}_k}\nonumber
        \end{multline}
        \ENDFOR
		\ENSURE $\lc\hat{\bs{x}}_{k\mid K}\rc_{k=1}^K$ and $\lc\bs{u}_{k}\rc_{k=1}^K$
	\end{algorithmic}
	\label{alg:RKS-BP}
\end{algorithm}
where the covariance of the noise component $\bs{\Psi}_1\bs{\Pi} \tilde{\bs{n}}_K $ is
\begin{equation}
\bar{\bs{Q}}_{K}=\bs{\Psi}_1\bs{\Pi}\tilde{\bs{Q}}_{K}\bs{\Pi}^{\T}\bs{\Psi}_1^{\mathsf{H}}.\label{eq:Q_n_1}
\end{equation}
Finally, we multiply the reduced measurements in \eqref{eq:reduced_mes} with the prewhitening matrix $\bar{\bs{Q}}_{K}^{-\frac{1}{2}}$ to make the noise uncorrelated. Hence, the new system of equations is
\begin{equation}
\bar{\bs{y}}_K = \bar{\bs{Q}}_{K}^{-\frac{1}{2}}\bs{\Psi}_1\bs{\Pi} \tilde{\bs{y}}_K = \bar{\bs{\Gamma}}_K\tilde{\bs{u}}_K+\bar{\bs{n}}_K,\label{eq:W}
\end{equation}
where the measurement matrix $\bar{\bs{\Gamma}}_K = \bar{\bs{Q}}_{K}^{-\frac{1}{2}} \bs{\Psi}_1\bs{\Pi}\bs{\Gamma}_{K}$ and the noise term 
$\bar{\bs{n}}_K= \bar{\bs{Q}}_{K}^{-\frac{1}{2}}\bs{\Psi}_1\bs{\Pi} \tilde{\bs{n}}_K$ follows the standard Gaussian distribution $\mathcal{N}(\bs{0},\bs{I})$.
Hence, the optimization problem to solve for the inputs becomes
\begin{equation}
\tilde{\bs{u}}^*_K=\underset{\tilde{\bs{u}}_{K}}{\arg\min}\Vert \tilde{\bs{u}}_{K}\Vert_{1}\text{ s.t. }\Vert\bar{\bs{y}}_K -\bar{\bs{\Gamma}}_K\tilde{\bs{u}}_K\Vert_2\leq \epsilon.\label{eq:BPQC}
\end{equation}
Here, $\epsilon>0$ is an algorithm parameter which is typically chosen as
\begin{equation}
\epsilon =\sqrt{R}\lb 1+2\sqrt{\frac{2}{R}}\rb.\label{eq:epsilon}
\end{equation}

Using the optimal solution $\tilde{\bs{u}}^*_K$ obtained by solving above problem \eqref{eq:BPQC}, we can compute the initial state using \eqref{eq:x_1*} and the other states using the Kalman smoothing algorithm.  
The resulting algorithm, which is referred to as BP-RKS, is summarized in \Cref{alg:RKS-BP}. 

When the control inputs share a common support $\tilde{\bs{u}}_K$, we can rearrange its entries to get a block sparse vector as explained in \Cref{sec:l1_joint_sparse}. Specifically, $\hat{\bs{u}}_K = \begin{bmatrix}
    \bs{{u}}(1,:)^{\T} & \bs{{u}}(2,:)^{\T} &\ldots & \bs{{u}}(m,:)^{\T}
\end{bmatrix}\in\bb{R}^{Km}$ is block sparse with block length $K$, where we use the definition in \eqref{eq:rearrange_block}. The corresponding columns of $\bar{\bs{\Gamma}}_K$ are also rearranged, which we denote as $\hat{\bs{\Gamma}}_K$, leading to the system of equations $\bar{\bs{y}}_K = \hat{\bs{\Gamma}}_K\hat{\bs{u}}_K+\bar{\bs{n}}_K$. We can now exploit the block sparsity structure by imposing an $\ell_1/\ell_2$ type penalty~\cite{eldar2009robust} on $\hat{\bs{u}}_K$ to arrive at the following optimization problem:
\begin{equation}
\hat{\bs{u}}_K^*=\underset{\hat{\bs{u}}_K}{\arg\min}\sum_{i=1}^m\Vert\bs{{u}}(i,:)\Vert_2\text{  s.t. }\Vert\bar{\bs{y}}_K-\hat{\bs{\Gamma}}_K\hat{\bs{u}}_K\Vert^2\leq \epsilon\label{eq:u_vec l1-l2 opt}.
\end{equation}
The problem \eqref{eq:u_vec l1-l2 opt} can be solved by rewriting it as a second-order cone program using an auxiliary variable $\bs{t}$ as given below:
\begin{multline}\label{eq:group_BP_opt}
\min_{\bs{t}\in\bb{R}^m,\hat{\bs{u}}_K\in\bb{R}^{Km}}\sum_{i=1}^m\bs{t}_i \text{\quad s.t.    } \bs{t}_i\geq \Vert\bs{{u}}(i,:)\Vert_2 \text{ for } i=1,2,\ldots, m  \\
\text{ and }
\Vert\bar{\bs{y}}_K-\hat{\bs{\Gamma}}_K\hat{\bs{u}}_K\Vert_2^2\leq \epsilon.
\end{multline}
The resulting algorithm, which is referred to as group BP-RKS, is identical to \Cref{alg:RKS-BP} except that in Step 6 we solve \eqref{eq:group_BP_opt} instead of \eqref{eq:BPQC}.


\bibliographystyle{IEEEtran}
\bibliography{SparseEst}
\begin{IEEEbiography}[{\includegraphics[width=1in,height=1.25in,clip,keepaspectratio]{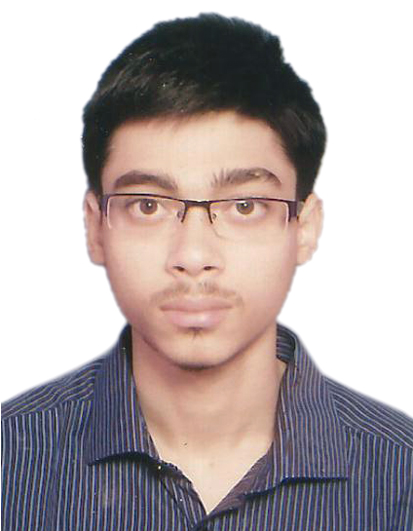}}] {Rupam Kalyan Chakraborty} received the B. Tech. degree in electronics and communication engineering from Institute of Engineering $\And$ Management, Kolkata, India, in 2016. He received M. Tech degree on Microwave Engineering from Institute of Radiophysics and Electronics, University of Calcutta, India and M. Tech degree in signal processing from Department of electrical communication engineering (ECE), Indian Institute of Science (IISc), Bangalore, in 2018 and 2021, respectively. From 2021-2023 he worked as algorithm design engineer in Signalchip Innovations, Bengaluru, on GNSS tracking algorithm development. Currently he is a Ph.D. student in the signal processing systems group at the Delft University of Technology (TU Delft), Netherlands. His research interests include statistical signal processing, adaptive filtering, and compressive sensing. 
\end{IEEEbiography}

\begin{IEEEbiography}[{\includegraphics[width=1in,height=1.25in,clip,keepaspectratio]{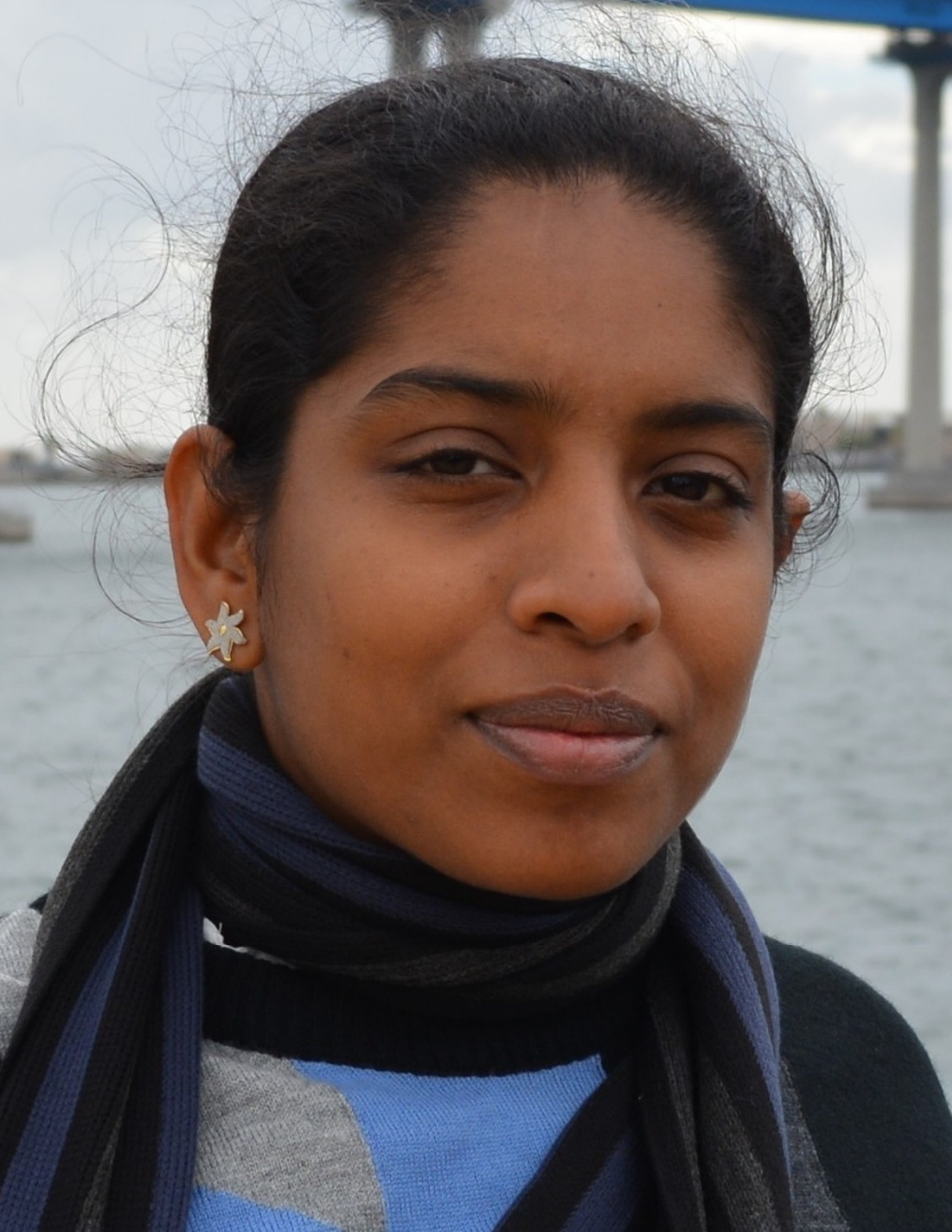}}] {Geethu Joseph} received the B. Tech. degree in electronics and communication engineering from the National Institute of Technology, Calicut, India, in 2011, and the M. E. degree in signal processing and the Ph.D. degree in electrical communication engineering (ECE) from the Indian Institute of Science (IISc), Bangalore, in 2014 and 2019, respectively. She was a postdoctoral fellow with the department of electrical engineering and computer science at Syracuse University, NY, USA, from 2019 to 2021. She is currently an assistant professor in the signal processing systems group at the Delft University of Technology, Delft, Netherlands. She was awarded the 2022 IEEE SPS best PhD dissertation award and the 2020 SPCOM best doctoral dissertation award.  She also secured the medals for the best Ph.D. thesis and the best M. E. (signal processing) of the ECE dept. at IISc for 2019-'20 and 2014, respectively. She is an associate  editor of the IEEE Sensors Journal.  Her research interests include statistical signal processing, network control, and machine learning. 
\end{IEEEbiography}

\vspace{-1cm}
\begin{IEEEbiography}[{\includegraphics[width=1in,height=1.25in,clip,keepaspectratio]{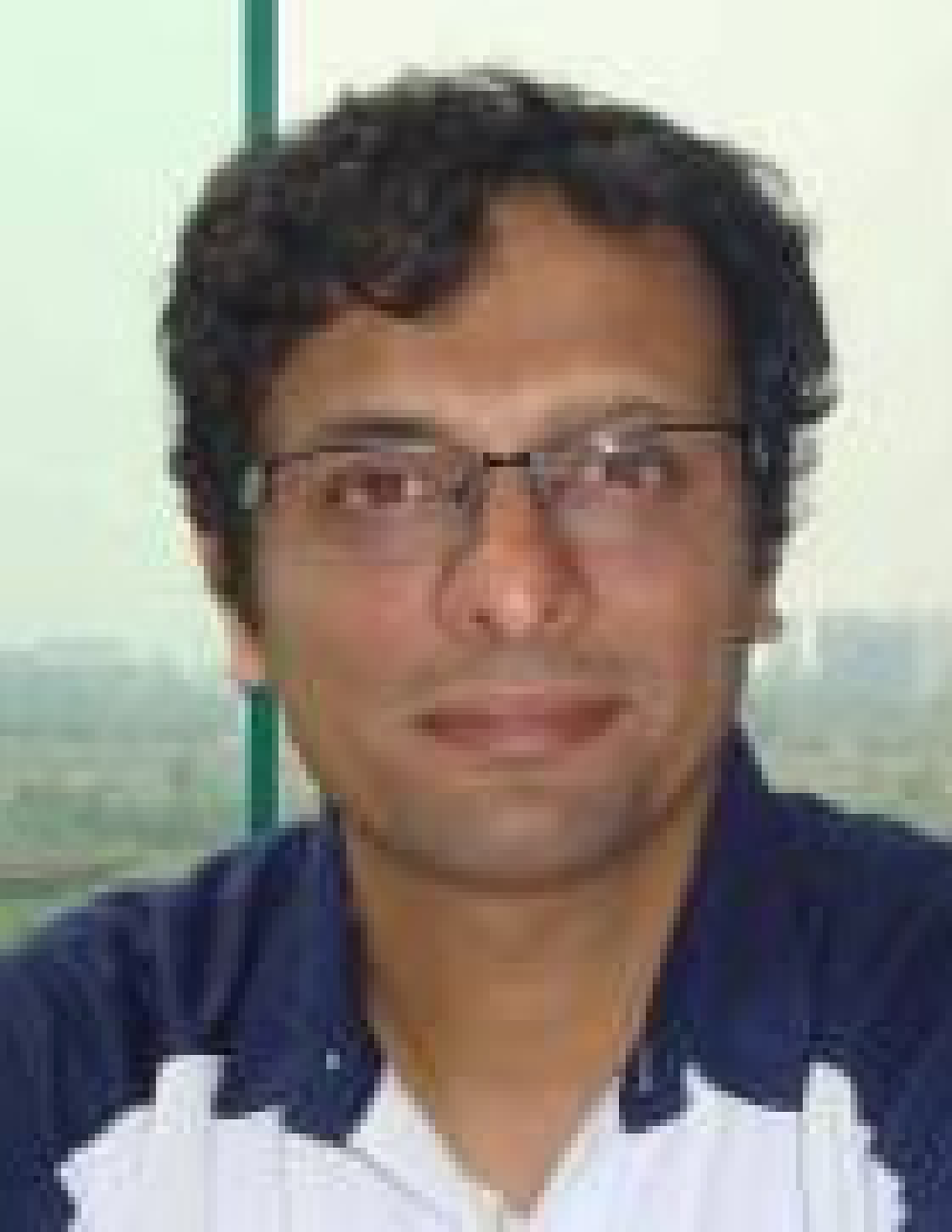}}]{Chandra R. Murthy} (S'03--M'06--SM'11--F'23) received the B.\ Tech. degree in Electrical Engineering from the Indian Institute of Technology, Madras in 1998, the M.\ S.\ and Ph.\ D.\ degrees in Electrical and Computer Engineering from Purdue University and the University of California, San Diego, in 2000 and 2006, respectively. From 2000 to 2002, he worked as an engineer for Qualcomm Inc., where he worked on WCDMA baseband transceiver design and 802.11b baseband receivers. From Aug. 2006 to Aug. 2007, he worked as a staff engineer at Beceem Communications Inc.\ on advanced receiver architectures for the 802.16e Mobile WiMAX standard. In Sept.\ 2007, he joined the Department of Electrical Communication Engineering at the Indian Institute of Science, Bangalore, India, where he is currently working as a Professor.

His research interests are in the areas of energy harvesting communications, 5G/6G technologies and compressed sensing. He has over 90 journal and 110 conference papers to his credit. He is a recipient of the MeitY Young Faculty Fellowship from the Govt.\ of India and the Prof.\ Satish Dhawan state award for engineering from the Karnataka State Government. Currently, he is serving as an area editor of the \textsc{IEEE Transactions on Information Theory}. He is a fellow of the IEEE.
\end{IEEEbiography}

\end{document}